\documentclass[12pt]{article}

\usepackage{graphicx}

\title{Observations and perspectives on the prebiotic sequence evolution} 
\author{Dirson Jian Li \footnote{E-Mail: dirson@mail.xjtu.edu.cn}\\ \normalsize{\it \small Department of Applied Physics, School of Science, Xi'an Jiaotong University, Xi'an 710049, China} }

\date{}
\begin{document}
\maketitle
\sloppy
\setcounter{section}{-1}
\baselineskip18pt
\setlength{\parskip}{12pt}

{\bf \begin{center}{Abstract}\end{center}
The post-genomic era has brought opportunities to bridge traditionally separate fields of early history of life and brought new insight into origin and evolution of biodiversity. According to distributions of codons in genome sequences, I found a relationship between the genetic code and the tree of life. This remote and profound relationship involves the origin and evolution of the genetic code and the diversification and expansion of genomes. Here, a prebiotic picture of the triplex nucleic acid evolution is proposed to explain the origin of the genetic code, where the transition from disorder to order in the origin of life might be due to the increasing stabilities of triplex base pairs. The codon degeneracy can be obtained in detail based on the coevolution of the genetic code with amino acids, or equivalently, the coevolution of tRNAs with aaRSs. This theory is based on experimental data such as the stability of triplex base pairs and the statistical features of genomic codon distributions. Several experimentally testable proposals have been developed. This study should be regarded as an exploratory attempt to reveal the early evolution of life based on sequence information in a statistical manner.} 


\section{Overview} 

\subsection{Motivation} 

There are large amounts of evolutionary information in genome sequences of contemporary species. Statistical analysis of codon distributions in genome sequences provides substantial evidence that there is a close relationship between the evolution of the genetic code and the tree of life, which means that rich primordial information on the origin of life is still stored in contemporary genomic data. Such an amazing and heuristic remote relationship gives us a chance to guess and verify the picture of prebiotic evolution in our post-genomic era. 

\subsection{Theoretical framework} 

The random movement of matter forms numerous phenomena at different scales in nature. Life is a complex phenomenon intervened between microcosmic and macroscopic systems, and originated from adaptation between microcosmic and macroscopic phenomena. The long-term evolution of life will continue, the biodiversity will develop, but the inner most unity of life will remain. The complexity of life can be studied at three levels: (I) at the molecular level, (II) at the sequence level, and (III) at the species level. In this series of articles, I seek rules at each of the three levels, and try to reveal the profound relationships among them, which are helpful to understand the phenomenon of life. The prebiotic picture of triplex nucleic acid evolution (or the triplex picture for short) is the key new idea in my theory. 

This series of articles consists of three parts. The present article, as the first part, is dedicated to explain the origin of the genetic code by a prebiotic sequence evolution model. The second part describes the universal genome format and the origin of the three domains of life (Li 2018-II). And the third part explains the Phanerozoic biodiversity curve and the adaptation strategy of life (Li 2018-III). In this series of articles, a theoretical framework has been developed in new scenarios, by analysing experimental data and by comparing literatures. Some new concepts and methods are invented for these fundamental problems, such as the picture of triplex nucleic acid evolution, the genetic code evolution roadmap, genomic codon distributions, the primordial translation mechanism, a stochastic model for genome size evolution, etc. I confirmed the close relationships between previously unrelated subjects at different levels based on detailed qualitative and quantitative analyses. A close relationship is confirmed between the genetic code evolution at the molecular level (to see the present article) and the three domain tree of life at the sequence or species level (to see Li 2018-II), based on the whole genome analysis and the genetic code evolution roadmap. Another close relationship is also confirmed between the growth trend of genome size evolution at the sequence level (to see Li 2018-II, III) and the growth trend of the Phanerozoic biodiversity curve at the species level (to see Li 2018-III), based on biological and geological data. Although much evidence and detailed explanations have been presented, the validity of my theory need to be proved by future experiments. Thus this study should only be regarded as a hypothesis. The driving force in the sequence evolution in the triplex picture, the assembly of tRNAs and generation of aaRS and the coevolution between them need to be verified experimentally. The structures of the three parts of this series of articles are similar: There is especially a biological picture section after introduction in each part, in consideration that these evolutionary pictures are crucial for understanding my theory. A nomenclature is listed as follows, the notions in which will be defined in their respective sections. These new concepts are essential for understanding my theory. It is helpful to grasp my theory quickly by perusing the main diagrams such as the genetic code evolution roadmap, the biodiversity space, and the reconstructed biodiversity curve etc. 

\subsection{Nomenclature}

The three levels: at the molecular level, at the sequence level and at the species level\\
Common features of life: the genetic code, the homochirality, the universal genome format\\
Triplex nucleic acid: triplex base pair; duplex DNA: base pair\\
The triplex picture (a picture of triplex nucleic acid evolution)\\
The triplex2duplex picture (a picture of transition from triplex nucleic acids to duplex nucleic acids)\\
The picture of genome evolution in the universal genome format \\
The genetic code evolution roadmap (or the roadmap for short): route 0-3, hierarchy 1-4\\
Initial subset, post-initiation-stage stagnation, expansion via biosynthetic families\\
Coevolution of tRNAs with aaRSs: pair connection, route duality\\
Primordial translation mechanism, origins of coding and non-coding DNA\\
Genomic codon distributions: 3-base fluctuation; distinguishing feature\\
Genome organisation, primordial genome assembly, incomplete subset of codons\\
Biodiversity space: route bias, hierarchy bias, fluctuation amplitude\\
Three domain tree of life, reconstructed tree of life based on the genetic code evolution\\
Sepkosky's biodiversity curve, reconstructed biodiversity curve\\
The trend of Sepkosky's biodiversity curve: trend in genome size evolution\\
The climato-eustatic curve: consensus climatic curve, consensus eustatic curve\\
Declining extinction and origination rates, adaptation strategy of homochiral living system\\

And the abbreviated notations are listed as follows:\\
$\circ$ Corresponding notations ($n=1\ to\ 20$): amino acid $No. n$ $\leftrightarrow$ $aaRSn$ $\leftrightarrow$ tRNA $tn$, $tn'$, $tn^+$, $tn^+{'}$, $tn^-$, $tn^-{'}$, where the amino acids from $No. 1$ to $No. 20$ are $1Gly$, $2Ala$, $3Glu$, $4Asp$, $5Val$, $6Pro$, $7Ser$, $8Leu$, $9Thr$, $10Arg$, $11Cys$, $12Trp$, $13His$, $14Gln$, $15Ile$, $16Met$, $17Phe$, $18Tyr$, $19Asn$, $20Lys$, respectively\\
$\circ$ Triplex DNAs ($D \cdot D*D$): $YR*R$, $YR*Y$ and the inverse triplex DNAs: $yr*r$, $yr*y$, where $Y$, $y$ stands for pyrimidine strands and $R$, $r$ purine strands\\
$\circ$ Triplex DNA$\cdot$DNA*RNA ($D \cdot D*R$): $yr*r_t$, $yr*y_t$, $YR*R_t$, $YR*Y_t$, where two types of tRNAs can be generated by linking the RNA strands $5'y_t + r_t3'$ or $5'R_t + Y_t3'$, and aaRSs can approach tRNAs from major groove side (M) or minor groove side (m)\\
$\circ$ Triplex base pairs: $CG*G$ etc.; codon pairs: $\#1\ GGG \cdot CCC$ etc.; pair connections: $\#1-Gly-\#2$ etc.; route dualities: $\#1-Gly-\#3 \sim \#2-Gly-\#6$ etc., where the numbers $\#m$ ($m=1\ to\ 32$) indicates the positions on the roadmap

\section{Introduction} 
\subsection{Problems} 

Life has experienced an evolution from the basic prebiotic form until such complex human beings ourselves. Nowadays, we look back on the origin of life. The crux of the problem is to guess correctly a detailed prebiotic picture. Pragmatically speaking, the genetic information forms a logically coherent thread that has run through the whole four billion year history of life on the earth. It shows that it is indeed feasible to obtain the prebiotic picture for the origins of nucleic acids and proteins based on the structures and sequences of the informative molecules of contemporary species. 

Both the genetic code and homochirality are common features of life, which can be regarded as the relics of prebiotic evolution. The origin of the genetic code and the origin of homochirality were generally explained separately in literatures. For example, frozen accident, error minimisation, stereochemical interaction, amino acid biosynthesis, expanding codons and so forth have been suggested to explain the origin of the genetic code (Woese et al. 1966; Crick 1968; Wong 1975; Yarus 1988; Di Giulio M 1989a, 1989b; Osawa and Jukes 1989; Root-Bernstein 2007; Rodin AS et al. 2009; Knight et al 2001; Sengupta and Higgs 2015; Sengupta et al. 2007), while biochemical, geochemical or interstellar processes may account for the origin of homochirality of life (Frank 1953; Soai et al. 1995; Hazen et al 2001; Bailey et al. 1998). However, there is a delicate yet profound relationship between the above two problems. Simply speaking, there is an evolutionary relationship between the permutations of bases among codons and the chiral structures of corresponding tRNAs and aaRSs. Deciphering this relationship is a breakthrough to guess the prebiotic picture. 

As far as the origin of the genetic code is concerned, the following problems are urgent to be explained: (1) the origin of triplet code, (2) the origin and evolution of tRNA, (3) the origin and evolution of aaRS, (4) the codon degeneracy, (5) the non-standard genetic code, (6) the primordial transcription and translation mechanisms, and (7) the recruitment order of amino acids and the corresponding codons. 

\subsection{Data and observations} 

The theory in this article is based on the following experimental data and observations in separate fields, concerning triplex nucleic acids, amino acid recruitment, the genetic code, genome sequences, etc. 
 
The stability of the 16 triplex base pairs in triplex DNA are listed from instability (-), weak (+) to strong (4+) as follows (Soyfer and Potaman 1996; Belotserkovskii et al. 1990):\\
\begin{eqnarray*}
(-)& & GC*A,\ AT*C,\ AT*A\\
(+) & & CG*G,\ TA*C,\ TA*A,\ TA*G,\ GC*C,\ GC*G,\ AT*T\\
(++) & & CG*A,\ CG*T,\ GC*T\\
(3+) & & AT*G\\
(4+) & & CG*C,\ TA*T
\end{eqnarray*}
The above stability order in experiments played a significant role in the primordial evolution of triplex DNA, based on which the genetic code evolution roadmap is constructed. The substitutions of triplex base pairs from weak to strong provided the principal driving force in the spontaneous transition from non-living non-chiral system to living chiral system.

There are $10$ phase I amino acids $Gly$, $Ala$, $Ser$, $Asp$, $Glu$, $Val$, $Leu$, $Ile$, $Pro$, $Thr$, which came from prebiotic synthesis, and $10$ phase II amino acids $Phe$, $Tyr$, $Arg$, $His$, $Trp$, $Asn$, $Gln$, $Lys$, $Cys$, $Met$, which came from biosynthesis (Wong 2005; Trifonov et al. 2006). 

There are rich observations on the genetic code: (1) codon degeneracy contains rich evolutionary information; (2) aaRSs approach the corresponding tRNAs from either major groove side or minor groove side; (3) an amino acid generally corresponds to several tRNAs but only one aaRS; (4) the evolution of aaRSs is closely related to the amino acid synthesis families and the evolution of the genetic code; (5) phase I amino acids can be synthesised naturally while phase II ones cannot; (6) the complementary codon pairs are closely related to each other respectively; (7) stop codons and non-standard genetic code contain evolutionary information; (8) there is a close relationship between the amino acid frequencies and the recruitment order of amino acids; and so on. A comprehensive explanation of all these observations is helpful to understand the origin of the genetic code. 

The primordial information of the evolution of the genetic code is still stored in the complete genome sequences of contemporary species. The distance of codons on the roadmap can be calculated by comparing the genomic codon distributions. And the recruitment orders of amino acids and codons along the roadmap can be obtained by comparing amino acid frequencies or codon position GC contents in genomes. There are intricate relationships between the observations at the molecular level in the present article and the observations at the sequence level in the second part of this series (Li 2018-II). 

\subsection{Main results} 

A picture of triplex nucleic acid evolution is proposed to try to explain the origins of the genetic code, based on experimental data and observations. In this prebiotic picture, the codon degeneracy can be obtained in detail by a genetic code evolution roadmap (Fig 1a, 7). It is possible to explain the above problems together in the same picture. In face of the immense difficulty to recreate and explain the significant primordial events, details determine success or failure in any effort. Taking the codon degeneracy as an example, comprehensive and detailed explanation of the codon degeneracy can be taken as a criteria to evaluate any candidate theories on the origin of the genetic code. In this paper, the intricate and ingenious recruitment of both amino acids and codons is explained step by step according to the coevolution of tRNAs with aaRSs. The roadmap provides perspectives on the origin of the genetic code considering sequence evolution of aaRSs and tRNAs. 

Before obtaining the roadmap, there was an inspired exploration of the genetic code evolution, when considering the role of complementary codons in the genetic code evolution. There are $32$ complementary codon pairs for the $64$ codons. A substitution rule for codon pairs is formulated as follows. Starting from the initial codon pair $GGG \cdot CCC$, only one base of a codon can be substituted in each step. It is also demanded for each substitution that there must exist a common amino acid that is encoded by one codon in the former codon pair as well as by one codon in the latter codon pair. According to the substitutions step by step, a relationship tree of the $64$ codons can be obtained. This tree indicates some evolutionary relationship for the genetic code, but the drawbacks of such an exploration are as follows. First, such obtained relationship trees are not unique; second, no physical driving force is provided in the base substitutions. In the following, a triplex picture for the prebiotic evolution is proposed to avoid these drawbacks, upon which a unique genetic code evolution roadmap is obtained based on the substitutions of triplex base pairs from weak stability to strong stability. 

The hypothetical triplex picture is indeed a physical picture rather than a superficial description. Although triplex nucleic acids are rare in contemporary organisms, some simple homologous triplex nucleic acids can be easily formed in the primordial surroundings through combining three oligonucleotide strands by triplex base pairing, each of which consists of same bases respectively. There are $8$ kinds of triplex nucleic acids $S \cdot S' * S''$, where the strands $S, S', S''$ can be either DNA or RNA, such as the triplex DNA $D \cdot D * D$ (`$\cdot$' represents a Watson-Crick base pair while `$*$' a Hoogsteen base pair) and the triplex nucleic acids mixed with DNA and RNA $D \cdot D * R$ ($D$ stands for DNA while $R$ for RNA), etc. The $YR*R$ ($Y$ denotes pyrimidine while $R$ for purine) triplex DNA $Poly\ C \cdot Poly\ G * Poly\ G$ can be formed, which is supposed here to be the initial physical conditions in the genetic code evolution. The $64$ codons can be recruited one by one in the sequence evolution by alternatively combining and separating for the three strands. Such sequence evolution in the prebiotic evolution was driven by the substitutions of triplex base pairs according to their relative stabilities. The sequence evolution of $D \cdot D * D$ led to the evolution of the genetic code, while the RNA strands separated from the coevolving $D \cdot D * R$ yielded tRNAs and the template RNAs for aaRSs. The tRNAs and aaRSs were generated in accompany with the recruitment of the corresponding codons, respectively. So the triplex picture gives a physical basis for the coevolution of the genetic code with the corresponding tRNAs and aaRSs. 

Especially, the stability of the triplex base pairs played a crucial role to establish the genetic code evolution roadmap. First, the driving force in prebiotic evolution came from the substitutions of triplex base pairs from weak to strong. This is the key physical requirement for the transition from disorder to order in the origin of life. Second, the layout of the roadmap is unique if all the unstable triplex base pairs have been avoided in the substitutions. So the standard genetic code resulted from maximum probability. 

The roadmap contains rich prebiotic evolutionary information. According to the substitution order of triplex base pairs, the genetic code evolution can be divided into the initiation stage, the midway stage and the ending stage. At the initiation stage, it is reasonable that all $9$ new recruits of amino acids belong to phase I amino acids. The recruitment order of amino acids obtained from the roadmap is furthermore supported by experimental data at the sequence level. At the midway stage, the genetic code expands according to the biosynthetic families of amino acids. And stop codons and non-standard genetic code often occur at the ending stage. 

In the triplex picture, tRNAs and aaRSs coevolved with the triplex DNAs along the roadmap. Each codon pair on the roadmap are situated at the corresponding complementary pyrimidine strand and purine strand in the $YR*R$ triplex DNA. The third DNA strand in the $D \cdot D * D$ can be replaced by an RNA strand so as to form a $D \cdot D * R$ triplex nucleic acid. A prototype tRNA can be assembled by linking two complementary RNA strands, one of which just came from the RNA strand of $D \cdot D * R$ and contains the corresponding anti-codon. A prototype aaRS was encoded by an RNA, which is merely homologous to one side of the corresponding prototype tRNA. Therefore, an aaRS can only combine tRNA from the corresponding either major groove side or minor groove side by the primordial translation mechanism. This is able to explain the observation that the aaRS generally recognises the corresponding tRNA from a certain side. It is pleasing to note that it is feasible to explain the complex relationship between aaRSs and tRNAs in detail by the roadmap. 

There is also a profound problem on the origin of the genetic code: why did the genetic code choose triplet codons? Actually, the number of bases in triplet codons as ``three'' might be due to the number of strands in triplex DNAs as ``three'', as will be explained in detail at the sequence level in the second part of this series (Li 2018-II).

\section{The picture at molecular level}

Guessing the right prebiotic picture is the key for understanding the origin of life. A qualified theory must be able to explain all the following problems: the origins of the genetic code, the driving force for the transition from disorder to order, the evolutionary relationship between DNA, RNA and proteins, the origin and evolution of the complexity of life, the origin of transcription and translation, and all that. The picture of triplex nucleic acid evolution might be able to explain these problems together. This is a coevolutionary picture of sequence evolution for both nucleic acids and proteins. 

\subsection{Triplex nucleic acid evolution}

In the triplex picture, as a hypothesis, the evolution of triplex nucleic acid is the physical basis in the process of the origins of the genetic code. There are mainly two kinds of nucleic acids $D \cdot D * D$ and $D \cdot D * R$ in the triplex picture. The stabilities of triplex nucleic acids vary with their sequences as well as with the kinds of triplex nucleic acids $S \cdot S' * S''$, where the strands can be either DNA or RNA (Escud{\'e} et al. 1993; Han and Dervan 1993; Wang and Kool 1995). When the stability of $D \cdot D * D$ is about equal to that of $D \cdot D * R$ (Han and Dervan 1993; Roberts and Crothers 1992), the initial triplex DNA can evolve to $D \cdot D * R$. The relationship between the genetic code and tRNA, aaRS depends on the third RNA strand. In the case of triplex DNA, there are $16$ kinds of triplex base pairs, whose stabilities range from weak to strong. The weak triplex base pairs can be substituted by strong ones. Such spontaneous base substitutions can be interpreted as a driving force in the primordial sequence evolution. The $64$ codons were generated step by step respectively, which determined certain stages of the triplex DNA sequence evolution. In each stage, one strand of the triplex DNA can be replaced by a homologous RNA strand, which carried the corresponding anti-codon, as well as a complementary RNA strand. The single RNA strand can combine with its complementary RNA strand, and consequently fold into a tRNA with certain anti-codon. Meanwhile, aaRS can be generated, which is encoded by a template RNA that is homologous to the corresponding side of the tRNA. Thus aaRS can recognise tRNA, in the way similar to the combination between the template RNA and the aaRS according to the primordial translation mechanism. The evolution of aaRS is closely related to biosynthesis of amino acids. The assignment of $64$ codons to the $20$ amino acids can be explained by the coevolution of tRNA with aaRS. 

The stabilities of the triplex base pairs determine the feasibilities of substitutions of triplex base pairs on the roadmap theory (Fig 1a, 2). A triplex DNA $Poly\ C \cdot Poly\ G * Poly\ G$ is bonded by triplex base pairs $CG * G$. An initial codon pair $GGG \cdot CCC$ locate respectively in the purine strand and the pyrimidine strand of the triplex DNA $Poly\ C \cdot Poly\ G * Poly\ G$. Since the stability of $CG * A$ ($++$) is greater than that of $CG * G$ ($+$), the third base $G$ in $CG * G$ can be substituted by $A$; namely $CG * G$ can spontaneously convert to $CG * A$. Similarly, the stability of $CG * C$ ($4+$) is greater than that of $CG * G$ ($+$), the third base $G$ in $CG * G$ can hence be substituted by C; namely $CG * G$ can also spontaneously convert to $CG * C$. And the stability of $GC * T$ ($++$) is greater than that of $GC * C$ ($+$), the third base $C$ in $GC * C$ can be substituted by $T$; namely $GC * C$ can furthermore spontaneously convert to $GC * T$. Thus, the genetic code can evolve step by step, from the initial codon pair $GGG \cdot CCC$ to the last codon pair $AAA \cdot TTT$, via the above base substitutions $G$ to $A$, $G$ to $C$ and $C$ to $T$. It must be emphasised that all the unstable triplex base pairs ($-$) $GC * A$, $AT * C$ and $AT * A$ have been elegantly avoided in the above process. There hence exists only one roadmap for the genetic code evolution, while all the other roadmaps are blocked and wiped out by the unstable triplex base pairs. 

Until the end of the initiation stage of the roadmap, conditions had been ripe for generating arbitrary finite sequences via the base substitutions $G$ to $A$, $G$ to $C$ and $C$ to $T$, which concerned primordial sequences of the prototype tRNA, the template RNA of prototype aaRS and the early ribozymes etc. There is a primordial translation mechanism for the origin of the earliest proteins, especially for that of aaRSs. A template RNA encoding aaRS can be combined by aminoacyl-tRNAs successively in a row. The directional angles of each aminoacyl-tRNA are fixed by the triple codon-anticodon bonds between template RNA and aminoacyl-tRNA, because of the three-point fixation principle. Hence, all the aminoacyl-tRNAs were aligned neatly along the contour of the template RNA so that a peptide formed by combining the aminoacyls on the opposites side of the approximately parallel neighbouring aminoacyl-tRNAs. Thus the earliest aaRS has been generated in absence of ribosomes required by the modern translation mechanism, namely without the help of additional proteins such as ribosomal proteins and elongation factors etc. The para-codons in tRNA originated from the homology between the tRNA and the corresponding template RNA that encodes aaRS. 

\subsection{Chiral informative molecules}

Diverse sequences can be generated by the base substitutions in triplex nucleic acids. The yielding rates of these sequences are different, due to the different probabilities of base substitutions. In the triplex picture, diverse sequences can be yielded in a non-random manner along the roadmap. Namely, some sequences with high yielding rates accumulated in the primordial surroundings, while some other sequences were seldom generated. The earliest functional molecules, such as prototype ribozyme, tRNA, aaRS, originated when certain non-random sequences happened to have the corresponding specific functions. Such functions once again promoted the accumulation of functional molecules in the surroundings. Recognition of tRNA by aaRS can be achieved by their coevolution along the roadmap; hence the functions of tRNA and aaRS provided biological meaning to the triplet codons in the triplex picture. 

Homochirality brought about an effect of particular selectivity between chiral molecules, which resulted in extraordinary enzyme reaction specificity in living chiral system rather than in non-living non-chiral system. Both homochirality and the genetic code helped to establish a network of informative molecules with specific functions. Interactions between RNAs and proteins played an important role in the prebiotic evolution. Most functions of informative molecules should be established via RNA-protein interactions. Ribonucleoproteins (RNPs) evolved along the roadmap in the triplex picture. When single strands evolved from the oligonucleotides consisted of same base to the oligonucleotides consisted of variety of bases, they tended to form duplex DNA rather than triplex DNA. This brings about the triplex2duplex picture. DNA played the role for storage of information, while the complex molecular functions came from the RNA-protein interactions. Thus numerous diverse elementary units of life had been generated, proliferated and evolved, which can assemble into diverse genomes in the biodiversification process. 

\section{Origin of the genetic code}
\subsection{The roadmap}

A hypothetical roadmap for the evolution of the genetic code (Fig 1a) has been constructed based on the relative stabilities of triplex base pairs (Soyfer and Potaman 1996; Belotserkovskii et al. 1990) in the base substitutions in triplex DNA, as shown below. 

At the beginning of the evolution of the genetic code, there existed single-stranded DNA $Poly\ G$ and $Poly\ C$, which tended to form a triplex DNA (Fig 1a, 1b) (Soyfer and Potaman 1996; Frank-Kamenetskii 1995). $Poly\ C \cdot Poly\ G * Poly\ G$ is a usual $YR*R$ triplex DNA, which is combined by triplex base pair $CG*G$ (Fig 1b). The sequences evolved via substitutions of triplex base pairs in the procedure of alternative combining and separating for the strands of triple-stranded DNA. Only three kinds of substitutions of triplex base pairs are practically required on the roadmap: (1) substitution of $(+)\ CG*G$ by $(++)\ CG*A$ (Soyfer and Potaman 1996; Belotserkovskii et al. 1990), with the transition from $G$ to $A$ in the third $R$ strand. This is of the most common substitution on the roadmap by which all the codons in $Route\ 0$ and most codons in $Route\ 1 \sim 3$ were recruited (Fig 1a); (2) substitution of $(+)\ CG*G$ by $(4+)\ CG*C$, with the transversion from $G$ to $C$ in the third $R$ strand, which blazed a new path at $\#2$, $\#7$, $\#10$ for the recruitment of codons in $Route\ 1 \sim 3$ respectively (Fig 1a); (3) substitution of $(+)\ GC*C$ by $(++)\ GC*T$, with the transition from $C$ to $T$ in the third $R$ strand at $\#6$, $\#19$, $\#12$ (Fig 1a, 2), by which the remaining codons in $Route\ 1 \sim 3$ were recruited (Fig 1a). Thus, all the $64$ codons have been recruited following the roadmap (Fig 1a, 1b). 

According to the base substitutions on the roadmap, the recruitment order of the codon pairs from $\#1$ to $\#32$ is as follows (Fig 1a): 
\begin{quote}
$\#1\ GGG \cdot CCC$, $\#2\ GGC \cdot GCC$, $\#3\ GGA
\cdot UCC$, $\#4\ GAG \cdot CUC$, $\#5\ GAC \cdot GUC$, $\#6\ GGU
\cdot ACC$, $\#7\ GCG \cdot CGC$, $\#8\ AGC \cdot GCU$, $\#9\ GCA
\cdot UGC$, $\#10\ CGG \cdot CCG$, $\#11\ AGG \cdot CCU$, $\#12\ UGG
\cdot CCA$, $\#13\ CGA \cdot UCG$, $\#14\ AGA \cdot UCU$, $\#15\ UGA
\cdot UCA$, $\#16\ ACG \cdot CGU$, $\#17\ AGU \cdot ACU$, $\#18\ ACA
\cdot UGU$, $\#19\ GUG \cdot CAC$, $\#20\ CAG \cdot CUG$, $\#21\ GAU
\cdot AUC$, $\#22\ AUG \cdot CAU$, $\#23\ GAA \cdot UUC$, $\#24\ GUA
\cdot UAC$, $\#25\ UAG \cdot CUA$, $\#26\ AAC \cdot GUU$, $\#27\ AAG
\cdot CUU$, $\#28\ CAA \cdot UUG$, $\#29\ AUA \cdot UAU$, $\#30\ AAU
\cdot AUU$, $\#31\ UAA \cdot AUU$, $\#32\ AAA \cdot UUU$;
\end{quote}
and the recruitment order of the amino acids from $No.1$ to $No.20$ is as follows (Fig 1a): 
\begin{quote}
$No.1\ Gly$, $No.2\ Ala$, $No.3\ Glu$,
$No.4\ Asp$, $No.5\ Val$, $No.6\ Pro$, $No.7\ Ser$, $No.8\ Leu$,
$No.9\ Thr$, $No.10\ Arg$, $No.11\ Cys$, $No.12\ Trp$, $No.13\ His$,
$No.14\ Gln$, $No.15\ Ile$, $No.16\ Met$, $No.17\ Phe$, $No.18\
Tyr$, $No.19\ Asn$, $No.20\ Lys$.
\end{quote}

The evolution of the genetic code can be divided into three stages (Fig 1a): the initiation stage ($\#1 \sim \#6$), the midway stage ($\#7 \sim \#20$, $\#24 \sim \#27$) and the ending stage ($\#21 \sim \#23$, $\#28 \sim \#32$). All the amino acids recruited in the initiation stage belong to phase I. The recruitment of amino acids along the roadmap is described step by step hereinafter, and the pair connections and route dualities on the roadmap will be explained according to the evolution of tRNAs and aaRSs in the following sections. 
\begin{description} 
{\footnotesize 
\item {\bf Initiation}
\item {\it step 1}: \hspace{3.15mm}{\bf 1Gly} {\tiny \underline{Vacant}}\#1
\item {\it step 2}: \hspace{3.15mm}1Gly {\tiny \underline{Vacant}}\#1\hspace{6.6mm}1Gly {\tiny \underline{Vacant}}\#2 
\item {\it step 3}: \hspace{3.15mm}1Gly {\tiny \underline{Vacant}}\#1\hspace{6.6mm}1Gly {\bf 2Ala} \#2
\item {\it step 4}: \hspace{3.15mm}1Gly {\tiny \underline{Vacant}}\#1\hspace{6.6mm}1Gly 2Ala \#2\hspace{6.6mm}1Gly {\tiny \underline{Vacant}}\#3
\item {\it step 5}: \hspace{3.15mm}1Gly {\tiny \underline{Vacant}}\#1\hspace{6.6mm}1Gly 2Ala \#2\hspace{6.6mm}1Gly {\tiny \underline{Vacant}}\#3\hspace{6.6mm}{\bf 3Glu} {\tiny \underline{Vacant}}\#4
\item {\it step 6}: \hspace{3.15mm}1Gly {\tiny \underline{Vacant}}\#1\hspace{6.6mm}1Gly 2Ala \#2\hspace{6.6mm}1Gly {\tiny \underline{Vacant}}\#3\hspace{6.6mm}3Glu {\tiny \underline{Vacant}}\#4\hspace{6.6mm}{\bf 4Asp} {\tiny \underline{Vacant}}\#5
\item {\it step 7}: \hspace{3.15mm}1Gly {\tiny \underline{Vacant}}\#1\hspace{6.6mm}1Gly 2Ala \#2\hspace{6.6mm}1Gly {\tiny \underline{Vacant}}\#3\hspace{6.6mm}3Glu {\tiny \underline{Vacant}}\#4\hspace{6.6mm}4Asp {\bf 5Val} \#5
\item {\it step 8}: \hspace{3.15mm}1Gly {\bf 6Pro} \#1\hspace{6.6mm}1Gly 2Ala \#2\hspace{6.6mm}1Gly {\tiny \underline{Vacant}}\#3\hspace{6.6mm}3Glu {\tiny \underline{Vacant}}\#4\hspace{6.6mm}4Asp 5Val \#5
\item {\it step 9}: \hspace{3.15mm}1Gly 6Pro \#1\hspace{6.6mm}1Gly 2Ala \#2\hspace{6.6mm}1Gly {\bf 7Ser} \#3\hspace{6.6mm}3Glu {\tiny \underline{Vacant}}\#4\hspace{6.6mm}4Asp 5Val \#5
\item {\it step 10}: \hspace{3.15mm}1Gly 6Pro \#1\hspace{6.6mm}1Gly 2Ala \#2\hspace{6.6mm}1Gly 7Ser \#3\hspace{6.6mm}3Glu {\bf 8Leu} \#4\hspace{6.6mm}4Asp 5Val \#5
\item {\it step 11}: \hspace{3.15mm}1Gly 6Pro \#1\hspace{6.6mm}1Gly 2Ala \#2\hspace{6.6mm}1Gly 7Ser \#3\hspace{6.6mm}3Glu 8Leu \#4\hspace{6.6mm}4Asp 5Val \#5\hspace{6.6mm}1Gly {\tiny \underline{Vacant}}\#6
\item {\it step 12}: \hspace{3.15mm}1Gly 6Pro \#1\hspace{6.6mm}1Gly 2Ala \#2\hspace{6.6mm}1Gly 7Ser \#3\hspace{6.6mm}3Glu 8Leu \#4\hspace{6.6mm}4Asp 5Val \#5\hspace{6.6mm}1Gly {\bf 9Thr} \#6
\item {\bf Midway \& ending}
\item {\it step 13:} (\#1 $\sim$ \#6 are fully filled by 1Gly to 9Thr, the same below for the following steps)\ 2Ala {\bf 10Arg}\ \#7
\item and {\it the following steps} (omitting the previously fully filled \#1 $\sim$ \#(n-1) codon pairs in step \#n, from \#8 to \#32): 7Ser 2Ala\ \#8; 2Ala {\bf 11Cys}\ \#9; 10Arg 6Pro\ \#10; 10Arg 6Pro\ \#11; {\bf 12Trp} 6Pro\ \#12; 10Arg 7Ser\ \#13; 10Arg 7Ser\ \#14; {\bf stop} 7Ser\ \#15; 9Thr 10Arg\ \#16; 7Ser 9Thr\ \#17; 9Thr 11Cys\ \#18; 5Val {\bf 13His}\ \#19; {\bf 14Gln} 8Leu\ \#20; 4Asp {\bf 15Ile}\ \#21; {\bf 16Met} 13His\ \#22; 3Glu {\bf 17Phe}\ \#23; 5Val {\bf 18Tyr}\ \#24; stop 8Leu\ \#25; {\bf 19Asn} 5Val\ \#26; {\bf 20Lys} 8Leu\ \#27; 14Gln 8Leu\ \#28; 15Ile 18Tyr\ \#29; 19Asn 15Ile\ \#30; stop 8Leu\ \#31; 20Lys 17Phe\ \#32. }
\end{description}

\subsection{Initiation}

In the beginning, there was an $R$ ($R$ denotes purine) single-stranded DNA $Poly\ G$ (Fig 1a, 1b $\#1$). By complementary base pairing formed a $YR$ ($Y$ denotes pyrimidine) double-stranded DNA $Poly\ C \cdot Poly\ G$. And by triplex base pairing $CG*G$ formed a $YR*R1$ triple-stranded DNA $Poly\ C \cdot Poly\ G*Poly\ G$ (Fig 1a, 1b $\#1$). The third $R1$ strand $Poly\ G$ separated out of this $YR*R1$ triple-stranded DNA, which then formed a new $Y1R1$ double-stranded DNA $Poly\ C \cdot Poly\ G$. So far, there was only initial codon pair $GGG \cdot CCC$ (Fig 1a, 1b $\#1$). 

In the initiation stage of the roadmap, the codon pairs from $\#1$ to $\#6$ were recruited along the roadmap, which constituted the initial subset of the genetic code: 
\begin{quote}
$\#1\ GGG (1Gly) \cdot CCC (6Pro)$, $\#2\ GGC (1Gly) \cdot GCC (2Ala)$, $\#3\ GGA (1Gly)
\cdot UCC (7Ser)$, \\
$\#4\ GAG (3Glu) \cdot CUC (8Leu)$, $\#5\ GAC (4Asp) \cdot GUC (5Val)$, $\#6\ GGU (1Gly) \cdot ACC (9Thr)$.\end{quote}
And in this stage were recruited the earliest $9$ amino acids in order: $1Gly$, $2Ala$, $3Glu$, $4Asp$, $5Val$, $6Pro$, $7Ser$, $8Leu$, $9Thr$, all of which belong to phase I amino acids (Wong 2005; Trifonov et al. 2006). For example, at codon pair position $\#6$ on the roadmap, $1Gly$ and $9Thr$ are encoded by the codon pair $5'GGT3'$ in $R6$ strand and $5'ACC3'$ in $Y6$ strand respectively. Although the initial subset is concise, two essential features of the roadmap, pair connection and route duality, had taken shape in this initiation stage (Fig 1a, 3a). 

Pair connection is an essential feature of the roadmap. A connected codon pair on the roadmap generally encode a common amino acid (Fig 1a, 3b). For instance, the pair connection $\#1-Gly-\#2$ indicates that both $GGG$ in $\#1$ and $GGC$ in $\#2$ encode the common amino acid $Gly$. Pair connections reveal the close relationship between recruitment of codons and recruitment of amino acids, which will be explained later according to the evolution of tRNAs. 

Route duality is another essential feature of the roadmap, which shows the relationship of pair connections between different routes (Fig 1a, 3b). For instance, the route duality $$\#1-Gly-\#3 \sim \#2-Gly-\#6$$ indicates that the pair connection $\#1-Gly-\#3$ in $Route\ 0$ and the pair connection $\#2-Gly-\#6$ in $Route\ 1$ are dual, which encodes a common amino acid $Gly$. Route dualities generally exist between $Route\ 0$ and $Route\ 3$, or between $Route\ 1$ and $Route\ 2$ (Fig 3b), which will be explained later according to the evolution of aaRSs.

In the initiation stage of the roadmap, the non-chiral $Gly$ helped to create the first pair connection $\#1-Gly-\#2$, recruiting chiral $Ala$ at $\#2$ (Fig 1a). And the non-chiral $Gly$ also helped to create the first route duality on the roadmap (Fig 1a): $$\#1-Gly-\#3 \sim \#2-Gly-\#6.$$ This route duality played a central role in the initiation stage; consequently the initial subset played a central role in the midway stage (Fig 3a). The chirality was required at the beginning of the roadmap by the triplex DNA itself (Fig 1a, 1b). Even so, there was still a transition period from non-chirality to chirality, in consideration of the special role of non-chiral $Gly$.

\subsection{Midway}

The genetic codes evolved along four routes $Route\ 0-3$ respectively, where $8$ codon pairs in each route evolved in the order of four hierarchies $Hierarchy\ 1 \sim 4$ respectively (Fig 1a). The roadmap can be divided into two groups: the early hierarchies $Hierarchy\ 1 \sim 2$ and the late hierarchies $Hierarchy\ 3 \sim 4$. It can also be divided into two groups: the initial route $Route\ 0$ (all-purine codons pairing with all-pyrimidine codons) and the expanded routes $Route\ 1 \sim 3$ (purine-pyrimidine-mixing codons) (to see Fig 3a, 7b in Li 2018-II). These groupings will be helpful to explain the origin of the three domains (Li 2018-II). 

In the midway stage of the roadmap, the genetic codes expanded spontaneously from the initial subset (Fig 1a, 3a). Each of the $6$ codon pairs in the initial subset expanded to three additional codon pairs, respectively, by route dualities. Details are as follows. The codon pair $\#2$ in the initial subset expanded to the three continual codon pairs $\#7$, $\#8$ and $\#9$ by route duality $$\#2-Ala-\#8 \sim \#7-Ala-\#9;$$ the codon pair $\#1$ in the initial subset expanded to the three continual codon pairs $\#10$, $\#11$ and $\#12$ by route duality $$\#1-Pro-\#11 \sim \#10-Pro-\#12;$$ the codon pair $\#3$ in the initial subset expanded to the three continual codon pairs $\#13$, $\#14$ and $\#15$ by route duality $$\#3-Ser-\#14 \sim \#13-Ser-\#15;$$ the codon pair $\#6$ in the initial subset expanded to the three continual codon pairs $\#16$, $\#17$ and $\#18$ by route duality $$\#6-Thr-\#17 \sim \#16-Thr-\#18;$$ the codon pair $\#5$ in the initial subset expanded to the three codon pairs $\#19$, $\#24$ and $\#26$ by route duality $$\#5-Val-\#26 \sim \#19-Val-\#24;$$ and the codon pair $\#4$ in the initial subset expanded to the three codon pairs $\#20$, $\#25$ and $\#27$ by route duality $$\#4-Leu-\#27 \sim \#20-Leu-\#25.$$ 

The recruitment order of the codon pairs and the recruitment order of the amino acids are intricately well organised and coherent, according to the subtle roadmap (Fig 1a, 3a). In the initiation stage, firstly, the amino acid $No.1$ was recruited with the codon pair $\#1$, remaining a vacant position. Subsequently, $No.1$ and $No.2$ were recruited with the codon pair $\#2$; $No.1$ was recruited with the codon pair $\#3$, remaining a vacant position; $No.3$ was recruited with the codon pair $\#4$, remaining a vacant position; $No.4$ and $No.5$ were recruited with the codon pair $\#5$; $No.6$ filled up the vacant position of $\#1$; $No.7$ filled up the vacant position of $\#3$; $No.8$ filled up the vacant position of $\#4$; $No.1$ and $No.9$ were recruited with the codon pair $\#6$ (Fig 3a). Thus the framework of the genetic code had been established at the end of the initiation stage. From $\#7$ on, the latecomer amino acids no longer jumped the queue in recruitment so that there were no more vacant positions in the recruited codon pairs. Details are as follows. $No.2$ and $No.10$ amino acids were recruited with the codon pair $\#7$; and subsequently, $No.2$ and $No.7$ were recruited with $\#8$; $No.2$ and $No.11$ were recruited with $\#9$; $No.6$ and $No.10$ were recruited with $\#10$; $No.6$ and $No.10$ were recruited with $\#11$; $No.6$ and $No.12$ were recruited with $\#12$; $No.7$ and $No.10$ were recruited with $\#13$; $No.7$ and $No.10$ were recruited with $\#14$; $No.7$ and $stop$ were recruited with $\#15$; $No.9$ and $No.10$ were recruited with $\#16$; $No.7$ and $No.9$ were recruited with $\#17$; $No.9$ and $No.11$ were recruited with $\#18$; $No.5$ and $No.13$ were recruited with $\#19$; $No.8$ and $No.14$ were recruited with $\#20$; $No.4$ and $No.15$ were recruited with $\#21$; $No.13$ and $No.16$ were recruited with $\#22$; $No.3$ and $No.17$ were recruited with $\#23$; $No.5$ and $No.18$ were recruited with $\#24$; $No.8$ and $stop$ were recruited with $\#25$; $No.5$ and $No.19$ were recruited with $\#26$; $No.8$ and $No.20$ were recruited with $\#27$; $No.8$ and $No.14$ were recruited with $\#28$; $No.15$ and $No.18$ were recruited with $\#29$; $No.15$ and $No.19$ were recruited with $\#30$; $No.8$ and $stop$ were recruited with $\#31$; $No.17$ and $No.20$ were recruited with $\#32$ (Fig 3a). 

Take for example from $\#1$ to $\#29$, the evolution of the genetic code along the roadmap can be described in details as follows (Fig 1a, 1b). Starting from the position $\#1$ (Fig 1b \#1), an $R$ single-stranded DNA brought about a $YR$ double-stranded DNA; next, the $YR$ double-stranded DNA brought about a $YR*R1$ triple-stranded DNA (the number $1$ denotes $\#1$, similar below); next, an $R1$ single-stranded DNA departed from the $YR*R1$ triple-stranded DNA; next, the $R1$ single-stranded DNA brought about a $R1Y1$ double-stranded DNA. Thus, the codon pair $GGG \cdot CCC$ were achieved at $\#1$. At the beginning of $\#7$ (Fig 1b \#7), the $R1Y1$ double-stranded DNA was renamed as $Y1R1$ double-stranded DNA, where the $180^{\circ}$ rotation in writing did not change the right-handed helix; next, the $Y1R1$ double-stranded DNA brought about a $Y1R1*R7$ triple-stranded DNA, through the transversion from $G$ to $C$, where the stability $(+)$ of $CG*G$ increased to the stability $(4+)$ of $CG*C$; next, an $R7$ single-stranded DNA departed from the $Y1R1*R7$ triple-stranded DNA; next, the $R7$ single-stranded DNA brought about a $R7Y7$ double-stranded DNA. Thus, the codon pair $GCG \cdot CGC$ were achieved at $\#7$. The case of $\#19$ is similar to $\#7$ (Fig 1b \#19); the codon pair $GTG \cdot CAC$ were achieved through the transition from $C$ to $T$, where the stability $(+)$ of $GC*C$ increased to the stability $(2+)$ of $GC*T$. The case of $\#24$ is also similar to $\#7$ (Fig 1b \#24); the codon pair $GTA \cdot TAC$ were achieved through the common transition from $G$ to $A$, where the stability $(+)$ of $CG*G$ increased to the stability $(2+)$ of $CG*A$. At the position $\#29$ (Fig 1b \#29), the codon pair $GCG \cdot CGC$ in $Y24R24$ are non-palindromic in consideration that both $GCG$ and $CGC$ do not read the same backwards as forwards. In this case, a reverse operation is necessary so that the obtained codon pair $CAT \cdot ATG$ in $y24r24$ read reversely the same as the codon pair $TAC \cdot GTA$ in $Y24R24$. The process from $y24r24$ to $R29Y29$ is still similar to the case of $\#7$; the codon pair $ATA \cdot TAT$ were achieved through the transition from $G$ to $A$, where the stability $(+)$ of $CG*G$ increased to the stability $(2+)$ of $CG*A$. Other processes on the roadmap are similar to the above example (Fig 1a, 1b). The reverse operation is unnecessary in the cases of $\#2$, $\#7$, $\#10$, $\#11$, $\#3$, $\#4$, $\#16$, $\#9$, $\#19$, $\#27$, $\#23$, $\#22$, $\#24$ after palindromic codon pairs and the last one $\#32$ (Fig 1a), whereas the reverse operation is necessary in the remaining cases of $\#5$, $\#6$, $\#8$, $\#12$, $\#13$, $\#14$, $\#15$, $\#17$, $\#18$, $\#20$, $\#21$, $\#25$, $\#26$, $\#28$, $\#29$, $\#30$, $\#31$ (Fig 1a).

\subsection{The ending}

So far, the genetic code table had been expanded from the $6$ codon pairs in the initial subset to the $6+18$ codon pairs by route duality; the remaining $8$ codon pairs were recruited into the genetic code table in the ending stage of the roadmap (Fig 1a, 3a). There were $2$ codon pairs remained in each of the four routes $Route\ 0-3$ respectively. They satisfied pair connections as follows: $\#23-Phe-\#32$, $\#21-Ile-\#30$, $\#22-Met/Ile-\#29$, $\#28-Leu-\#31$ (Fig 3a). Two of them satisfied route duality (Fig 3a): $$\#21-Ile-\#30 \sim \#22-Met/Ile-\#29.$$ the last two stop codons appeared in the pair connection $\#25-stop-\#31$ (Fig 1a, 3a). When the last two amino acids were recruited through the base pairs $\#26-Asn-\#30$ and $\#27-Lys-\#32$, the codon $UAG$ at $\#25$ had to be selected as a stop codon. The codon $UAA$ at $\#31$ was selected as the last stop codon, due to lack of corresponding tRNA.

The non-standard codons also satisfy codon pairs and route dualities on the roadmap (Fig 1a). The codon pairs pertaining to non-standard codons are as follows: $\#11-Arg\ (Ser, stop)-\#14$, $\#4-Leu\ (Thr)-\#27$ in $Route\ 0$; none in $Route\ 1$; $\#22-(Met)-\#29$ in $Route\ 2$; $\#20-Leu\ (Thr, Gln)-\#25$, $\#12-(Trp)-\#15$, $\#25-stop\ (Gln)/Leu-\#31$, $\#28-Leu\ (Gln)-\#31$ in $Route\ 3$. Majority of non-standard codons appear in the last $Route\ 3$ (Fig 1a). Route dualities of non-standard codons exist between $Route\ 0$ and $Route\ 3$ (Fig 1a):
\begin{eqnarray*}
\#4-Leu\ (Thr)-\#27 &\sim& \#20-Leu\ (Thr)-\#25\\
\#11-(stop)-\#14 &\sim& \#12-Trp/stop-\#15,
\end{eqnarray*}
where the first stop codon $UGA$ at $\#15$ is dual to the non-standard stop codons in $Route\ 0$.

The choice of the genetic code was by no means random, which resulted from the increasing stabilities of triplex base pairs in the substitutions. It had been emphasised that the roadmap followed the strict rule that the stabilities of triplex base pairs monotonically increase (Fig 2). Also note that the roadmap had tried its best to avoid the unstable triplex DNA. The roadmap is the only possible one that has avoided the unstable triplex base pairs ($-$) $GC*A$, $AT*C$ and $AT*A$ as show in the following table, while other eliminated possible roadmaps cannot avoid. 
\begin{center}{\scriptsize
\begin{tabular}{|p{1.1cm}||p{3.3cm}|p{3.3cm}|p{3.3cm}|p{3.3cm}|}\hline
{\bf stability}&\multicolumn{1}{|c|}{\bf CG*N}&\multicolumn{1}{|c|}{\bf GC*N}&\multicolumn{1}{|c|}{\bf TA*N}&\multicolumn{1}{|c|}{\bf AT*N}\\ \hline \hline
(-)&&GC*A&&AT*C AT*A\\
(+)&CG*{\bf G}&GC*C GC*G&TA*C TA*G TA*A&AT*T\\
(++)&CG*{\bf A} CG*{\em T}&GC*{\bf T}&&\\
(3+)&&&&AT*{\em G}\\
(4+)&CG*{\bf C}&&TA*{\em T}&\\ \hline \hline
&(+)CG*G $\rightarrow$ (++)CG*A increase in stability&(+)GC*C$\rightarrow$ (-)GC*A \hspace{1mm}{\it unstable}&(+)TA*A $\rightarrow$ (+)TA*G {\it no increase in stability}&(+)AT*T $\rightarrow$ (3+)AT*G\\
&(+)CG*G $\rightarrow$ (4+)CG*C increase in stability&(+)GC*C $\rightarrow$ (+)GC*G {\it no increase in stability}&(+)TA*A $\rightarrow$ (4+)TA*T&(+)AT*T $\rightarrow$ (-)AT*A \hspace{1mm} {\it unstable}\\
&(+)GC*C $\rightarrow$ (++)GC*T increase in stability&(+)CG*G $\rightarrow$ (++)CG*T&(+)AT*T $\rightarrow$ (+)AT*C {\it no increase in stability}&(+)TA*A $\rightarrow$ (+)TA*C {\it no increase in stability}\\
&\multicolumn{1}{|c|}{\bf POSSIBLE (Roadmap)}&\multicolumn{1}{|c|}{\bf Impossible}&\multicolumn{1}{|c|}{\bf Impossible}&\multicolumn{1}{|c|}{\bf Impossible}\\\hline
&(+)CG*G $\rightarrow$ (++)CG*T&(+)GC*C $\rightarrow$ (++)GC*T&(+)TA*A $\rightarrow$ (+)TA*C {\it no increase in stability}&(+)AT*T $\rightarrow$ (-)AT*C \hspace{1mm} {\it unstable}\\
&(+)CG*G $\rightarrow$ (4+)CG*C&(+)GC*C $\rightarrow$ (+)GC*G {\it no increase in stability}&(+)TA*A $\rightarrow$ (4+)TA*T&(+)AT*T $\rightarrow$ (-)AT*A \hspace{1mm} {\it unstable}\\
&(+)GC*C $\rightarrow$ (-)GC*A \hspace{1mm} {\it unstable}&(+)CG*G $\rightarrow$ (++)CG*A&(+)AT*T $\rightarrow$ (3+)AT*G&(+)TA*A $\rightarrow$ (+)TA*G {\it no increase in stability}\\
&\multicolumn{1}{|c|}{\bf Impossible}&\multicolumn{1}{|c|}{\bf Impossible}&\multicolumn{1}{|c|}{\bf Impossible}&\multicolumn{1}{|c|}{\bf Impossible}\\\hline
\end{tabular}}
\end{center}

Among the $16$ possible triplex base pairs, there are three relatively unstable triplex base pairs. So the statistical ratio of instability for the triplex base pairs is $3/16$. However, the ratio of instability for the triplex base pairs on the roadmap is much smaller. There are $49$ triplex DNAs through $\#1$
to $\#32$ on the roadmap, which involve $3 \times 49 = 147$ triplex base pairs (Fig 1a). The relatively unstable triplex base pairs $GC*A$ and $AT*C$ have not appeared on the roadmap; only the relatively unstable triplex base pair $AT*A$ has appeared inevitably for $7$ times in the reverse operations so as to fulfil all the permutations of $64$ codons (Fig 1a). The ratio of instability $7/147$ on the roadmap is much smaller than the ratio of instability $3/16$ by the statistical requirement. When the relatively unstable $AT*A$ appears at the positions $\#15$, $\#17$, $\#21$, $\#25$, $\#29$, $\#30$ and $\#31$, both stabilities of the other two triplex base pairs in the triplex DNA are $(4+)$ (Fig 1a), which compensates the instability of the triplex DNA to some extent. The amino acid $Ile$, whose degeneracy uniquely is three, occupied three positions $\#21$, $\#29$ and $\#30$ among those $7$ positions. And the three stop codons occupied other three neighbour positions $\#15$, $\#25$ and $\#31$ (Fig 1a). The first stop codon $UGA$ appeared at the position $\#15$, where the relatively unstable $AT*A$ appeared firstly (Fig 1a). According to the primordial translation mechanism, the weak combination of $AT*A$ might help to assign stop codons. The route dualities played significant roles in the midway stage, where the remnant codons were chosen as the stop codons (Fig 1a, 3a). The stop codon appeared as early as the midway of the evolution of the genetic code (Fig 1a, 3a), which indicates that the genetic code had been taken shape around the midway to promote the formation of the primitive life. Not until the fulfilment of the genetic code, did the translation efficiency increase notably by recognising all the $64$ codons.

\section{Origin of tRNA}

The roadmap illustrates the coevolution of the genetic code with the amino acids, where tRNAs and aaRSs play an intermediary role. The expansion of the genetic code along the roadmap can be explained by the coevolution of tRNAs with aaRSs (Fig 5c, 6b, 7). The cloverleaf shape of tRNA can be explained by assembling the complementary RNA strands separated from triplex nucleic acid $D \cdot D * R$ in the triplex picture (Fig 6a). The origin of aaRS will be explained in the next section. 

\subsection{Anti-codon} 

When studying the evolution of the genetic code, we were focused on only three bases in the triplex DNA. But when studying the origin of tRNAs, it is necessary to study the evolution of entire sequences of both triplex DNA and triplex nucleic acid $D \cdot D * R$, where the third RNA strands in $D \cdot D * R$ can be used to assemble tRNAs (Fig 5a, 5b, 6a). According to the order of the relative stabilities of $YR*Y$ for the $8$ kinds of triplex nucleic acids: $D \cdot D*D$, $D \cdot D*R$, $R \cdot D*R$, $R \cdot D*D$ $>$ $D \cdot R*R$, $R \cdot R*R$ $>>$ $R \cdot R*D$, $D \cdot R*D$ (Han and Dervan 1993, Roberts and Crothers 1992), the relative stabilities of $D \cdot D * D$ and $D \cdot D * R$ are greater than the relative stabilities of other kinds of triplex nucleic acids. The choice of triplex DNA for the roadmap and the choice of $D \cdot D * R$ for the origin of tRNAs are based on the observed relative stabilities. And the other kinds of triplex nucleic acids can be neglected due to their less probabilities to appear. 

There are four types of RNA strands for assembling tRNAs that were generated by the triplex base pairing of triplex nucleic acids $D \cdot D * R$: via the triplex nucleic acid $yr*y_t$, via the triplex nucleic acid $yr*r_t$ (Fig 5a, 5c), and via the triplex nucleic acid $YR*Y_t$, via the triplex nucleic acid $YR*R_t$ (Fig 5b, 5c), where the subscript $t$ indicates that theses RNA strands $y_t$, $r_t$ and $Y_t$, $R_t$ are used to assemble tRNA (Fig 5a, 5b, 6a). The sequences $Y_t$, $R_t$ are the respective reverse sequences of $y_t$ and $r_t$. There is a difference in the sequence evolution along the roadmap between purine strands and pyrimidine strands. The pyrimidine sequences $Y_t$, $y_t$ and the purine sequences $R_t$, $r_t$ are complementary respectively, owing to the triplex pairing with the purine DNA strand and the pyrimidine DNA stand in the triplex nucleic acids $D \cdot D * R$ respectively. These tRNA strands coevolved with the triplex DNA along the roadmap. Therefore the evolution of the anti-codons on tRNAs can also be explained according to the evolution of the genetic code along the roadmap. In addition, aaRS evolution should be considered in the next section. After separating from the triplex nucleic acids $D \cdot D * R$, the pair of complementary single RNA strands $y_t$ and $r_t$, or $R_t$ and $Y_t$, can concatenate and fold into a cloverleaf-shaped tRNA (Di Giulio 1992, 1995, 1999, 2004, 2006), whose anticodon corresponds to the codon of the triplex DNA on the roadmap (Fig 6a). Owing to the different positions of anti-codons in the RNA strands, either near to $3'$-ends or near to $5'$-ends, it must be seriously considered for the different reading directions between $Y_t$, $R_t$ and $y_t$, $r_t$ (Fig 6a). There were two types of tRNAs: the type $5'y_tr_t3'$ tRNA and the type $5'R_tY_t3'$ tRNA (Fig 5a, 5b), where the anti-codons are near to the $3'$-end of the RNA strand $y_t$ and the $3'$-end of the RNA strand $R_t$ respectively. The other concatenated RNA strands $5'r_ty_t3'$ and $5'Y_tR_t3'$ cannot evolve together with the above two types of tRNAs, because the corresponding triplets would be on the acceptor arms rather than on the anti-codon loops. 

It is possible to explain the sequence evolution of tRNAs in detail along the roadmap (Fig 5a, 5b, 5c, 6a). For example, the tRNA $t2$ for $2Ala$ can form by concatenating $y_t7$ and $r_t7$, which are generated by triplex base parings $y7r7*y_t7$ and $y7r7*r_t7$ at the branch node $\#7$. The anti-codon $CGC$ near the $3'$-end of the strand $y_t7$ is palindromic. The two complementary strands $y_t7$ and $r_t7$ can combine into a cloverleaf-shaped type $5'y_tr_t3'$ tRNA $t2$ by concatenating, pairing and folding (Fig 6a). Thus anticodon arm of $t2$ contains the anticodon $CGC$, which corresponds to $Ala$, with the help of aaRS; consequently the codon $GCG$ at the $R$ DNA strand in $\# 7$ is assigned to $Ala$. The sequences evolve from $\#7$ to $\#16$ along the roadmap. As another example, the codons at the position $\#16$ is non-palindromic, where the type $5'y_tr_t3'$ tRNA $t9$ and the type $5'R_tY_t3'$ tRNA $t11$ are assembled by concatenating $y_t16 $ and $r_t16$ for $t9$ and by concatenating $R_t16 $ and $Y_t16$ for $t11$ respectively (Fig 6a). Hence the codon $ACG$ at $\#16$ and the reversely complimentary codon $UGC$ at $\#9$ are assigned to $9Thr$ and $11Cys$ respectively. 

There are $4$ pairs of palindromic codons: $\#1\ CCC \cdot GGG$, $\#4\ CUC \cdot GAG$, $\#7\ CGC \cdot GCG$, $\#19\ CAC \cdot GUG$ in the $16$ branch nodes of the roadmap (Fig 1a). Accordingly there are $12$ non-palindromic codons among the branch nodes at the positions $\#2$, $\#5$, $\#6$, $\#10$, $\#11$, $\#12$, $\#16$, $\#20$, $\#21$, $\#23$, $\#24$ and $\#25$. The sets of complementary pairs of RNA strands are same for the two routes, because of the bijection between $Route\ 1$ and $Route\ 3$ in the sense of reverse relationship (Fig 1a). Thus, there are totally $4 + (12 - 4) \times 2 = 20$ pairs of complementary single RNA strands ($4$ palindromic codons, and the $12$ non-palindromic codons minus $4$ identities between $Route\ 1$ and $Route\ 3$), which can assemble into $20$ groups of cognate tRNAs respectively. This could be among the reasons why there are $20$ canonical amino acids. 

There is another reason at the sequence level for the number ``$20$'' of the canonical amino acids (Fig 6b). There are $64$ triple permutations for the $4$ bases, which accounts for the number $64$ of the codons. However, little attention has been paid to the $20$ triple combinations for the $4$ bases (Fig 7a in Li 2018-II), which is actually related to the number $20$ of the canonical amino acids. The products $p(i)*p(j)*p(k)$ ($i, j, k=G,\ C,\ A,\ T$) are same respectively for the $20$ groups of combinations for the $4$ bases (Fig 6b), owing to the multiplication exchange law, where $p(i)$ denotes the base compositions for $i=G,\ C,\ A,\ T$. The products determine the average interval distances of codons in genome sequences. There are therefore $20$ classes of genomic codon distributions according to the $20$ combinations rather than the $64$ permutations of the $4$ bases (Fig 7a in Li 2018-II). Consequently, there are $20$ cognate tRNA-synthetase systems so as to improve the translation efficiency for tRNAs to recognise the corresponding codons, considering the $20$ average interval distances of codons. So the number ``$20$'' of the canonical amino acids actually should be attributed to a statistical origin at the sequence level. The $20$ combinations of the $4$ bases can be divided into $4$ groups $<G>$, $<C>$, $<A>$, $<T>$ (Fig 7a in Li 2018-II). $Hierarchy\ 1$ and $Hierarchy\ 2$ corresponds $<G>$ and $<C>$; $Hierarchy\ 3$ and $Hierarchy\ 4$ corresponds to $<A>$ and $<T>$. Their positions on the roadmap are $Hierarchy\ 1 \sim 2\ Y:\ <G>$, $Hierarchy\ 1 \sim 2\ R:\ <C>$, $Hierarchy\ 3 \sim 4\ Y:\ <A>$, $Hierarchy\ 3 \sim 4\ R:\ <T>$. Each group can be divided into $5$ combinations, which correspond to $Route\ 0$ or $Route\ 1 \sim 3$ respectively. In the case $<G>$, $<G,\ G,\ G>$ and $<G,\ G,\ A>$ belong to $Route\ 0$; $<G,\ G,\ C>$, $<G,\ G,\ T>$ and $<G,\ C,\ A>$ belong to $Route\ 1 \sim 3$, and it is similar for the other cases $<C>$, $<A>$, $<T>$ (Fig 7a in Li 2018-II). These $20$ combinations roughly correspond to the $20$ cognate tRNAs (Fig 6b). This rough correspondence shows that the codons, especially those in $Hierarchy\ 1 \sim \ 3$ are assigned to the tRNAs based on the combinations, considering that the codons in $Hierarchy\ 4$ are $AT$-rich and the context sequences tend to form $AT$-rich repeats. Concretely speaking, the group of codons in the combinations $<GGG>$, $<GGC>$, $<GGA>$, $<GGU>$, $<GCA>$, $<GCU>$, $<GAA>$, $<GAU>$, $<CCC>$, $<CCA>$, $<CCU>$, $<CAA>$, $<CAU>$, $<CUU>$, $<AAU>$ are assigned respectively to $t1$, $t2$ and $t10$, $t3$, $t5$ and $t12$, $t4$ and $t9$ and $t14$, $t8$ and $t11$, $t20$, $t16$, $t6$, $t13$, $t7$, $t19$, $t18$, $t17$, $t15$ (Fig 6b). And the first stop codon appeared halfway in the evolution of tRNAs (Fig 6b). The order of combinations are simply organised by the bases in the order ``$G$'', ``$C$'', ``$A$'', ``$U$'' (Fig 6b), considering the substitutions ``$G$ to $C$'', ``$G$ to $A$'', ``$C$ to $U$'' on the roadmap (Fig 1a). And the amino acids are in the recruitment order. Then, a rough diagonal distribution of tRNAs has been obtained (Fig 6b), which is due to the evolutionary relationship between the genetic code and amino acids. 

\subsection{Evolution of tRNA}

There was a post-initiation-stage stagnation (Fig 1a) between the initiation stage and the midway stage of the roadmap. Such a stagnation in the prebiotic evolution was just to await the birth of functional macromolecules. In this period, oligonucleotides with arbitrary finite sequences can be generated via the base substitutions $G$ to $A$, $G$ to $C$ and $C$ to $T$ in the triplex picture. The primordial sequences of the prototype tRNAs and the template RNAs of prototype aaRSs can be generated along the roadmap (Fig 5a, 5b, 8a). In the light of complicated interactions between oligonucleotides and amino acids, some early tRNAs with certain anti-codons can be generated in the sequence evolution along the roadmap so as to carry the corresponding prebiotically synthetized phase I amino acids, respectively. These tRNAs were not necessarily homologous, as long as they were capable of fulfilling their respective tasks. There are two independent codon systems for tRNAs: the anti-codons and the para-codons. The anti-codons evolved along the roadmap, while the para-codons evolved with aaRSs (Fig 5c, 7). When the para-codons did not evolve but the anti-codons evolved, only cognate tRNAs originated. But when both the para-codons and the anti-codons evolved, more new tRNAs originated to carry the remaining amino acids. 

There exists an assignment scheme for the genetic code. The $64$ codons can be assigned to the $20$ amino acids and stop codons with the help of approximate four dozens of tRNAs: $t1$, $t1'$, $t1^+$, $t2$, $t2'$, $t2^+$, $t3$, $t3'$, $t4$, $t5$, $t5'$, $t5^+$, $t6$, $t6^+$, $t6^+{'}$, $t7$, $t7^+$, $t7^-$, $t7^-{'}$, $t8$, $t8'$, $t8^+$, $t8^-$, $t8^-{'}$, $t9$, $t9'$, $t9^+$, $t10$, $t10'$, $t10^+$, $t10^-$, $t10^-{'}$, $t11$, $t12$, $t13$, $t14$, $t14'$, $t15$, $t15^+$, $t16$, $t17$, $t18$, $t19$, $t20$, $t20'$ (Fig 5c, 6b). The naming rules for tRNAs are as follows. The tRNA series numbers are named after the recruitment order of the respective canonical amino acids. The prime tRNAs $t1$ $\sim$ $t20$ are the early recruited tRNAs that coevolve with the corresponding aaRSs. The derivative tRNAs $tn^+$ are the cognate tRNAs expanded within the codon boxes, namely with the same first two bases in codons. The derivative tRNAs $tn^-$ are the cognate tRNAs expanded outside the codon boxes. The derivative tRNAs $tn{'}$, $n^+{'}$ and $tn^-{'}$ are the cognate tRNAs needed by wobble pairing rules. The bracket in ``$(tn)$'' indicates the same tRNA $tn$. It is also possible to generate more or less new tRNAs in the triplex picture for different species, so the numbers of tRNAs are different among species. 

On one side, the tRNAs can recognise the respective codons according to the genetic code evolution along the roadmap. On the other side, they can recognise the respective aaRSs to combine with the respective aminoacyls. Among the $20$ prime tRNAs $t1$ $\sim$ $t20$, there are $13$ type $5{'}y_tr_t3{'}$ tRNAs ($t1$, $t2$, $t3$, $t4$, $t5$, $t9$, $t10$, $t12$, $t14$, $t15$, $t16$, $t19$, $t20$) and $7$ type $5{'}R_tY_t3{'}$ tRNAs ($t6$, $t7$, $t8$, $t11$, $t13$, $t17$, $t18$) (Fig 5c). The codons for the type $5{'}y_tr_t3{'}$ prime tRNAs are situated in the purine strand on the roadmap, whose first base are purine except $t10$, $t12$, $t14$. And the codons for the type $5{'}R_tY_t3{'}$ prime tRNAs are situated in the Y strand on the roadmap, whose first base are pyrimidine. And there are $6$ prime tRNAs ($t1$, $t3$, $t6$, $t7$, $t17$, $t20$) in $Route\ 0$, $3$ prime tRNAs ($t4$, $t8$, $ t19$) in $Route\ 1$, $8$ prime tRNAs ($t2$, $t5$, $t9$, $t11$, $t13$, $t15$, $t16$, $t18$) in $Route\ 2$, and $3$ prime tRNAs ($ t10$, $t12$, $t14$) in $Route\ 3$ (Fig 5c). The majority of prime tRNAs situated in the branch nodes, except $t15$, $t17$, $t19$, $t20$ (Fig 5c). For each amino acid, several cognate tRNAs can be generated at certain steps of the roadmap as follows. 
\begin{center}{\footnotesize
\begin{tabular}{ll}
$1Gly$: $t1(GGG)$, $t1'(GGA)$, $t1^+(GGC, GGU)$&$11Cys$: $t11(UGC, UGU)$\\
$2Ala$: $t2(GCG)$, $t2'(GCA)$, $t2^+(GCC, GCU)$&$12Trp$: $t12(UGG)$\\
$3Glu$: $t3(GAG)$, $t3'(GAA)$&$13His$: $t13(CAC, CAU)$\\
$4Asp$: $t4(GAC, GAU)$&$14Gln$: $t14(CAG)$, $t14'(CAA)$\\
$5Val$: $t5(GUG)$, $t5'(GUA)$, $t5^+(GUC, GUU)$&$15Ile$: $t15(AUA)$,$t15^+(AUC,AUU)$\\
$6Pro$: $t6(CCC, CCU)$, $t6^+(CCG)$, $t6^+{'}(CCA)$&$16Met$: $t16(AUG)$\\
$7Ser$: $t7(UCC, UCU)$, $t7^+(UCG)$, $t7^+{'}(UCA)$, $t7^-(AGC, AGU)$&$17Phe$: $t17(UUC, UUU)$\\
$8Leu$: $t8(CUG)$, $t8'(CUA)$, $t8^+(CUC, CUU)$, $t8^-(UUG)$, $t8^-{'}(UUA)$&$18Tyr$: $t18(UAC, UAU)$\\
$9Thr$: $t9(ACG)$, $t9'(ACA)$, $t9^+(ACC, ACU)$&$19Asn$: $t19(AAC, AAU)$\\
$10Arg$:$t10(CGG)$,$t10'(CGA)$,$t10^+(CGC, CGU)$,$t10^-(AGG)$,$t10^-{'}(AGA)$&$20Lys$: $t20(AAG)$, $t20'(AAA)$
\end{tabular}}
\end{center}
 
The following evolution of derivative tRNAs can be explained by the base substitution $G$ to $A$ along the roadmap (Fig 5c): $t1(GGG)$ to $t1'(GGA)$, $t2(GCG)$ to $t2'(GCA)$, $t3(GAG)$ to $t3'(GAA)$, $t5(GUG)$ to $t5'(GUA)$, $t6^+(CCG)$ to $t6^+{'}(CCA)$, $t7^+(UCG)$ to $t7^+{'}(UCA)$, $t8(CUG)$ to $t8'(CUA)$, $t8^-(UUG)$ to $t8^-{'}(UUA)$, $t9(ACG)$ to $t9'(ACA)$, $t10(CGG)$ to $t10'(CGA)$, $t10^-(AGG)$ to $t10^-{'}(AGA)$, $t14(CAG)$ to $t14'(CAA)$, $t20(AAG)$ to $t20'(AAA)$. And the following evolution of derivative tRNAs can be explained by the base substitution $G$ to $C$ along the roadmap (Fig 5c): $t1(GGG)$ to $t1^+(GGC, GGU)$, $t2(GCG)$ to $t2^+(GCC, GCU)$, $t5(GUG)$ to $t5^+(GUC, GUU)$, $t6^+(CCG)$ to $t6(CCC, CCU)$, $t8(CUG)$ to $t8^+(CUC, CUU)$, $t9(ACG)$ to $t9^+(ACC, ACU)$, $t10(CGG)$ to $t10^+(CGC, CGU)$. However, the following tRNAs can recognise the respective two codons whose third bases are $C$ or $U$, owing to the wobble pairing (Fig 5c): $t1^+(GGC, GGU)$, $t2^+(GCC, GCU)$, $t4(GAC, GAU)$, $t5^+(GUC, GUU)$, $t6(CCC, CCU)$, $t7(UCC, UCU)$, $t7^-(AGC, AGU)$, $t8^+(CUC, CUU)$, $t9^+(ACC, ACU)$, $t10^+(CGC, CGU)$, $t11(UGC, UGU)$, $t13(CAC, CAU)$, $t15^+(AUC,AUU)$, $t17(UUC, UUU)$, $t18(UAC, UAU)$, $t19(AAC, AAU)$. 

The wobble pairing rules can be explained by the origin and evolution of tRNAs in the triplex picture. The transition from $C$ to $T$ occurred at the position $\#6$ on the roadmap, which resulted in the wobble pairing rule $G:U\ or\ C$. Taking $y2r2$ as a template, $y_t2$ with $GCC$ is formed by the triplex base pairing, while $r_t2$ with $GGC$ and $r'_t2$ with $GGU$ are formed, where the transition from $C$ to $U$ occurred in the formation of $r'_t2$. The complementary strands $y_t2$ and $r'_t2$ combine into a tRNA with anticodon $GCC$, where $G$ at the first position of the anticodon of the tRNA is paired with $U$ at the third position of the triple code of an additional single strand $r'_t2$. It implies that the wobble pairing rule $G:U$ had been established as early as the end of the initiation stage of the roadmap. The transition from $C$ to $T$ occurred at the position $\#12$, which resulted in the wobble pairing rule $U:G\ or\ A$. Taking $y10r10$ as a template, $y_t10$ with $CCG$ is formed by the triplex base pairing, and $r_t10$ with $CGG$ and $r'_t10$ with $UGG$ are also formed, where the transition from $C$ to $U$ occurred in the formation of $r'_t10$. The complementary strands $y_t10$ and $r'_t10$ combine into a tRNA with anticodon $UGG$, where $U$ at the first position of the anticodon of the tRNA is paired with $G$ at the third position of the triple code of an additional single strand $y_t10$. The above explanation of the wobble pairing rules by tRNA mutations is supported by the observations of nonsense suppressor. For instance, the wobble pairing rule $C:A$ for a $UGA$ suppressor can be established by a transition from $G$ to $A$ at the $24th$ position of $tRNA^{Trp}$. The wobble pairing rules $G:U\ or\ C$ and $U:G\ or\ A$ had been established early in the evolution of the genetic code, which continued to flourish so as to make full use of the short supply tRNAs. 

The evolutionary relationship between tRNAs that correspond to pairs of different amino acids can also be explained according to the evolution of tRNAs along the roadmap. For example, based on the substitution $G$ to $A$, $t16(AUG, Met)$ can evolve to $t15(AUA, Ile)$, and based on the substitution $G$ to $C$, $t3(GAG, Glu)$ can evolve to $t4(GAC, GAU, Asp)$, and so on (Fig 5c). However, this kind of evolution of tRNAs involves not only anti-codons but also para-codons, because it inevitably need extra help from aaRSs. There is a close relationship between the evolution of tRNAs and the biosynthetic families of amino acids, so the sequences of tRNAs coevolved with the sequences of aaRSs at each step of the roadmap. The recognition between tRNAs and aaRSs will be explained in the next section, where there are many technical details and each step need to be straightened out in order to draw a comprehensive conclusion. 

The evolution of tRNAs played significant roles to implement the number of canonical amino acids as $20$. There is an important difference between the early prime tRNAs $tn$ and the late derivative tRNAs $tn^+$. Generally speaking, the wobble pairing rules apply to the late derivative tRNAs $tn^+$ rather than to the early prime tRNAs $tn$ (Fig 6b). The early prime tRNAs need not wobble pairings so as to accurately implement the number of bases in codons as $3$, whereas the late derivative tRNAs need wobble pairings so as to improve translation efficiency via codon degeneracy. This was a dynamic process to achieve that the number of canonical amino acids equals to the combination number of bases, which can hardly be fulfilled in lack of tRNAs but can be adjusted by choosing among the numerous candidates of tRNAs. 

\subsection{Palindrome}

Palindromic sequences play significant roles not only in contemporary molecular biology but also in the prebiotic evolution. Palindromic or non-palindromic codons on the roadmap can produce different effects in the origin and evolution of informative macromolecules. The cloverleaf secondary structure of tRNAs can be explained by the complementary palindrome in assembling tRNAs. And the evolution of aaRSs also depended strongly on the evolution of palindromic para-codons along the roadmap, which will be explained in the next section. 

There are two types of tRNAs: type $5'y_tr_t3'$ and type $5'R_tY_t3'$, where the two single RNA strands $y_t$ and $r_t$, $Y_t$ and $R_t$ are complementary to each other. A D-loop and an anti-codon loop situate in the $5'$-end RNA strand ($y_t$ for type $5'y_tr_t3'$ and $R_t$ for type $5'R_tY_t3'$), while a T$\Psi$C loop and a missing loop situate in the $3'$-end RNA strand ($r_t$ for type $5'y_tr_t3'$ or $Y_t$ for type $5'R_tY_t3'$) (Fig 6a). The strand pair $y_t$ and $r_t$ or $Y_t$ and $R_t$ can form two pairs of hairpins in the complementary double-stranded RNA, where the D-loop and the T$\Psi$C loop constitute a pair of hairpins, and the anti-codon loop and the missing complementary loop constitute another pair of hairpins (Fig 6a). When the missing loop has been deleted, the three other loops form a cloverleaf-shaped tRNA (Fig 6a). A palindromic nucleotide sequence can form a hairpin, and palindromic complementary double RNA sequences can form a pair of hairpins, which can account for the cloverleaf secondary structure of tRNAs (Fig 6a, 8a). If there are palindromic sequence intervals in the $5'$-end RNA strand, there will also be the corresponding palindromic sequence intervals in the complementary $3'$-end RNA strand. A D-loop and an anti-codon loop can form in the $5'$-end RNA strand, owing to the complementarity in the palindromic sequence intervals. Accordingly, a T$\Psi$C loop and a missing loop can also form in the $3'$-end RNA strand, which correspond to the D-loop and the anti-codon loop respectively. After deleting the missing loop, a catenated RNA strand with three loops can form a cloverleaf secondary structure, and consequently, a stable tertiary structure can form. Therefore, palindromic sequences contribute to the formation of stable RNA structures in the prebiotic evolution. It is easy to generate palindromic oligonucleotides according to the base substitutions along the roadmap (Fig 5a, 5b). So it tended to generate pairs of palindromic single RNA strands so as to assemble cloverleaf-shaped tRNA candidates. Numerous tRNA candidates can be produced by such an assembly line during the prebiotic evolution, where several qualified tRNAs with proper anti-codons and para-codons can be selected to carry the respective amino acids. Although it is difficult for the origin of aaRSs in the prebiotic evolution (Fig 8a), it is not too difficult for the origin of tRNAs and amino acids. The early aaRSs had chance to adapt by choosing among the numerous tRNA candidates and amino acid candidates. Thus, the degree of difficulty for the origin of life can be reduced to some extent. Yet if both tRNAs and aaRSs had been rare, there would have been little opportunity to establish the correspondence relationship between aaRSs and tRNAs. 

\section{Origin of aaRS}
\subsection{Para-codon}

On one hand, an aaRS is able to recognise cognate tRNAs by para-codons (Fig 6b, 8a). On the other hand, the aaRS is able to catalyse the esterification of proper amino acid to its cognate tRNA (Fig 8a). The origin of aaRS is one of the most difficult events in the origin of life, because a primordial mechanism must be invented to generate the earliest proteins in absence of ribosome, and meanwhile aaRSs have to possess both para-codons and enzyme activity. It should be a rare critical event for the emergence of the first aaRS with enzyme activity in primordial sequence evolution. Following this process, the enzyme activity can transmit from the common ancestor of aaRSs to all the descendant aaRSs, either to the class I or class II aaRSs. Thus, the evolution of para-codons became to play a leading role in the evolution of aaRSs. The evolution of aaRS closely related to both the evolution of tRNA and the biosynthesis families of amino acids. The evolution of para-codons can be explained in the triplex picture. The para-codons of aaRSs coevolved with the sequences of tRNAs along the roadmap. And the abilities to recognise certain amino acids came from the coevolution within the biosynthetic families of amino acids. According to the sequence evolution in the triplex picture, the recognition of tRNA by aaRS can be explained by the sequence homology between the template RNA of aaRS and the corresponding major or minor groove side sequence of tRNA. The recognition between aaRS and its template RNA led to the recognition between aaRS and the corresponding tRNA. 

There are two types of tRNA according to the generation process of tRNA along the roadmap: type $5'y_tr_t3'$ and type $5'R_tY_t3'$ (Fig 5a, 5b), where the $5'$ side corresponds to the minor groove while the $3'$ side to the major groove. Additionally, the aaRSs can combine with the two types of tRNAs from either minor groove or major groove (Fig 5c, 8a). Thus, there are four classes of aaRSs: class $y_t\mbox{-}m$ aaRS, class $r_t\mbox{-}M$ aaRS, class $R_t\mbox{-}m$ aaRS, class $Y_t\mbox{-}M$ aaRS (Fig 5c, 7). The four symbols indicate that aaRSs combine with tRNAs, respectively, from the minor groove ($m$) side $5'y_t$ ($y$) of type $5'y_tr_t3'$ tRNA, from the major groove ($M$) side $r_t3'$ ($r$) of type $5'y_tr_t3'$ tRNA, from the minor groove ($m$) side $5'R_t$ ($R$) of type $5'R_tY_t3'$ tRNA, and from the major groove ($M$) side $Y_t3'$ ($Y$) of type $5'R_tY_t3'$ tRNA. 

The evolution of aaRSs occurred between the four classes of aaRSs (Fig 7). The sequences of para-codon can evolved between the homologous strands, and it can also evolve between the complementary strands when the sequences of para-codons are palindromic (Fig 7). According to the evolution of palindromic para-codons and the origin of the template RNA of aaRS (Fig 8a), the class $y_t\mbox{-}m$ aaRS can be complementary with the class $r_t\mbox{-}M$ aaRS owing to the complementary two strands $5'y_t$ and $r_t3'$ that combine into the type $5'y_tr_t3'$ tRNA (Fig 5a), and the class $R_t\mbox{-}m$ aaRS can be complementary with the class $Y_t\mbox{-}M$ aaRS owing to the complementary two strands $5'R_t$ and $Y_t3'$ that combine into the type $5'R_tY_t3'$ tRNA (Fig 5b). According to the evolution of palindromic para-codons and the coevolution of the template RNAs of aaRSs with tRNAs (Fig 7, 8a), the class $r_t\mbox{-}M$ aaRS can be complementary with the class $Y_t\mbox{-}M$ aaRS, and the class $R_t\mbox{-}m$ aaRS can be complementary with the class $y_t\mbox{-}m$ aaRS. The class $y_t\mbox{-}m$ aaRS can be homologous to the class $Y_t\mbox{-}M$ aaRS, and the class $r_t\mbox{-}M$ aaRS can be homologous to the class $R_t\mbox{-}m$ aaRS. These relationships are useful for studying the evolution of aaRS along the roadmap. 

The aaRSs are denoted as $aaRS1$ to $aaRS20$, according to the recruitment order of the corresponding amino acids from $No.1$ $Gly$ to $No.20$ $Lys$ respectively. The ancestor of aaRSs, namely the major groove $aaRS1$, belongs to the class $r_t\mbox{-}M$ aaRS, which catalysed pairing between the amino acid $1Gly$ and the tRNA $t1$ and which approaches to the type $5'Y_tR_t3'$ tRNA $t1$ from the major groove side $R_t3'$ (Fig 7). The $aaRS1$ evolved into the same class $aaRS2$ and the $Y_t\mbox{-}M$ class $aaRS7$ (Fig 7). The $aaRS2$ evolved into $aaRS3$. According to the evolution of the $Glu$ biosynthesis family, $aaRS3$ evolved into $aaRS6$, $aaRS10$, $aaRS13$ and furthermore $aaRS14$, and $aaRS3$ evolved into $aaRS4$ (Fig 7). According to the evolution of the $Asp$ biosynthesis family, $aaRS4$ evolved into $aaRS9$, $aaRS19$, and furthermore $aaRS15$, $aaRS16$ and $aaRS20$ (Fig 7). According to the evolution of the $Ser$ biosynthesis family, $aaRS7$ evolved into $aaRS11$ and $aaRS12$. According to the evolution of the $Val$ biosynthesis family, $aaRS2$ evolved into $aaRS5$, $aaRS8$. According to the evolution of the $Phe$ biosynthesis family, $aaRS8$ evolved into $aaRS17$ and $aaRS18$. In general, the evolutions via the $Glu$ and $Ser$ biosynthesis families took place in $Hierarchy\ 1$ and $Hierarchy\ 2$, corresponding to the codons whose second bases are $G$ or $C$, while the evolutions via the $Asp$, $Val$ and $Phe$ biosynthesis families took place in $Hierarchy\ 3$ and $Hierarchy\ 4$, corresponding to the codons whose second bases are $A$ or $U$ (Fig 5c). This result accounts for the observation that the second bases of codons relate to the biosynthesis families of amino acids (Fig 4c). 
 
The evolution of aaRSs depends strongly on the para-codon evolution (Fig 7, 8a). Some para-codons of aaRS are homologous but not complementary to the previous para-codons. But the para-codons of aaRSs that are complementary to the previous para-codons had to be palindromic. Some evolutions occurred between the same classes, which includes $aaRS1$ to $aaRS2$, $aaRS3$ to $aaRS10$, $aaRS15$ to $aaRS16$, $aaRS4$ to $aaRS9$, $aaRS4$ to $aaRS19$, $aaRS8$ to $aaRS17$ (Fig 7). Some evolutions of palindromic para-codons occurred between class $y_t\mbox{-}m$ and class $r_t\mbox{-}M$, which includes $aaRS2$ to $aaRS3$, $aaRS2$ to $aaRS5$, $aaRS3$ to $aaRS4$, $aaRS9$ to $aaRS15$, $aaRS19$ to $aaRS20$ (Fig 7). Some evolutions of palindromic para-codons occurred between class $R_t\mbox{-}m$ and class $Y_t\mbox{-}M$, which includes $aaRS7$ to $aaRS11$, $aaRS17$ to $aaRS18$ (Fig 7). And $aaRS1$ to $aaRS7$ occurred between class $r_t\mbox{-}M$ and class $Y_t\mbox{-}M$; $aaRS2$ to $aaRS8$ occurred between class $r_t\mbox{-}m$ and class $R_t\mbox{-}m$; $aaRS3$ to $aaRS6$, $aaRS13$ and $aaRS13$ to $aaRS14$ occurred between class $y_t\mbox{-}m$ and class $Y_t\mbox{-}M$; $aaRS11$ to $aaRS12$ occurred between class $R_t\mbox{-}m$ and class $y_t\mbox{-}m$ (Fig 7). 

The evolution of aaRSs along the roadmap helps to clarify the traditional classifications of aaRSs in the literatures (Fig 4c), such as the major groove ($M$), minor groove ($m$) classification (Eriani et al. 1990) or the class $I$ ($IA$, $IB$, $IC$), class $II$ ($IIA$, $IIB$, $IIC$) classification (Gesteland et al. 2006). The four classes $y_t\mbox{-}m$, $r_t\mbox{-}M$, $R_t\mbox{-}m$, $Y_t\mbox{-}M$ classification here makes clear some confused ideas in the above classifications. The majority of class $r_t\mbox{-}M$ aaRSs correspond to class $IIA$ aaRSs, and the majority of class $R_t\mbox{-}m$ aaRSs correspond to class $IA$ aaRSs, which indicates an evolution from $IIA$ to $IA$ due to the reverse sequence relationship between the RNA templates of class $r_t\mbox{-}M$ aaRS and class $R_t\mbox{-}m$ aaRS (Fig 7). The majority of $Y_t\mbox{-}M$ aaRSs correspond to class $IIA$ aaRSs, which were from the homologous $r_t\mbox{-}M$ aaRSs. And the majority of class $y_t\mbox{-}m$ aaRSs correspond to class $IA$ or $IB$ aaRSs, which were from the complementary $r_t\mbox{-}M$ aaRSs due to evolution of palindromic para-codons (Fig 7). The traditional classification of aaRSs by the major groove and minor groove are reasonable in practice because the template RNAs of aaRSs are complementary between the major groove class and the minor groove class, where the para-codons are palindromic to link the two classes. And the traditional classification of aaRS by classes $A$, $B$ and $C$ reflects some reasonable evolutionary relationships between aaRSs based on the evolution of the biosynthetic families. 

\subsection{Coevolution of tRNA with aaRS}

A comprehensive study of the evolution of the genetic code inevitably involves the origins of tRNAs and aaRSs. The intricate evolutionary relationships between tRNAs and aaRSs can be explained step by step for each codon in the triplex picture (Fig 7). The initiation stage on the roadmap played a fundamental role. At the end of the initiation stage, arbitrary finite sequences can be generated, which provided opportunities to generate complex RNAs such as tRNAs, the template RNAs for aaRSs, ribozymes and the prototype of rRNAs, coding and non-coding RNAs etc. The primordial translation mechanism were invented during the evolution of the genetic code. There were a junior stage and a senior stage of the primordial translation mechanism (Fig 8a, 8b). The ancestor of aaRSs originated in the junior stage when no tRNAs involved (Fig 8a). While the tRNAs and ribosomes were indispensable in the senior stage of the primordial translation mechanism (Fig 8b) as well as in the modern translation mechanism. Certainly, the translation efficiency was low in the junior stage, and was medium in the senior stage and was high in the modern translation mechanism. These exists non-standard translation in experiments, such as direct translation from DNA to protein (McCarthy and Holland 1965; Uzawa et al. 2002). 

The benefits to explain the origins of tRNAs and aaRSs in the triplex picture are as follows. First, the ancestors of tRNAs and aaRSs did not originate from the random sequences; the sequence evolution along the roadmap was recurrent so the informative molecules were generated recurrently and accumulated in the prebiotic surroundings. Second, the evolutionary relationships between tRNAs and aaRSs can be naturally explained by the relationships of the homologous strands of the evolving triplex DNAs. The sequence of the template of the ancestor aaRS can be generated in the triplex picture by the junior stage of the primordial translation mechanism; meanwhile the sequence of ribozyme can also be generated by the other strand of the same triplex nucleic acid. Thus the earliest proteins such as the ancestor of aaRSs can be generated by the complex consisting of the ribozyme, the RNA template of aaRS as well as a triplex DNA. Such a complex itself was the product of sequence evolution of triplex nucleic acids based on specific substitutions of triplex base pairs, where both the sequence for ribozyme and the sequence for the template of ancestor aaRS with enzyme activity were generated in different strands of the same triplex DNA by chance. Although the efficiency to produce proteins was low in this junior stage, it was feasible to generate a small number of proteins by this complex consisting only nucleic acids. The ancestor of aaRS with enzyme activity can be generated by this complex, which naturally tends to combine with the corresponding RNA template.

If the sequence of tRNA is homologous to the above RNA template, the ancestor aaRS also tends to combine with the tRNA. Furthermore, the above requirement can be reduced to homologous para-codons. Thus, in the triplex picture, the aaRSs coevolved with the para-codons, while the tRNAs coevolved with the codons. When considering the homologous or complementary sequence relationships, the reverse sequence relationships and the base substitution relationships in the strands of triplex nucleic acids, the intricate evolutionary relationships between tRNAs and aaRSs can be revealed in detail (Fig 5c, 7). It is more difficult to generate aaRSs than to generate tRNAs, so there existed numerous tRNAs candidates in the prebiotic surroundings. Only the tRNAs that were recognised by aaRSs can be recruited into the living system. For example, the RNA $5'\mbox{-}y_t1r_t1\mbox{-}3'$ were recognised by the class $r_t\mbox{-}M$ $aaRS1$, so it was chosen as the first tRNA $t1$ to transport $1Gly$. And the prime RNAs $tn$ were recognised by $aaRSn$, so they were chosen as the tRNAs to transport $No.\ n$ amino acids (Fig 5c, 7), respectively. Similarly, the derivative RNAs $tn{'}$, $tn^+$, $tn^+{'}$, $tn^-$, $tn^-{'}$, with non-palindromic or palindromic para-codons homologous to the para-codons of $tn$, were recognised by $aaRSn$, so they became the tRNAs to transport $No.\ n$ amino acids, respectively. Para-codons are the key factors for the recognition between tRNAs and aaRSs. The types of tRNAs are not necessarily same for the cognate tRNAs. Generally, the aaRSs combine with the cognate tRNAs from the same side. For example, $aaRS8$ combines with the $5'R_tY_t3'$ type cognate tRNAs $t8$, $t8{'}$, $t8^+$, $t8^-$ and $t8^-{'}$ from the minor groove side, where the para-codons can be non-palindromic (Fig 7); $aaRS7$ combines with the $5'R_tY_t3'$ type tRNAs $t7$, $t7^+$, $t7^-$ and the $5'y_tr_t3'$ type tRNAs $t7^-{'}$ from the major groove side, where the para-codons of the two types of tRNAs have to be palindromic (Fig 7). But $aaRS10$ combines with the $5'y_tr_t3'$ type tRNAs $t10$, $t10{'}$ and the the $5'R_tY_t3'$ type tRNA $t10^+$ from the minor groove side, while combine with the $5'y_tr_t3'$ type tRNAs $t10^-$ and $t10^-{'}$ from the major groove side, where the para-codons also need to be palindromic (Fig 7). 

The biosynthetic families played significant roles in the evolution of aaRSs when both anti-codon and para-codon had changed (Fig 7). There were far more than $20$ amino acids in the prebiotic surroundings. Only the amino acids that were recognised by aaRSs can be recruited into the living system. When $aaRS1$ involved to $aaRS2$, $aaRS2$ recognised $2Ala$ as well as $t2$ from the major groove side, which inherited from $aaRS1$ that recognised $1Gly$ as well as $t1$ from the major groove side. When $aaRS2$ involved to $aaRS3$, $aaRS3$ recognised $3Glu$ as well as $t3$ from the minor groove side owing to the palindromic para-codons, which inherited from $aaRS2$ that recognised $2Ala$ as well as $t2$ from the major groove side. When aaRSs involved in the same biosynthetic families: $Glu$ family, $Asp$ family, $Val$ family, $Ser$ family and $Phe$ family, the new aaRSs tended to recruit the new amino acids with the similar chemical properties in the same biosynthetic family. When aaRSs evolved from $aaRS1$ to $aaRS20$, the enzyme activity transmitted between the aaRSs, and the recognised tRNAs $t1$ to $t20$ and the recognised amino acids $No.1\ Gly$ to $No.20\ Lys$ were recruited, where the evolving non-palindromic or palindromic para-codons linked these evolutions. 

\subsection{Origin of ribosome}

The junior stage must be boosted to the senior stage for the primordial translation mechanism so as to increase the efficiency to generate proteins. In the senior stage, tRNAs and the prototype of ribosome participated in generating proteins. 

Owing to the easy production method of tRNAs, numerous tRNAs can be generated and be aminoacylated. These aminoacyl-tRNAs can combine in turn with the templates of aaRSs, or pre-mRNAs of any early proteins. The choice of the number of bases in codons as three is an important event in the prebiotic evolution, which will be explained at the sequence level in the second part of this series (Li 2018-II). Briefly speaking, the ``three'' in triplet codons originated statistically from the ``three'' in the triplex DNAs. In the primordial translation mechanism, the number of bases that participated in the combination between pre-mRNA and tRNAs varied around three (Fig 8b). The benefit of ``three'' is that the arrangement directions of aminoacyl-tRNAs tended to be parallel, because of the three-point fixation principle. The directions of aminoacyl-tRNAs can also be restricted by certain RNAs generated by the triplex DNA, which furthermore evolved to the small subunit of ribosome (Fig 8b). And the ribozyme in the junior stage evolved to the large subunit of ribosome. The early pre-ribosomal small and large subunits consisted purely of RNAs. The pre-ribosomal small subunit participated in combining aminoacyl-tRNAs with pre-mRNA neatly; while the pre-ribosomal large subunit played the role of peptidyl transferase. The ribosomal subunits can recognise each other due to their homologous relationship (Fig 8b); the small ribosomal subunit can recognise the pre-mRNA due to their homologous segment sequences (Fig 8b). 

When dozens of aminoacyl-tRNAs were arranged neatly along the pre-mRNA, they formed an asymmetric periodic potential along the pre-mRNA (Fig 8b). The pre-ribosome in this periodic potential can be driven by the random forces. As more proteins were generated, some of them combined with the rRNAs so as to start and accelerate the motion of pre-ribosome. Thus, the efficiency increased at the senior stage so as to generate more and more complex proteins. 

The pre-ribosome gradually became more mature so that the modern translation mechanism can be established. The modern ribosome can combine with tRNAs one by one. And the elongation factor furthermore increased the efficiency to generate larger proteins. Mature ribosomes appeared after the fulfilment of the evolution of the genetic code, so there existed different types of ribosomes for the three domains. 

The benefits to explain the origin of proteins in the triplex picture are as follows. First, the junior stage mechanism to generate early proteins do not need proteins themselves (Fig 8a). Second, it is not random for the sequence evolution of early proteins based on the sequence evolution of the triplex DNAs along the roadmap, where the triplex DNA played a role in recording the prebiotic evolutionary information in the triplex picture (Fig 8a, 8b). Third, a comprehensive process of the evolution of ribosome from simple to complex can be explained in the triplex picture (Fig 8a, 8b). Even though these pre-ribosomes were destroyed in the environment, they can be rebuilt according to the sequence evolution along the roadmap. 

\section{Recruitment of codons} 

The roadmap in section $3$ only provided a logical substitution relationship of the $64$ codons based on the stabilities of triplex base pairs (Fig 1a). It was the tRNAs and aaRSs that gave the genetic significance to the $64$ codons (Fig 5c). The pair connections and route dualities observed in the recruitment of codons along the roadmap should be explained based on the coevolution of tRNAs with aaRSs (Fig 5b, 7). The standard genetic code table can be comprehended in a biological context. Incidentally, the non-standard codons can also be explained. 

\subsection{Pair connection} 

The pair connections can be explained by the coevolution of tRNAs with aaRSs when $aaRSn$ recognise, respectively, both the prime tRNAs $tn$ and the corresponding derivative tRNAs $tn'$, $tn^+{'}$ and $tn^-{'}$, where the anti-codons of tRNAs change but the para-codons of tRNAs do not change, or when $tn$ have the efficient ability to recognise similar codons by wobble pairings (Fig 5c, 7). Taking $\#1-1Gly-\#3$ as an example, the $5'y_tr_t3'$ type tRNA $t1$ and the class $r_t\mbox{-}M$ $aaRS1$ originated at $\#1$ on the roadmap and the same type tRNA $t1'$ appeared at $\#3$ on the roadmap. The $aaRS1$ for $1Gly$ can recognise both the same type tRNAs $t1$ and $t1'$ via the same para-codon. Namely tRNAs $t1$ and $t1'$ recognise, respectively, the codons $GGG$ at $\#1$ and $GGA$ at $\#3$ on the purine stands ($R$) on the roadmap (Fig 5c). 

The following pair connections are due to wobble pairings or the tRNA evolution from $tn$ to $tn'$, both of which can be recognised by the respective same $aaRSn$ (Fig 5c, 6b, 7). 
\begin{center}{\small
\begin{tabular}{ll}
1Gly, aaRS1, {\bf t1}$\rightarrow$t1': \ {\bf \#1 R}-Gly-\#3 R
& 2Ala, aaRS2, {\bf t2}$\rightarrow$t2': \ {\bf \#7 R}-Ala-\#9 R\\
3Glu, aaRS3, {\bf t3}$\rightarrow$t3': \ {\bf \#4 R}-Glu-\#23 R
& 4Asp, aaRS4, {\bf t4} wobbling: \ {\bf \#5 R}-Asp-\#21 R\\
5Val, aaRS5, {\bf t5}$\rightarrow$t5': \ {\bf \#19 R}-Val-\#24 R
& 6Pro, aaRS6, {\bf t6} wobbling: \ {\bf \#1 Y}-Pro-\#11 Y\\
7Ser, aaRS7, {\bf t7} wobbling: \ {\bf \#3 Y}-Ser-\#14 Y
& 8Leu, aaRS8, {\bf t8}$\rightarrow$t8': \ {\bf \#20 Y}-Leu-\#25 Y\\
9Thr, aaRS9, {\bf t9}$\rightarrow$t9': \ {\bf \#16 R}-Thr-\#18 R
& 10Arg, aaRS10, {\bf t10}$\rightarrow$t10': \ {\bf \#10 R}-Arg-\#13 R\\
11Cys, aaRS11, {\bf t11} wobbling: \ {\bf \#9 Y}-Cys-\#18 Y
& 12Trp, aaRS12, {\bf t12} wobbling:\ {\bf \#12 R}-Trp-\#(15 R)\\
13His, aaRS13, {\bf t13} wobbling: \ {\bf \#19 Y}-His-{\bf \#22 Y}
& 14Gln, aaRS14, {\bf t14}$\rightarrow$t14': \ {\bf \#20 R}-Gln-\#28 R\\
{\footnotesize 15Ile/16Met,aaRS15/16,{\bf t15}/{\bf t16}:{\bf\#29R}-Ile/Met-{\bf\#22R}}
& 17Phe, aaRS17, {\bf t17} wobbling: \ {\bf \#23 Y}-Phe-\#32 Y\\
18Tyr, aaRS18, {\bf t18} wobbling: \ {\bf \#24 Y}-Tyr-\#29 Y
& 19Asn, aaRS19, {\bf t19} wobbling: \ {\bf \#26 R}-Asn-\#30 R\\
20Lys, aaRS20, {\bf t20}$\rightarrow$t20': \ {\bf \#27 R}-Lys-\#32 R
& stop, no aaRS, no tRNA: \ {\bf \#25 R}-stop-\#31 R
\end{tabular}}
\end{center}
Especially, in the pair connection ${\bf \#29 R}-Ile/Met-{\bf \#22 R}$, $aaRS15$ for $15Ile$ evolved to $aaRS16$ for $16Met$, and the corresponding $t15$ evolved to $t16$ by changing both anti-codon and para-codon. 

The following pair connections are due to wobble pairings or the tRNA evolution from $tn^+$ to $tn^+{'}$, both of which can be recognised by the respective same $aaRSn$ (Fig 5c, 6b, 7).
\begin{center}{\small
\begin{tabular}{ll}
1Gly, aaRS1, $t1^+$ wobbling: \ \#2 R-Gly-\#6 R
& 2Ala, aaRS2, $t2^+$ wobbling: \ \#2 Y-Ala-\#8 Y\\
5Val, aaRS5, $t5^+$ wobbling: \ \#5 Y-Val-\#26 Y
& 6Pro, aaRS6, $t6^+$ $\rightarrow$$t6^+{'}$: \ \#10 Y-Pro-\#12 Y\\
7Ser, aaRS7, $t7^+$ $\rightarrow$$t7^+{'}$: \ \#13 Y-Ser-\#15 Y
& 8Leu, aaRS8, $t8^+$ wobbling: \ \#4 Y-Leu-\#27 Y\\
9Thr, aaRS9, $t9^+$ wobbling: \ \#6 Y-Thr-\#17 Y
& 10Arg, aaRS10, $t10^+$ wobbling: \ \#7 Y-Arg-\#16 Y\\
15Ile, aaRS15, $t15^+$ wobbling: \ \#21 Y-Ile-\#30 Y
& 
\end{tabular}}
\end{center}

The following pair connections are due to wobble pairings or the tRNA evolution from $tn^-$ to $tn^-{'}$, both of which can be recognised by the respective same $aaRSn$ (Fig 5c, 6b, 7). 
\begin{center}{\small
\begin{tabular}{ll}
7Ser, aaRS7, $t7^-$ wobbling: \ \#8 R-Ser-\#17 R
& 8Leu, aaRS8, $t8^-$ $\rightarrow$ $t8^-{'}$: \ \#28 Y-Leu-\#31 Y\\
10Arg, aaRS10, $t10^-$ $\rightarrow$ $t10^-{'}$: \ \#11 R-Arg-\#14 R
& 
\end{tabular}}
\end{center}

The pair connections between non-standard codons are also due to the non-standard tRNA evolution. The non-standard tRNAs $tn*$ with non-standard anti-codons can also be recognised by $aaRSn$. The existence of non-standard codons indicates a variety of possibilities to choose tRNAs among the candidate tRNAs by the aaRSs during the evolution of the genetic code. The non-standard genetic code system can exist in case of certain metabolic cycle (Fig 5c, 7). 
\begin{center}{\small
\begin{tabular}{ll}
7Ser, aaRS7, $t7^*$ $\rightarrow$ $t7^*{'}$: \ \#11 R-Ser-\#14 R
& stop, no aaRS, no tRNA: \ \#11 R-Ser-\#14 R\\
9Thr, aaRS9, $t9^*$ wobbling: \ \#4 Y-Thr-\#27 Y
& 9Thr, aaRS9, $t9^{*+}$ $\rightarrow$$t9^{*+}{'}$:\ \#20 Y-Thr-\#25 Y\\
14Gln, aaRS14, $t14^*$ $\rightarrow$ $t14^*{'}$:\ \#25 R-Gln-\#31 R
& 
\end{tabular}}
\end{center}

\subsection{Route duality}

Route duality refers to the relationships between pair connections in different routes. The route duality can also be explained by the coevolution of tRNAs with aaRSs when $aaRSn$ recognise both the prime tRNAs $tn$ and the corresponding derivative tRNAs $tn^+$ and $tn^-$, respectively. Taking the route duality $\# 7-Ala-\# 9$ $\sim$ $\# 2-Ala-\# 8$ for example, there were two pair connections: $\# 7-Ala-\# 9$ connecting via the $5'y_tr_t3'$ type tRNA $t2$, $t2'$ and $\# 2-Ala-\# 8$ connecting via the $5'R_tY_t3'$ type tRNA $t2^+$. The route duality between $\# 7-Ala-\# 9$ in $Route\ 2$ and $\# 2-Ala-\# 8$ in $Route\ 1$ is due to that $aaRS2$ for $2Ala$ recognise both the tRNAs $t2$, $t2'$ and the different type tRNAs $t2^+$ by same para-codon. 

The following route dualities are due to the tRNA evolution from $tn$ to $tn^+$ or $tn^-$, all of which can be recognised by the respective same $aaRSn$ (Fig 5c, 6b, 7).
\begin{center}{\small
\begin{tabular}{ll}
1Gly, aaRS1, {\bf t1} $\rightarrow$ $t1^+$& {\bf \#1}-Gly-\#3 (Route 0) $\sim$ \#2-Gly-\#6 (Route 1)\\
2Ala, aaRS2, {\bf t2} $\rightarrow$ $t2^+$& {\bf \#7}-Ala-\#9 (Route 2) $\sim$ \#2-Ala-\#8 (Route 1)\\
5Val, aaRS5, {\bf t5} $\rightarrow$ $t5^+$& {\bf \#19}-Val-\#24 (Route 2) $\sim$ \#5-Val-\#26 (Route 1)\\
6Pro, aaRS6, {\bf t6} $\rightarrow$ $t6^+$& {\bf \#1}-Pro-\#11 (Route 0) $\sim$ \#10-Pro-\#12 (Route 3)\\
7Ser, aaRS7, {\bf t7} $\rightarrow$ $t7^+$& {\bf \#3}-Ser-\#14 (Route 0) $\sim$ \#13-Ser-\#15 (Route 3)\\
\hspace{15mm} and {\bf t7} $\rightarrow$ $t7^-$& {\bf \#3}-Ser-\#14 (Route 0) $\sim$ \#8-Ser-\#17 (Route 1)\\
8Leu, aaRS8, {\bf t8} $\rightarrow$ $t8^+$& {\bf \#20}-Leu-\#25 (Route 3) $\sim$ \#4-Leu-\#27 (Route 0)\\
\hspace{15mm} and {\bf t8} $\rightarrow$ $t8^-$& {\bf \#20}-Leu-\#25 (Route 3) $\sim$ \#28-Leu-\#31 (Route 3)\\
9Thr, aaRS9, {\bf t9} $\rightarrow$ $t9^+$& {\bf \#16}-Thr-\#18 (Route 2) $\sim$ \#6-Thr-\#17 (Route 1)\\
10Arg, aaRS10, {\bf t10} $\rightarrow$ $t10^+$& {\bf \#10}-Arg-\#13 (Route 3) $\sim$ \#7-Arg-\#16 (Route 2)\\
\hspace{15mm} and {\bf t10} $\rightarrow$ $t10^-$& {\bf \#10}-Arg-\#13 (Route 3) $\sim$ \#11-Arg-\#14 (Route 0)
\end{tabular}}
\end{center}

The relationship between pair connections via aaRS evolution can be regarded as quasi route dualities (Fig 5c, 6b, 7). 
\begin{center}{\small
\begin{tabular}{ll}
3Glu/4Asp, $t3$/$t4$, aaRS3 $\rightarrow$ aaRS4 & {\bf \#4}-Glu-\#23 (Route 0) $\sim$ {\bf \#5}-Asp-\#21 (Route 1)\\
7Ser/10Arg, $t7^-$/$t10^-$, aaRS7 / aaRS10 & \#8-Ser-\#17 (Route 1) $\sim$ \#11-Arg-\#14 (Route 0)\\
11Cys/12Trp, $t11$/$t12$, aaRS11 $\rightarrow$ aaRS12 & {\bf \#9}-Cys-\#18 (Route 2) $\sim$ {\bf \#12}-Trp-(\#15) (Route 3)\\
13His/14Gln, $t13$/$t14$, aaRS13 $\rightarrow$ aaRS14 & {\bf \#19}-His-\#22 (Route 2) $\sim$ {\bf \#20}-Gln-\#28 (Route 3)\\
15Ile/16Met,$t15$,$t16$/$t15^+$,aaRS15$\rightarrow$aaRS16 & {\bf \#29}-Ile/Met-{\bf \#22} (Route 2) $\sim$ \#21-Ile-\#30 (Route 1)\\
8Leu/17Phe, $t8^-$/$t17$, aaRS8 $\rightarrow$ aaRS17 & \#28-Leu-\#31 (Route 3) $\sim$ {\bf \#23}-Phe-\#32 (Route 0)\\
18Tyr/stop, t18, aaRS18 & {\bf \#24}-Tyr-\#29 (Route 2) $\sim$ {\bf \#25}-stop-\#31 (Route 3)\\
19Asn/20Lys, $t19$/$t20$, aaRS19 $\rightarrow$ aaRS20 & {\bf \#26}-Asn-\#30 (Route 1) $\sim$ {\bf \#27}-Lys-\#32 (Route 0)
\end{tabular}}
\end{center}

The route dualities between non-standard pair connections are also due to the non-standard tRNA evolution. The non-standard tRNAs $tn^*$ and $tn^{*+}$ with non-standard anti-codons can also be recognised by the respective same $aaRSn$ (Fig 5c, 7). The phenomenon of non-standard genetic code is due to alternative choice of tRNAs by aaRSs as small probability events in the fulfilment of the genetic code.
\begin{center}{\small
\begin{tabular}{ll}
7Ser, aaRS7, $t7^-$ $\rightarrow$ $t7^*$& \#8-Ser-\#17 (Route 1) $\sim$ \#11-(Ser)-\#14 (Route 0)\\
9Thr, aaRS9, $t9^*$ $\rightarrow$ $t9^{*+}$& \#4-(Thr)-\#27 (Route 0) $\sim$ \#20-(Thr)-\#25 (Route 3)\\
stop & \#11-(stop)-\#14 (Route 0) $\sim$ \#15-stop-\#31 (Route 3)
\end{tabular}}
\end{center}

The $4 \times 4$ codon boxes in the standard genetic code table come from the $8$ route dualities and the $8$ quasi route dualities (Fig 4a, 4b), where the pair connections are from $hierarchy\ 1$ to $hierarchy\ 2$, from $hierarchy\ 2$ to $hierarchy\ 3$, and from $hierarchy\ 3$ to $hierarchy\ 4$, only. And the route dualities only exist between $route\ 0$ and $route\ 1$, between $route\ 2$ and $route\ 3$, between $route\ 0$ and $route\ 3$, and between $route\ 1$ and $route\ 2$, but not between $route\ 0$ and $route\ 2$ and $route\ 1$ and $route\ 3$ (Fig 4a, 4b).
\begin{center}{\tiny
\begin{tabular}{|c||c|c|c|c||c|c|c|c|c|c|c|c||c|c|c|c|}\hline
&\multicolumn{4}{c||}{} &\multicolumn{8}{c||}{} &\multicolumn{4}{c|}{}\\
&\multicolumn{4}{c||}{\bf Hierarchy 1 to Hierarchy 2} &\multicolumn{8}{c||}{\bf Hierarchy 2 to Hierarchy 3} &\multicolumn{4}{c|}{\bf Hierarchy 3 to Hierarchy 4}\\
&\multicolumn{4}{c||}{} &\multicolumn{8}{c||}{} &\multicolumn{4}{c|}{}\\\hline
&&&&&&&&&&&&&&&&\\
{\bf Route 0}&1Gly&&6Pro&&3Glu&&7Ser&8Leu&&10Arg&&&&17Phe&&20Lys\\
&&&&&&&&&&&&&&&&\\
{\bf Route 1}&1Gly&2Ala&&&4Asp&5Val&&&9Thr&7Ser&&&15Ile&&&19Asn\\
&&&&&&&&&&&&&&&&\\
{\bf Route 2}&&2Ala&&10Arg&&5Val&&&9Thr&&11Cys&13His&16Met&&18Tyr&\\
&&&&&&&&&&&&&&&&\\
{\bf Route 3}&&&6Pro&10Arg&&&7Ser&8Leu&&&12Trp&14Gln&&8Leu&stop&\\
&&&&&&&&&&&&&&&&\\ \hline
{\bf Codon box}&GGN&GCN&CCN&CGN&GAN&GUN&UCN&CUN&ACN&AGN&UGN&CAN&AUN&UUN&UAN&AAN\\ \hline
\end{tabular}}
\end{center}

\section{Codon degeneracy}
\subsection{Explanation}

The degeneracies $6$, $4$, $3$, $2$ or $1$ for the $20$ amino acids can be explained one by one according to pair connections and route dualities on the roadmap based on the coevolution of tRNAs with aaRSs in the triplex picture (Fig 5c, 6b, 7). Especially, the evolution of aaRSs based on the biosynthetic families played significant roles in the expansion of the genetic code. The degeneracy $2$ mainly results from pair connections. The degeneracy $4$ or $6$ mainly result from the expansion of the genetic code from the initial subset by route dualities for $Ser$, $Leu$, $Ala$, $Val$, $Pro$ and $Thr$ (Fig 3a, 3b).

The degeneracy $6$ for $Ser$, $Leu$ and $Arg$ can be explained by pair connections and route dualities (Fig 1a, 3b, 5c, 6b, 7), where $Ser$ and $Leu$ belong to the initial subset and $Arg$ was recruited immediately after the initial subset. And all of them have appeared in $Route\ 0$. The $6$ codons of $Ser$ satisfy both the the route duality and pair connection
        $$\#3-Ser-\#14 \sim \#13-Ser-\#15\mbox{ and }\#8-Ser-\#17.$$
The $6$ codons of $Leu$ satisfy both the the route duality and pair connection
        $$\#20-Leu-\#25 \sim \#4-Leu-\#27\mbox{ and }\#28-Leu-\#31.$$
The $6$ codons of $Arg$ satisfy both the the route duality and pair connection
        $$\#10-Arg-\#13 \sim \#7-Arg-\#16\mbox{ and }\#11-Arg-\#14.$$
The degeneracy $4$ for $Gly$, $Ala$, $Val$, $Pro$ and $Thr$ can be explained by route dualities (Fig 1a, 3b). All of them belong to the initial subset. The degeneracy $4$ for $Gly$ satisfy the route duality:
        $$\#1-Gly-\#3 \sim \#2-Gly-\#6.$$
The degeneracy $4$ for $Ala$ satisfy the route duality:
        $$\#2-Ala-\#8 \sim \#7-Ala-\#9.$$
The degeneracy $4$ for $Val$ satisfy the route duality:
        $$\#5-Val-\#26 \sim \#19-Val-\#24.$$
The degeneracy $4$ for $Pro$ satisfy the route duality:
        $$\#1-Pro-\#11 \sim \#10-Pro-\#12.$$
The degeneracy $4$ for $Thr$ satisfy the route duality:
        $$\#6-Thr-\#17 \sim \#16-Thr-\#18.$$
The degeneracy $2$ for $Glu$, $Asp$, $Cys$, $His$, $Gln$, $Phe$, $Tyr$, $Asn$ and $Lys$ can be explained by pair connections (Fig 1a, 3b). They satisfy the following pair connections respectively: $\#4-Glu-\#23$, $\#5-Asp-\#21$, $\#9-Cys-\#18$, $\#19-His-\#22$, $\#20-Gln-\#28$, $\#23-Phe-\#32$, $\#24-Tyr-\#29$,
$\#26-Asn-\#30$, $\#27-Lys-\#32$. The degeneracy $3$ for $Ile$ and the degeneracy $1$ for $Met$ satisfies the route duality (Fig 1a, 3b, 5c, 6b, 7)
        $$\#21-Ile-\#30 \sim \#22-Met/Ile-\#29.$$
The degeneracy $1$ for $Trp$ satisfies the pair connection for nonstandard genetic code $\#12-Trp/stop(Trp)-\#15$. This pair connection includes a stop codon; the other stop codons satisfy the pair connection: $\#25-stop-\#31$ (Fig 1a, 3b, 5c, 6b, 7). 

It is convenient to demonstrate pair connections and route dualities by introducing a new cubic roadmap (Fig 3b), where the vertices of the cubes represent codon pairs and the edges of the cubes represent substitution
relationships (Fig 3b). The sets of codon pairs in different routes satisfy (Fig 3b):
\begin{eqnarray*}
Route\ 0: <N{'}N{''}N{'''} \cdot n{'''}n{''}n{'}> &\leftrightarrow& Route\ 1: <N{'}N{''}n{'''} \cdot N{'''}n{''}n{'}>\\
Route\ 0: <N{'''}N{'}N{''} \cdot n{''}n{'}n{'''}> &\leftrightarrow& Route\ 3: <n{'''}N{'}N{''} \cdot n{''}n{'}N{'''}>\\
Route\ 1: <N{'''}N{'}N{''} \cdot n{''}n{'}n{'''}> &\leftrightarrow& Route\ 2: <n{''}n{'}N{'''} \cdot n{'''}N{'}N{''}>\\
Route\ 2: <N{'}N{''}N{'''} \cdot n{'''}n{''}n{'}> &\leftrightarrow& Route\ 3: <N{''}N{'}N{'''} \cdot n{'''}n{'}n{''}>, 
\end{eqnarray*}
where $N{'}$, $N{''}$, $N{'''}$ refer to any bases and $n{'}$, $n{''}$, $n{'''}$ the corresponding complementary bases. The pair connections for $Pro$, $Ser$ and $Leu$ in $Route\ 0$ $Y$-strand are dual to the pair connections in $Route\ 3$ $Y$-strand, respectively (Fig 3a, 3b); the pair connections for $Ala$, $Thr$ and $Val$ in $Route\ 1$ $Y$-strand are dual to the pair connections in $Route\ 2$ $R$-strand, respectively (Fig 3a, 3b). There are relationships between codons in different facets of the cubes: $Route\ 0$ and
$Facet\ u$ are dual to $Route\ 3$ and $Facet\ s$; $Route\ 0$ and
$Facet\ d$ are dual to $Route\ 3$ and $Facet\ n$; $Route\ 1$ and
$Facet\ s$ are dual to $Route\ 2$ and $Facet\ u$; $Route\ 1$ and
$Facet\ n$ are dual to $Route\ 2$ and $Facet\ d$ (Fig 3b). 
So, there is a rough overall duality between $Route\ 0,\ 1$ and $Route\ 3,\ 2$ (Fig 1a, 3b).

It is also convenient to demonstrate the biosynthetic families by introducing a new $GCAU$ genetic code table, where the first two bases of codons are arranged in an evolutionary order $G$, $C$, $A$, $U$ rather than the traditional order $U$, $C$, $A$, $G$ (Fig 4b). The codons and amino acids are recruited roughly from left to right and from up to down in the new table (Fig 4b). The codons can be classified by the purine $R$-strands and pyrimidine $Y$-strands in different routes as follows:
$Route\ 0\ R: RRR$, $Route\ 0\ Y: YYY$,
$Route\ 1\ R: RRY$, $Route\ 1\ Y: RYY$,
$Route\ 2\ R: RYR$, $Route\ 2\ Y: YRY$,
$Route\ 3\ R: YRR$, $Route\ 3\ Y: YYR$.
Hence, the $GCAU$ genetic code table can be divided into codon boxes (Fig 4a, 4b). $Route\ 0$ $R$ $GRR$ and $Route\ 1$ $GNY$ correspond to the earliest amino acids $Gly$, $Ala$, $Glu$, $Asp$, $Val$ (Fig 4b); $Route\ 3$ $CNR$ mainly correspond to the biosynthetic family $Glu$ ($Arg$, $Pro$, $Gln$); $Route\ 1$ $ANY$ mainly correspond to the biosynthetic family $Asp$ ($Thr$, $Asn$, $Ile$) (Fig 4b). An in-depth explanation of codon boxes requires a dynamic understanding of the evolution of the genetic code in the triplex picture. 

\subsection{Evidence} 

The present study provides perspectives on the origins of the genetic code. Frankly speaking, it is not easy to test this hypothesis experimentally. However, there might be some indirect or weak evidence to support my hypothesis. The explanation of the codon degeneracy is based on the hypothetical roadmap. This roadmap theory can be supported indirectly by some experimental data. The roadmap itself is based on the experimental results on the relative stabilities of the base triplexes (Fig 1a, 2). The delicate roadmap has narrowly avoided the unstable base triplexes. 

The roadmap of the genetic code evolution obtained in the triplex picture might be verified by the biological data at the sequence level as well as at the species level, namely the three-domain tree of life can be reconstructed (Fig 7h in Li 2018-II) according to the evolutionary relationships of codons on the roadmap based on the complete genome sequences, which will be explained in detail in the second part of this series (Li 2018-II). An evolutionary tree of codons is obtained based on the genomic codon distributions, where the four hierarchies of the roadmap are distinguished clearly (Fig 3a in Li 2018-II). This is straightforward agreement with the roadmap. There might be weak evidence to support the roadmap. The major groove or minor groove classification of aaRSs can be explained by the coevolution between aaRSs and tRNAs in the triplex picture (Fig 7). And according to the sequence evolution in the triplex picture, there is a complementary relationship between the pyrimidine strand $y_t$ of the $5'y_tr_t3'$ type tRNAs and the purine strand $R_t$ of the $5'R_tY_t3'$ type tRNAs. This prediction of the roadmap is supported by the corresponding complementary relationship between $G$ and $C$ of the second bases of the consensus tRNA sequences (Fig 5c) (Rodin et al. 1996) that evolved early in $Route\ 0$ and $Hierarchy\ 1$ of the roadmap. A heuristic evidence for the roadmap might be that the origin of homochirality of life can be explained by the winner-take-all principle between the opposite chiral roadmap systems. Homochirality can be chosen during the evolution of the genetic code. Both chiral roadmap systems are competing for the non-chiral pyrimidines, purines and $No.1\ Gly$. The origins of the genetic code, homochirality and the three domains is possible to be explained together in the same triplex picture, where borrowed ideas from each other can enhance the respective explanations themselves. 

Furthermore, the roadmap might also be supported by the following evidence. The roadmap predicts the recruitment order of the $64$ codons, the recruitment order of the $3$ stop codons and the recruitment order of the $20$ amino acids. Evaluation of these recruitment orders may verify the roadmap theory. 

Concretely speaking, the recruitment order of the $32$ codon pairs can be obtained from $\#1$ to $\#32$ by the roadmap (Fig 3a, 9a). The declining $GC$ content indicates the evolution direction because of the substitutions from $G$ to $A$, from $G$ to $C$ and from $C$ to $T$ on the roadmap (Fig 9a). According to the roadmap, the total $GC$ content and the position specific $GC$ content for the $1st$, $2nd$ and $3rd$ codon positions are calculated for each step from $\#1$ to $\#32$. Then the relationship between the total $GC$ content and the position specific $GC$ content is obtained (Fig 9b). The $1st$ position $GC$ content is higher than the $2nd$ position $GC$ content. And the $3rd$ position $GC$ content declined rapidly from the highest to the lowest in the evolution direction when the total $GC$ content declines. The $GC$ content variation in the simulation agrees with the observation (to compare Fig 9b with Figure 2 in Muto and Osawa 1987 and Figure 5 in Gorban 2005), so the recruitment order obtained by the roadmap is reasonable. The recruitment order of the codons by the roadmap generally agree with the order in the literatures (Trifonov et al. 2006; Trifonov et al. 2001; Trifonov 2000; Trifonov 2004). In fact, the roadmap was conceived by studying the substitution relationships based on the recruitment order in (Trifonov et al. 2006).

The recruitment order for the three stop codons are $\#15\ UGA$, $\#25\ UAG$, $\#31\ UAA$ (Fig 1a, 3a), which results in the different variations of the stop codon usages. Along the evolution direction as the declining $GC$ content, the usage of the first stop codon $UGA$ decreases; the usage of the second stop codon $UAG$ remains almost constantly; the usage of the third stop codon $UAA$ increases (Figure 1 in Povolotskaya et al. 2012). The observations of the variations of stop codon usages can be simulated (to compare Fig 9c with Figure 1 in Povolotskaya et al. 2012) according to the recruitment order of the codon pairs (Fig 3a, 9a) and the variation range of the stop codon usages. Especially, the detailed features in observation can be simulated that the usage of $UGA$ jump downwards greatly; $UAA$, upwards greatly, around half $GC$ content (to compare Fig 9c with Figure 1 in Povolotskaya et al. 2012).

The recruitment order of the $20$ amino acids from $No.1$ to $No.20$ can be obtained by the roadmap (Fig 3a, 9a), which meets the basic requirement that Phase I amino acids appeared earlier than the Phase II amino acids (Wong 1975; Wong and Lazcano 2009). The species with complete genome sequences are sorted by the order $R_{10/10}$ according to their amino acid frequencies, where the order $R_{10/10}$ is defined as the ratio of the average amino acid frequencies for the last $10$ amino acids to that for the first $10$ amino acids (Li and Zhang 2009). Along the evolutionary direction indicated by the increasing $R_{10/10}$, the amino acid frequencies vary in different monotonous manners for the $20$ amino acids respectively (Fig 9d). For the early amino acids $Gly$, $Ala$, $Asp$, $Val$, $Pro$, the amino acid frequencies tend to decrease greatly, except for $Glu$ to increase slightly (Fig 9d); for the midterm amino acids $Ser$, $Leu$, $Thr$, $Cys$, $Trp$, $His$, $Gln$, the amino acid frequencies tend to vary slightly, except for $Arg$ to decrease greatly (Fig 9d); for the late amino acids $Ile$, $Phe$, $Tyr$, $Asn$, $Lys$, the amino acid frequencies tend to increase greatly, except for $Met$ to increase slightly (Fig 9d). In the recruitment order from $No.1$ to $No.20$, the variation trends of the amino acid frequencies increase in general (Fig 9e); namely, the later the amino acids recruited, the more greatly the amino acid frequencies tend to increase (Fig 9d, 9e). The recruitment order of the amino acids from $No.1$ to $No.20$ is supported not only by the previous roadmap theory but also by this pattern of amino acid frequencies based on genomic data.

\section{Conclusion and discussion}

A prebiotic picture based on the evolution of triplex nucleic acids is proposed in this article to try to explain the early evolution of life. It is indeed feasible to explain the origins of the genetic code and the informative molecules via studying the sequence evolution and comparing genome sequences. The codon degeneracy can be obtained according to the coevolution of tRNAs with aaRSs. It must be emphasised again that this study should be regarded as a hypothesis on the origins of the genetic code. It is of significance to evaluate this hypothesis by future experiments.

\section*{Acknowledgements} My warm thanks to Jinyi Li for valuable discussions. I wish to thank the contributors of the biological data used in this study. Supported by the Fundamental Research Funds for the Central Universities.

\clearpage
\begin{figure}
  \centering
   \includegraphics[width=18.3cm]{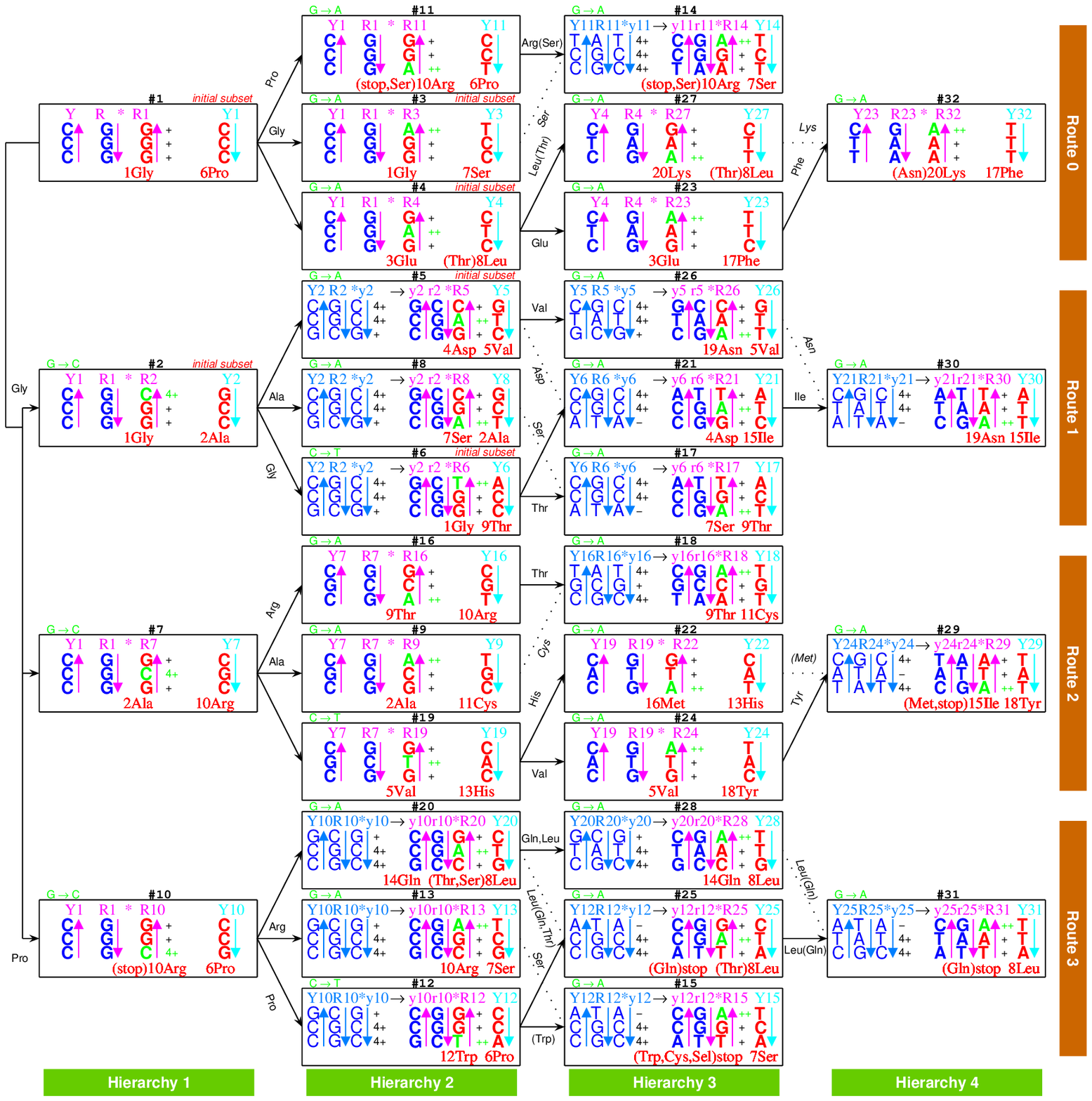}\\
  \vspace{1cm}{\small \bf a}
\end{figure}

\clearpage 
\begin{figure}
  \centering
  \includegraphics[width=18.3cm]{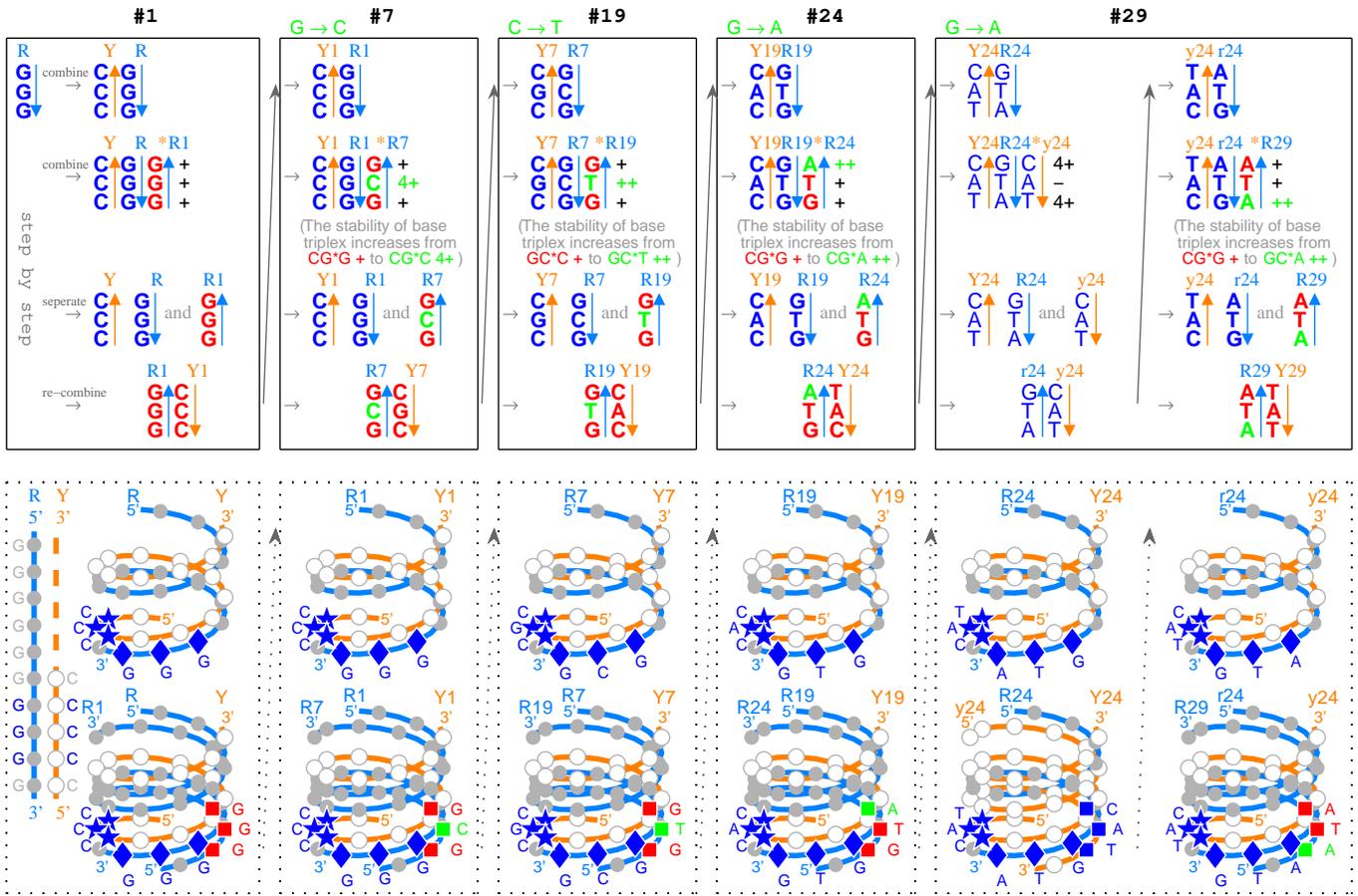}\\
  {\small \bf b}
  \caption{The origin of the genetic code. {\bf a} The roadmap for the evolution of the genetic code. The $64$ codons formed from base substitutions in triplex DNAs are in red. Only three-base-length segments of the triplex DNAs are shown explicitly; the whole length right-handed triplex DNAs are indicated in Fig 1b. In each position $\#n$ ($n=1, 2, ..., 32$), the $\#n$ codon pair on $Rn$ and $Yn$ is in red. The relative stabilities of the triplex base pairs (-, +, ++, 4+) are written to the right of the base triplexes, where the increased relative stabilities of triplex base pairs in base substitutions are indicated in green. Each triplex DNA is denoted by three arrows, whose directions are from 5' to 3'. The $YR*R$ triplex DNAs are in pink, and the $YR*Y$ triplex DNAs in azure. The recruitment order of codon pairs are from $\#1$ to $\#32$, and the recruitment order of the $20$ amino acids are to the left of them respectively. Non-standard genetic codes are indicated by brackets beside the corresponding amino acids. The $Route\ 0-3$ and $Hierarchy\ 1 \sim 4$ are indicated to the right of and below the roadmap respectively. The evolution of the genetic code are denoted by black arrows, beside which pair connections are indicated by the corresponding amino acids. Refer to an example in Fig 1b to understand details of the roadmap; refer to Fig 2 to understand the critical role of relative stabilities of triplex base pairs in achieving the real genetic code; refer to Fig 5a, 5b to see the origin of tRNAs; refer to Fig 3a to see the coherent relationship between the recruitment orders of codons and amino acids; refer to Fig 3b to see the codon degeneracy in the symmetric roadmap; and refer to Fig 10a to see the origin of homochirality of life. {\bf b} A detailed description of the roadmap. Taking for example from $\#1$ to $\#29$, the evolution of the genetic code from $\#1$, to $\#7$, to $\#19$, to $\#24$, and at last to $\#29$ are explained in detail in the upper boxes, and the corresponding right-handed single-stranded, double-stranded and triple-stranded DNAs are shown in the lower boxes, respectively.}
\end{figure}

\clearpage 
\begin{figure}
 \centering
 \includegraphics[width=16cm]{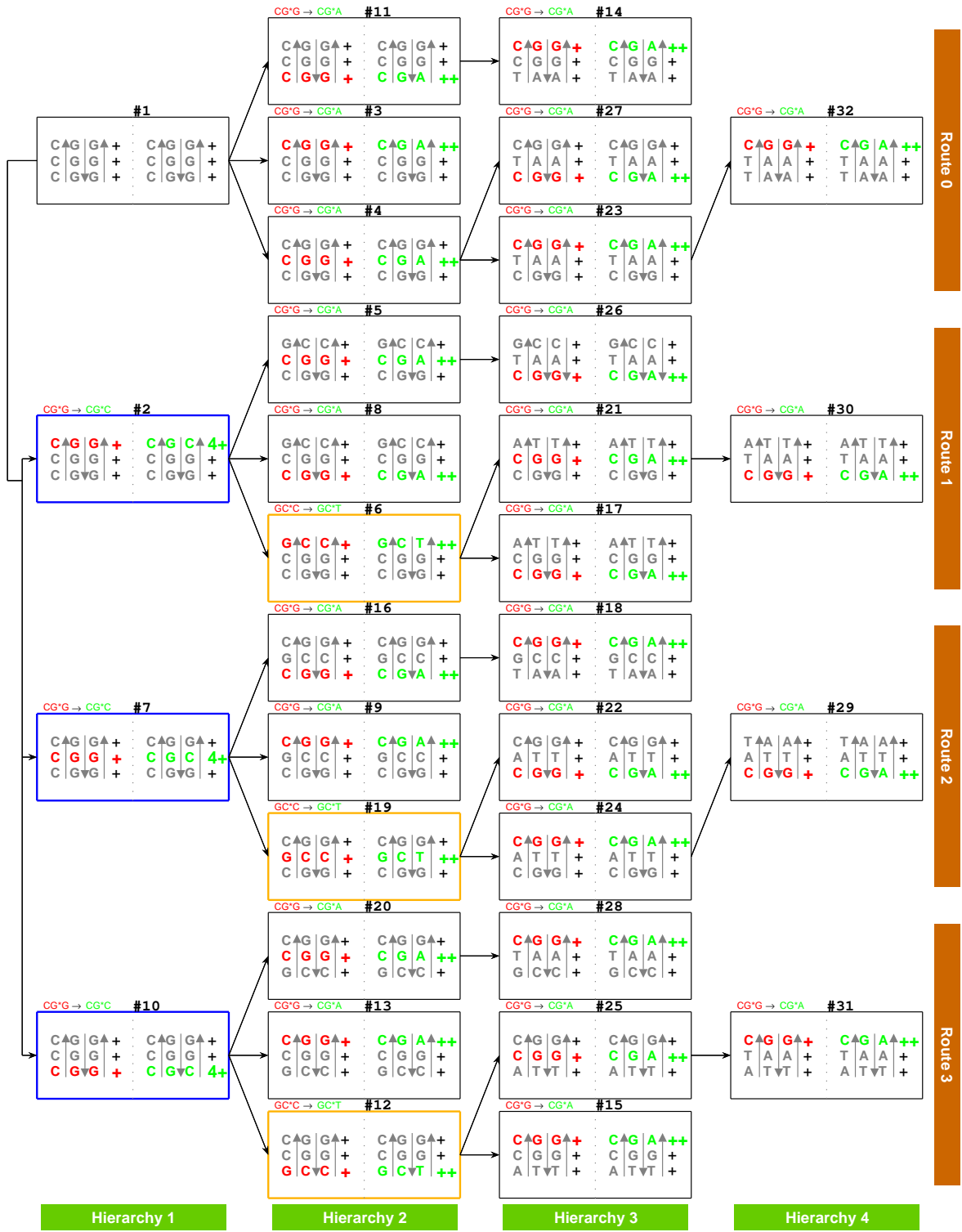}
 \caption{The driving force in the evolution of the genetic code based on the relative stabilities of triplex base pairs. The base substitutions on the roadmap occur when the relative stabilities of triplex base pairs increase. The roadmap is the best result to avoid the unstable triplex base pairs. So, the universal genetic code is a narrow choice by the relative stabilities of triplex base pairs. The relative stability increases from (+) of the triplex base pair $CG*G$ to (4+) of the triplex base pair $CG*C$ at $\#2$, $\#7$ and $\#10$ that initiates $Route\ 1 \sim 3$ respectively. $GC*C$ (+) changes to $GC*T$ (++) at $\#6$, $\#19$ and $\#12$, and $CG*G$ (+) changes to $CG*A$ (++) at other positions on the roadmap.}
\end{figure}

\clearpage 
\begin{figure}
  \centering
  \includegraphics[width=18.3cm]{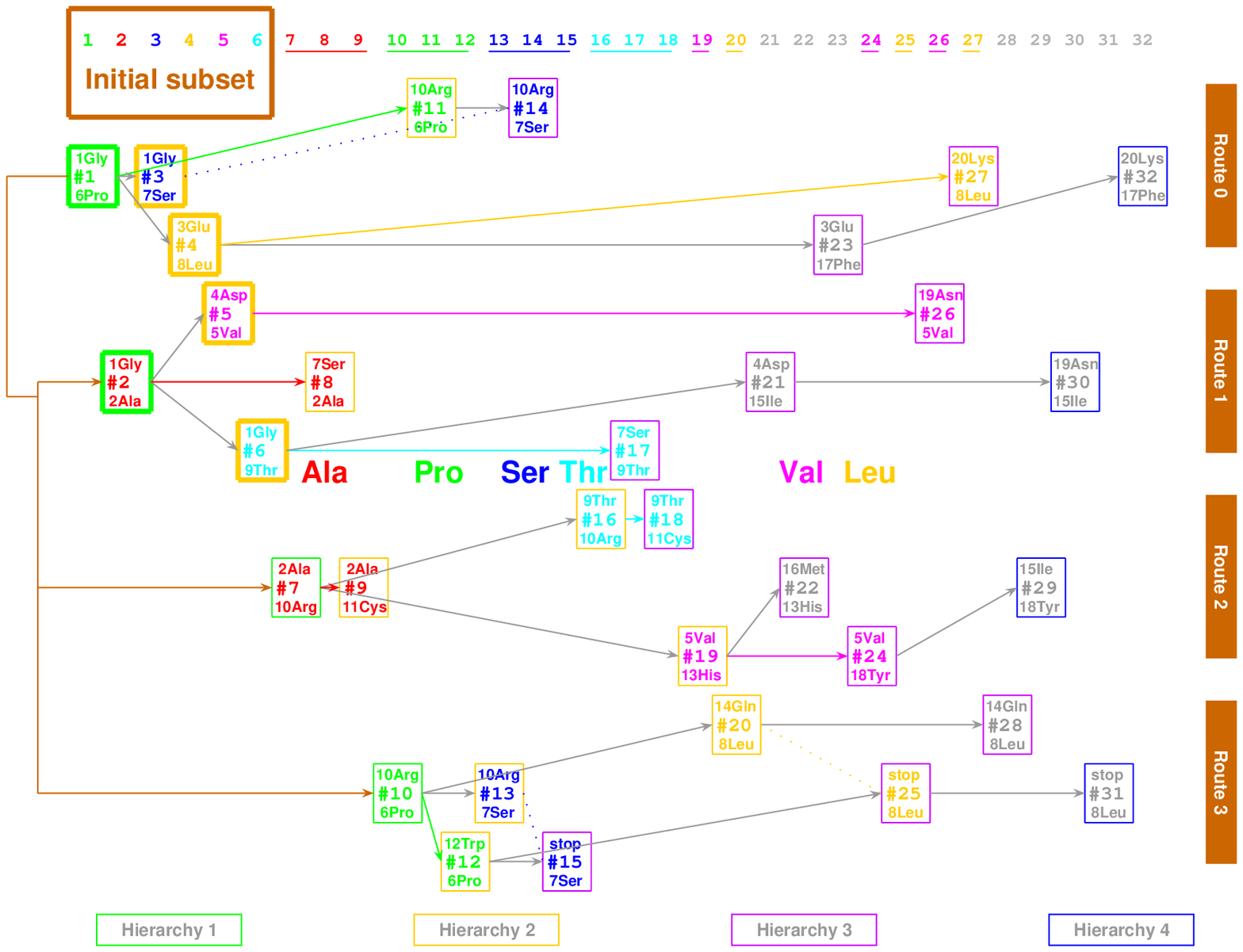}\\
  \vspace{1cm}{\small \bf a}
\end{figure}

\clearpage 
\begin{figure}
  \centering
  \includegraphics[width=18.3cm]{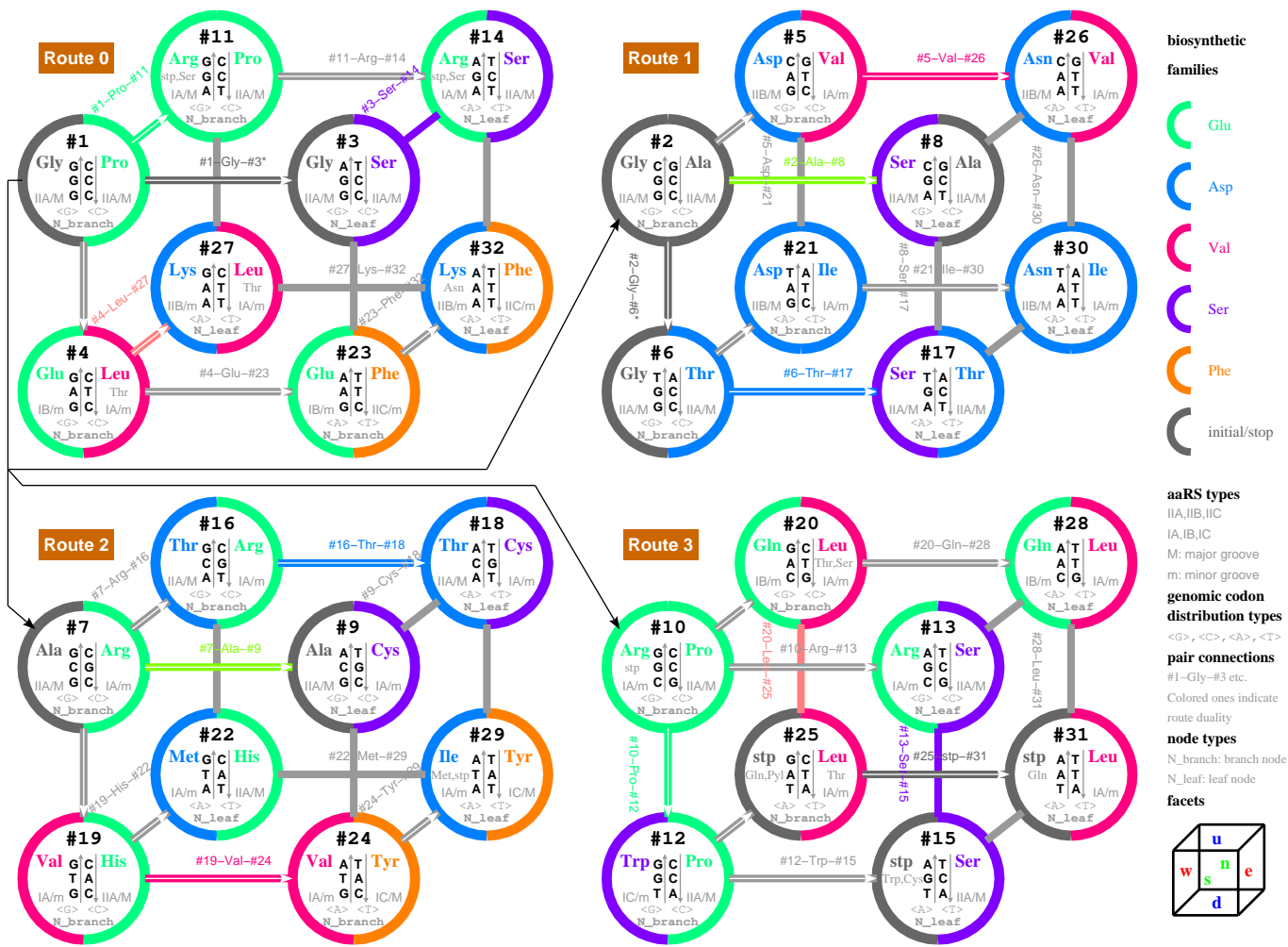}\\
  \vspace{1cm}{\small \bf b}
  \vspace{1cm}\caption{{\bf a} Cooperative recruitment of codons and amino acids. The codon pairs are plotted from left to right according to their recruitment order. The initial subset plays a crucial role in the expansion of the genetic code along the roadmap. The $6$ biosynthetic families of the amino acid are distinguished by different colours. {\bf b} The cubic roadmap. This is a revised plot of the roadmap Fig 1a to indicate the symmetry in the evolution of the genetic code, where the four routes are represented by four cubes respectively. Pair connections are marked besides the evolutionary arrows on the roadmap. Route dualities are indicated by same colours for the corresponding pair connections. The biosynthetic families of amino acids are denoted by coloured semicircles. The types of the aaRSs are besides the codons. Branch nodes and leaf nodes are distinguished. The $6$ facets for each route are indicated on a cube at bottom right.}
\end{figure}

\clearpage 
\begin{figure}
 \centering
 \includegraphics[width=9cm]{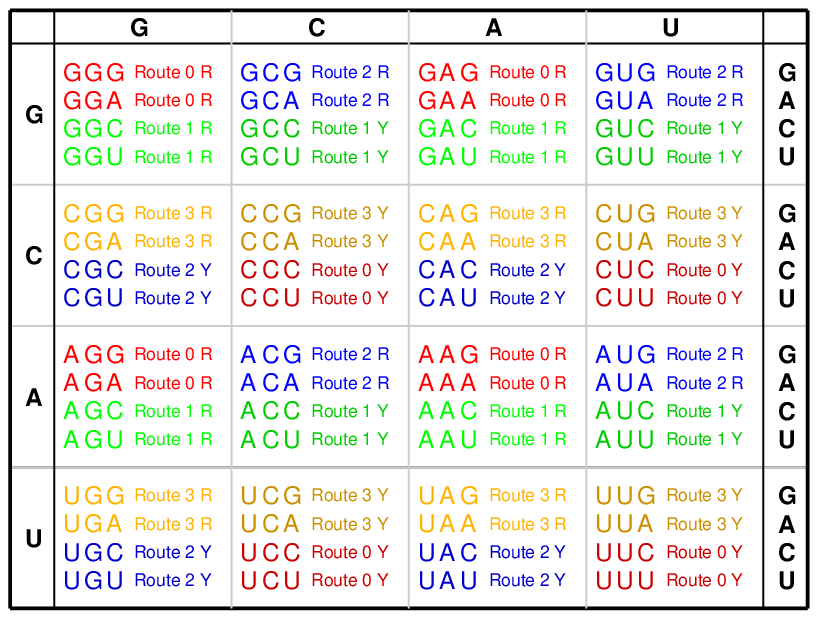}
 \includegraphics[width=9cm]{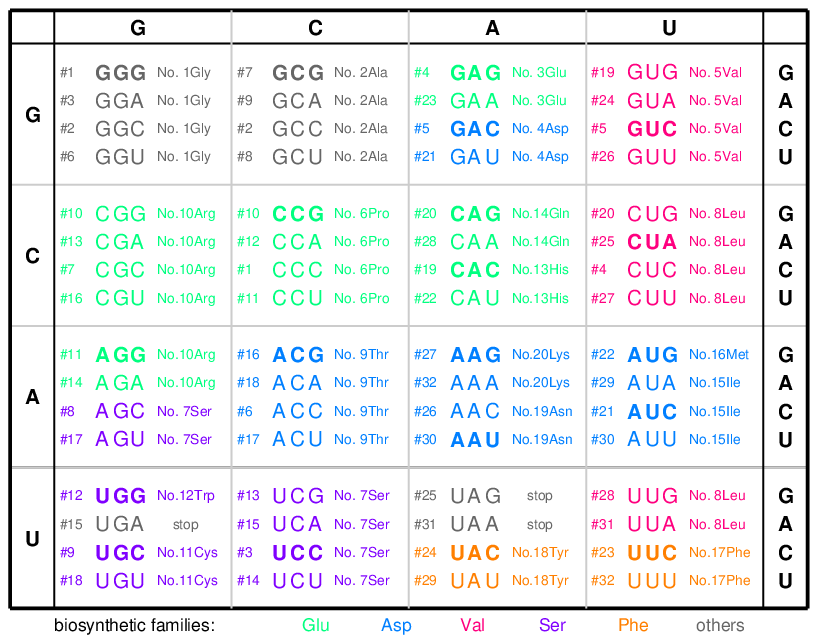}\\
 {\small \bf a}\hspace{9cm} {\small \bf b}\\
 \includegraphics[width=9cm]{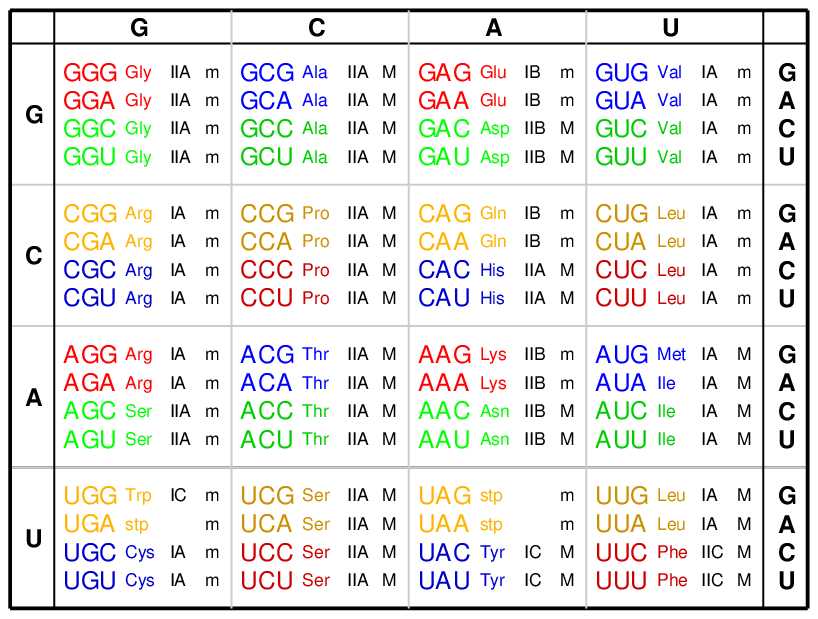}\\
 {\small \bf c}\\
 \vspace{1cm}\caption{{\bf a} The distribution of codons from R- and Y-strands of $Route\ 0-3$ in the $GCAU$ genetic code table. The pattern of the $4 \times 4$ codon boxes for the degenerate codons relates to such a distribution of the four routes, owing to the evolution of the genetic code along the roadmap. {\bf b} The $GCAU$ genetic code table. The clusterings of biosynthetic families (Glu, Asp, Val, Ser, Phe) in the $GCAU$ genetic code table. Such nice clusterings are correspondingly observed in the R- and Y-strands of $Route\ 0-3$ in Fig 3b (denoted in the same group of colour as in the present figure). The clusterings of biosynthetic families in the present figure are closely related to the distribution of codons from R- and Y-strands of $Route\ 0-3$, owing to the recruitment of amino acids along the roadmap. Generally speaking, the amino acids are arranged properly in the recruitment order from $No.1$ to $No.20$ along the direction from $G$, $C$ to $A$, $U$ in the $GCAU$ genetic code table. {\bf c} The distribution of types of aaRSs in the $GCAU$ genetic code table. The aaRSs can be divided into $Class\ II$ and $Class\ I$, which can be divided into subclasses $IIA$, $IIB$, $IIC$, and $IA$, $IB$, $IC$, respectively. And the aaRSs can also be divided into minor groove ones ($m$) and major groove ones ($M$).}
\end{figure}

\clearpage 
\begin{figure}
  \centering
  \includegraphics[width=18.3cm]{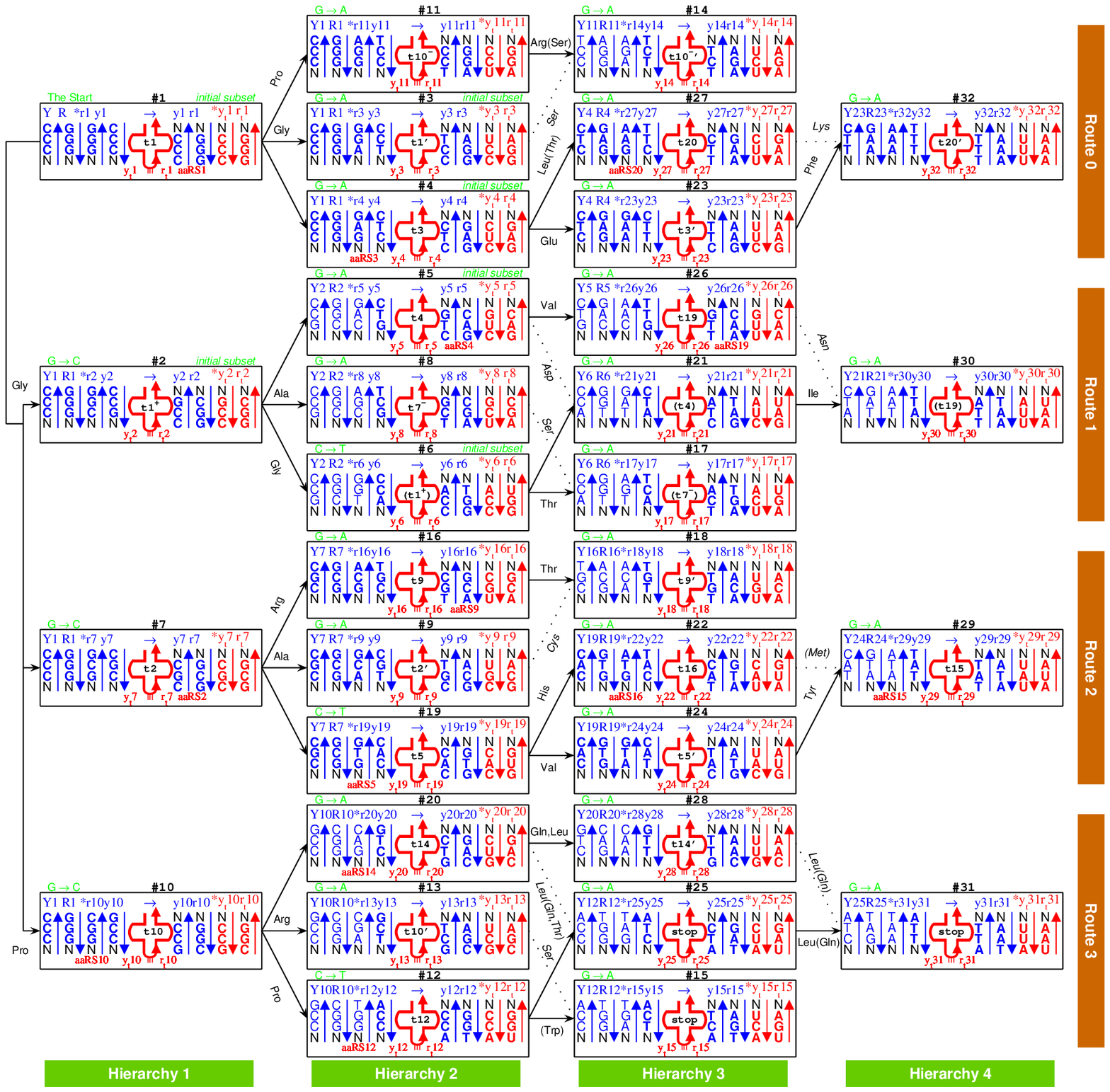}\\
  \vspace{1cm}{\small \bf a}
\end{figure}

\clearpage 
\begin{figure}
  \centering
  \includegraphics[width=18.3cm]{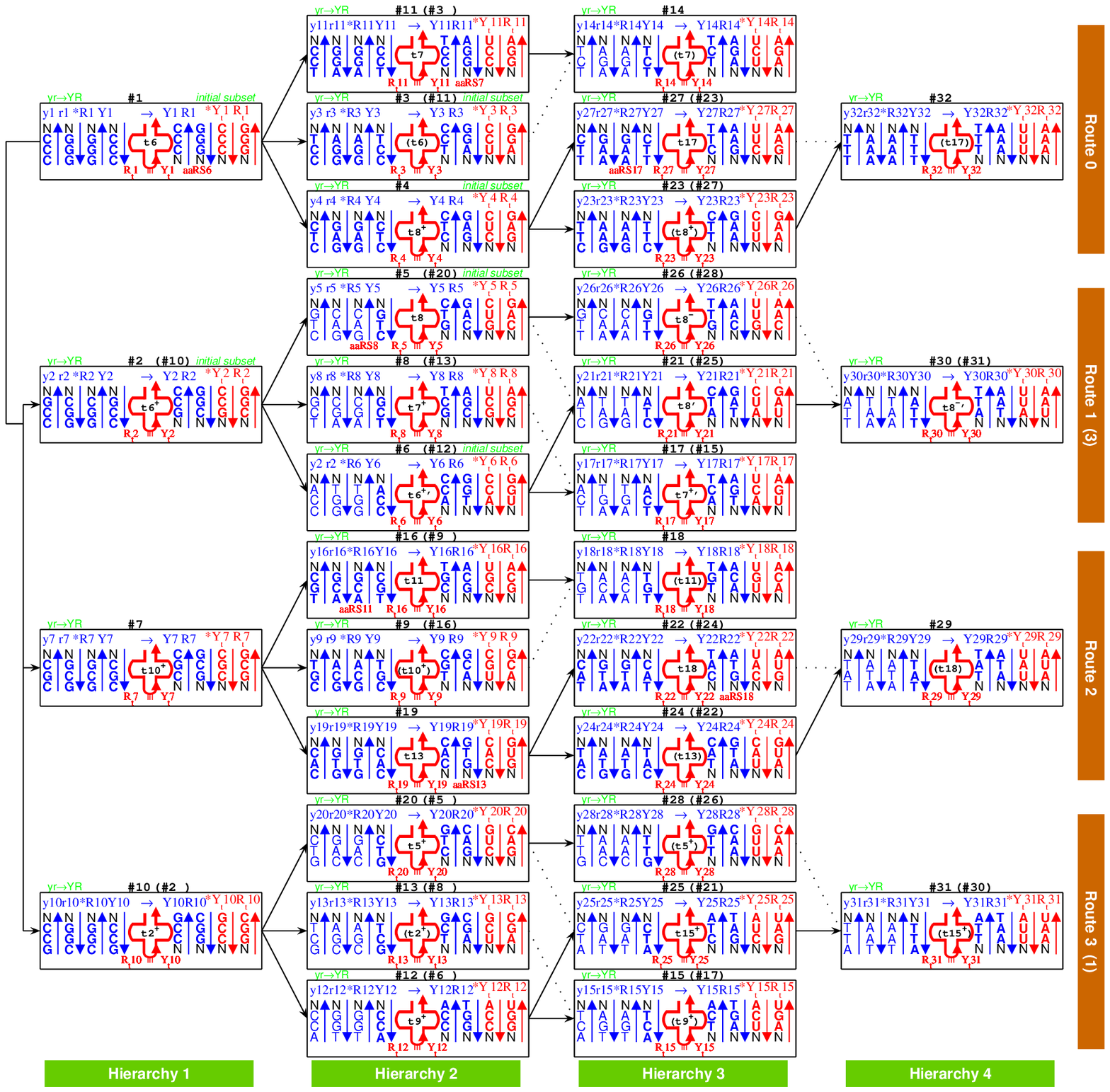}\\
  \vspace{1cm}{\small \bf b}
\end{figure}

\clearpage 
\begin{figure}
  \centering
  \includegraphics[width=18.3cm]{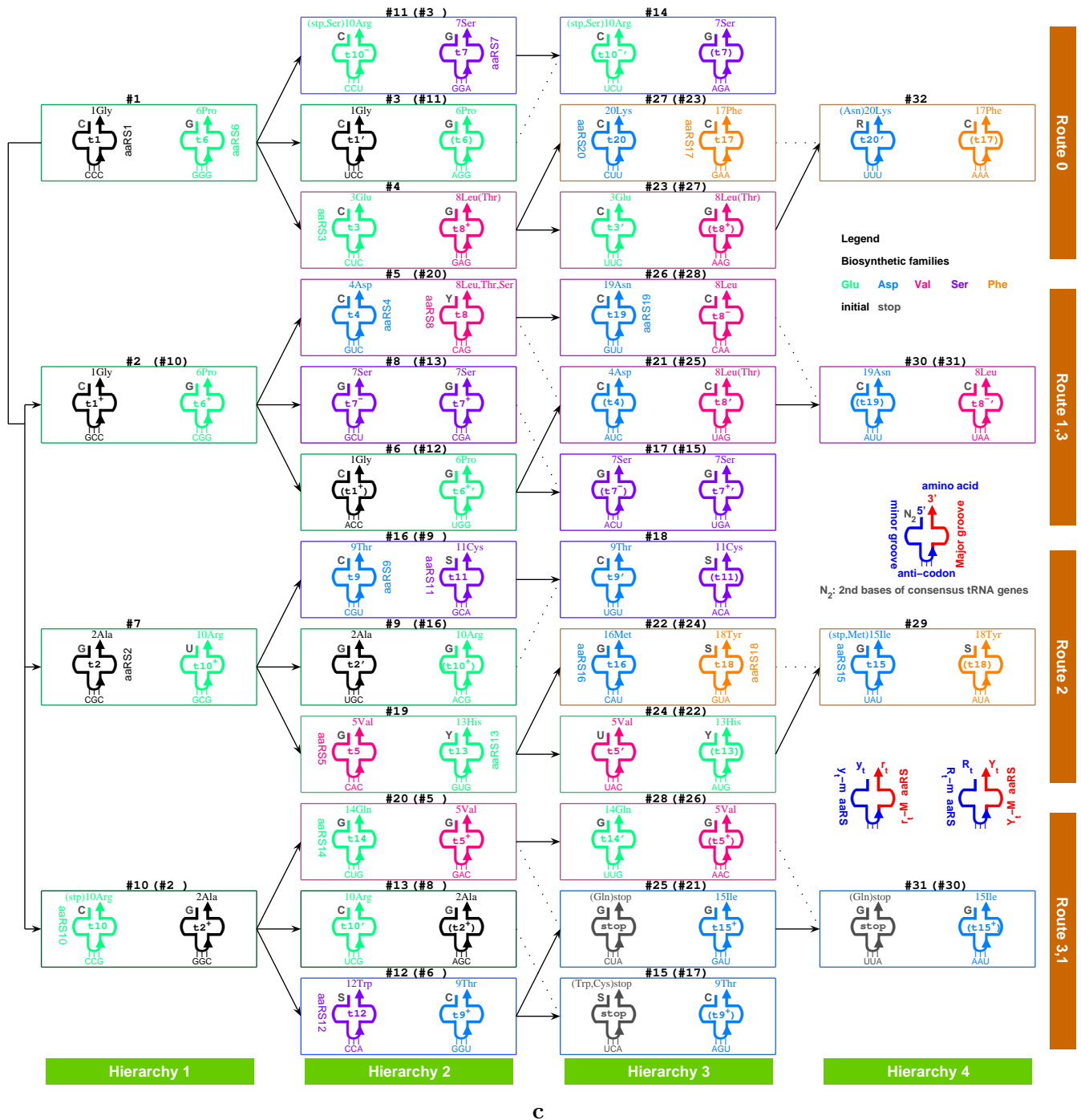}\\
  {\small \bf c}
  \caption{The origin and evolution of tRNAs along the roadmap. {\bf a} The evolution of the $5'y_tr_t3'$ type tRNAs by the triplex base pairings $yr*y_t$ and $yr*r_t$. {\bf b} The evolution of the $5'R_tY_t3'$ type tRNAs by the triplex base pairings $yr*R$, $yr*Y$ and $YR*Y_t$ and $YR*R_t$. The node numbers $\#n$ on the roadmap may exchange within or between routes because the sequences of $Y$ and $R$ are reverse to the sequences of $y$ and $r$ respectively. {\bf c} The coevolution of tRNAs with aaRSs along the roadmap, which determines the pair connections and route dualities. The aaRSs $aaRS1$ to $aaRS20$ combine respectively with the tRNAs $t1$ to $t20$ from certain major/minor groove side. The complementary relationship between the pyrimidine $y_t$ strand of the $5'y_tr_t3'$ type tRNAs and the purine $R_t$ strand of the $5'R_tY_t3'$ type tRNAs agrees with the complementary relationship between $G$ and $C$ for the second bases of the consensus genes of tRNAs especially for the early tRNAs in $Route\ 0$ and in $Hierarchy\ 1$.}
\end{figure}

\clearpage 
\begin{figure}
  \centering
  \includegraphics[width=16cm]{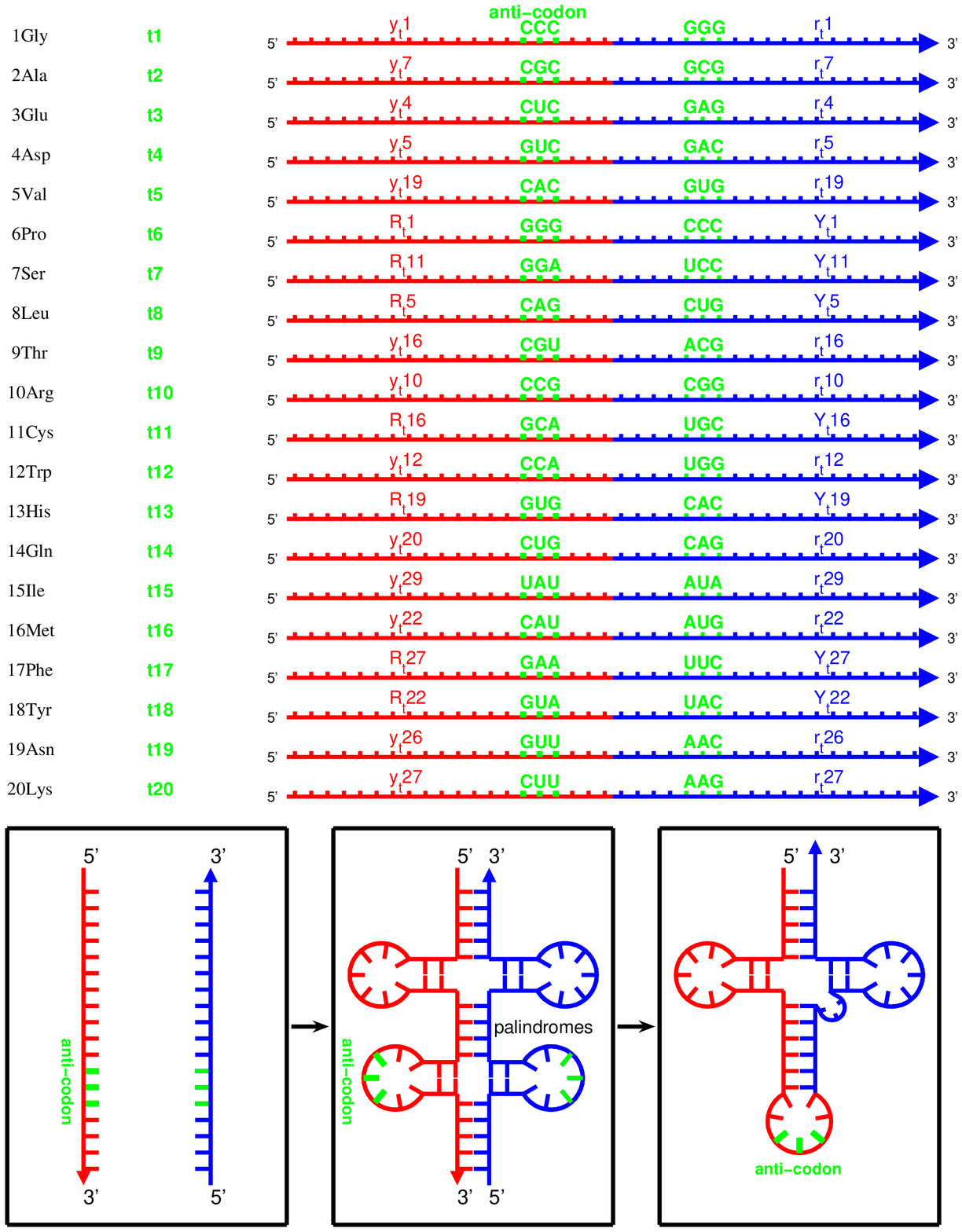}\\
  \vspace{1cm}{\small \bf a}
\end{figure}

\clearpage 
\begin{figure}
  \centering
  \includegraphics[width=18.3cm]{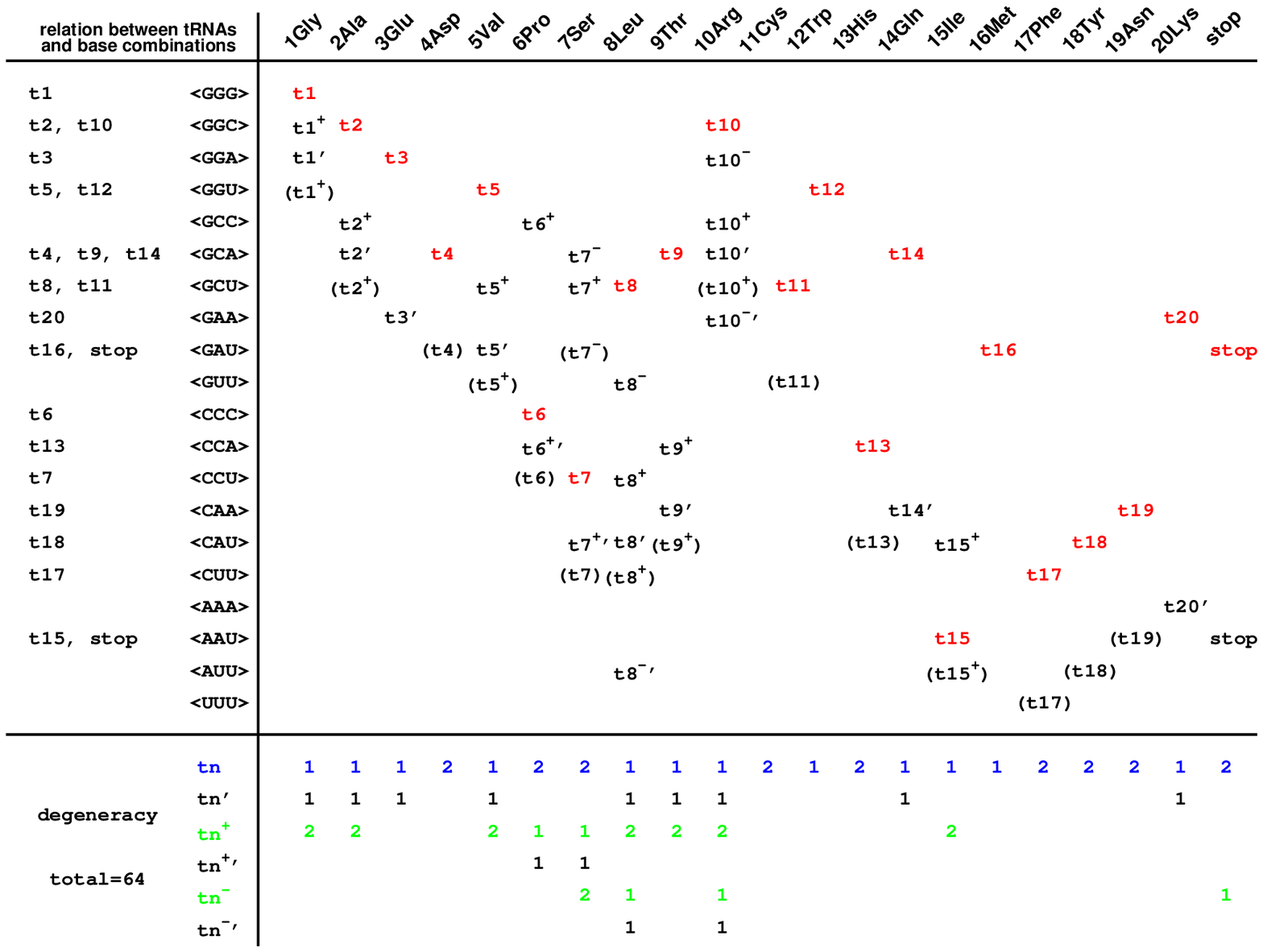}\\
  \vspace{0.8cm}{\small \bf b}
  \vspace{0.8cm}\caption{{\bf a} The assembly of tRNAs. The tRNAs $t1$-$t20$ with anti-codons (Fig 5c) are listed here to carry the amino acids from $No.1$ to $No.20$ respectively. The two complimentary single-stranded RNAs for each tRNA join together, and fold into a cloverleaf shape by taking advantage of the complementarity between the two strands. The joining position of the two strands is near to the 3' side of the anti-codon loop, which agrees with the position of introns in tRNA genes in observations. The anti-codons situate in the $3'$-ends of the $y_t$ strand or $R_t$ strand. The palindromic sequences tend to form loops of the tRNAs. And the para-codon of tRNA are non-palindromic or palindromic which adapt to the aaRSs (Fig 7, 8a). {\bf b} The cognate tRNAs. Explanation of the number of canonical amino acids as $20$ based on the relationship between the types of cognate tRNAs and the $20$ types of base combinations. The primer tRNAs generally appeared earlier than the derivative tRNAs. The primer tRNAs generally distribute along the diagonal line due to the chronological arrangements for both the $20$ amino acids and the $20$ base combinations, considering the substitution order $G$, $C$, $A$, $U$ along the roadmap. The codon degeneracies $6$, $4$, $3$, $2$ and $1$ are due to the tRNA evolution from $tn$ to $tn^+$ and $tn^-$, as well as from $tn$ to $tn'$ etc, all of which can be recognised by the corresponding $aaRSn$. }
\end{figure}

\clearpage 
\begin{figure}
 \centering
 \includegraphics[width=18.3cm]{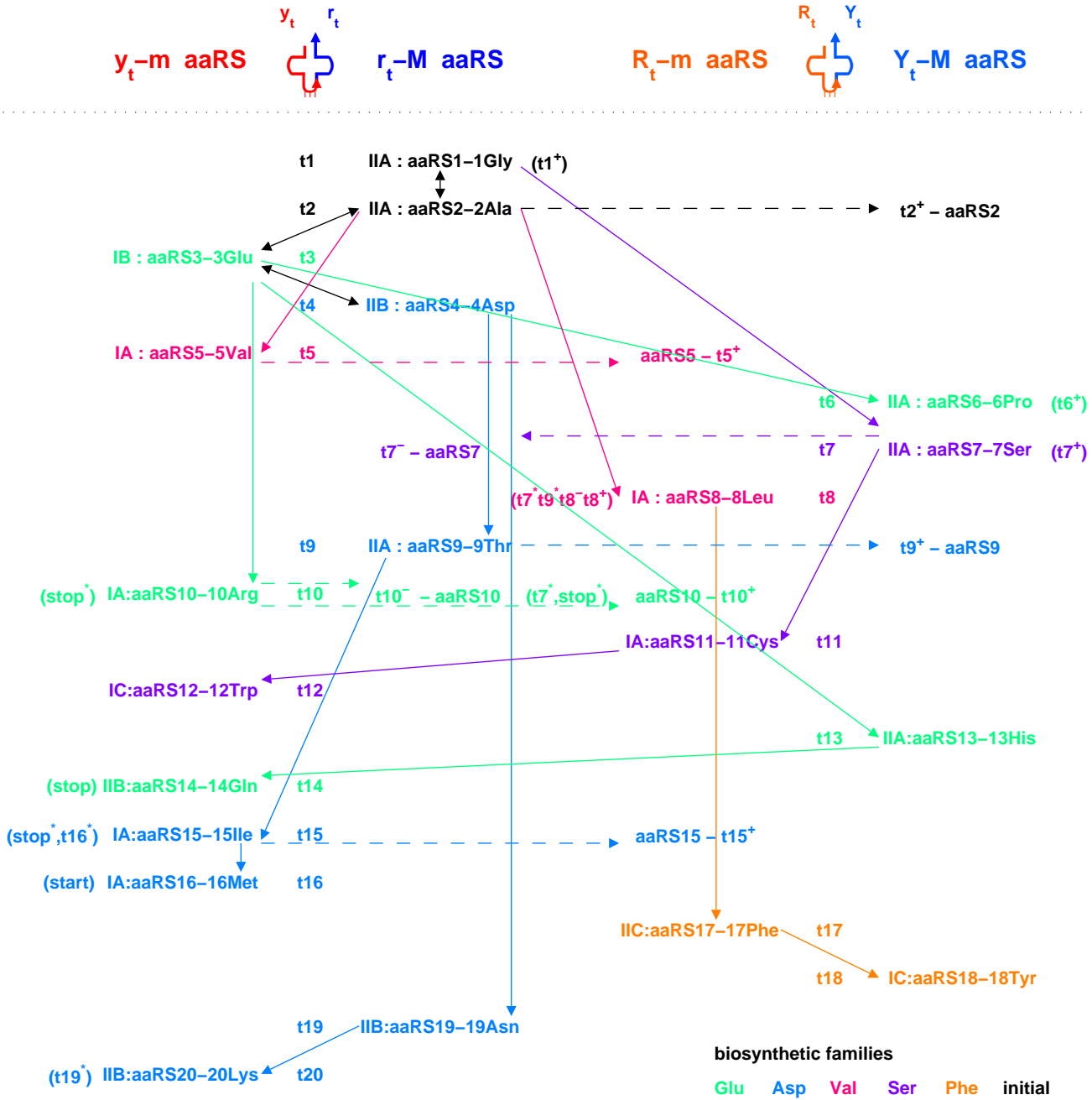}
 \caption{The coevolution of tRNAs with aaRSs. The coevolution of the four classes of aaRSs and the corresponding two types of tRNAs in accordance with the biosynthetic families indicated in certain colours. The ancestor of aaRS, namely $aaRS1$ corresponding to the non-chiral amino acid $1Gly$, belongs to the $r_t-M$ class. The codon degeneracy are due to the coevolution of tRNAs with aaRSs, where the surplus tRNAs were chosen by the rare aaRSs. There are some truths in the traditional classifications of aaRSs, but the evolutionary relationships of aaRSs are so intricate, as shown here. The start and stop codons generally appear in the positions corresponding to $y_t-m$ class. The non-standard codons also evolved as alternative choices of tRNAs by aaRSs.}
\end{figure}

\clearpage 
\begin{figure}
 \centering
 \includegraphics[width=18.3cm]{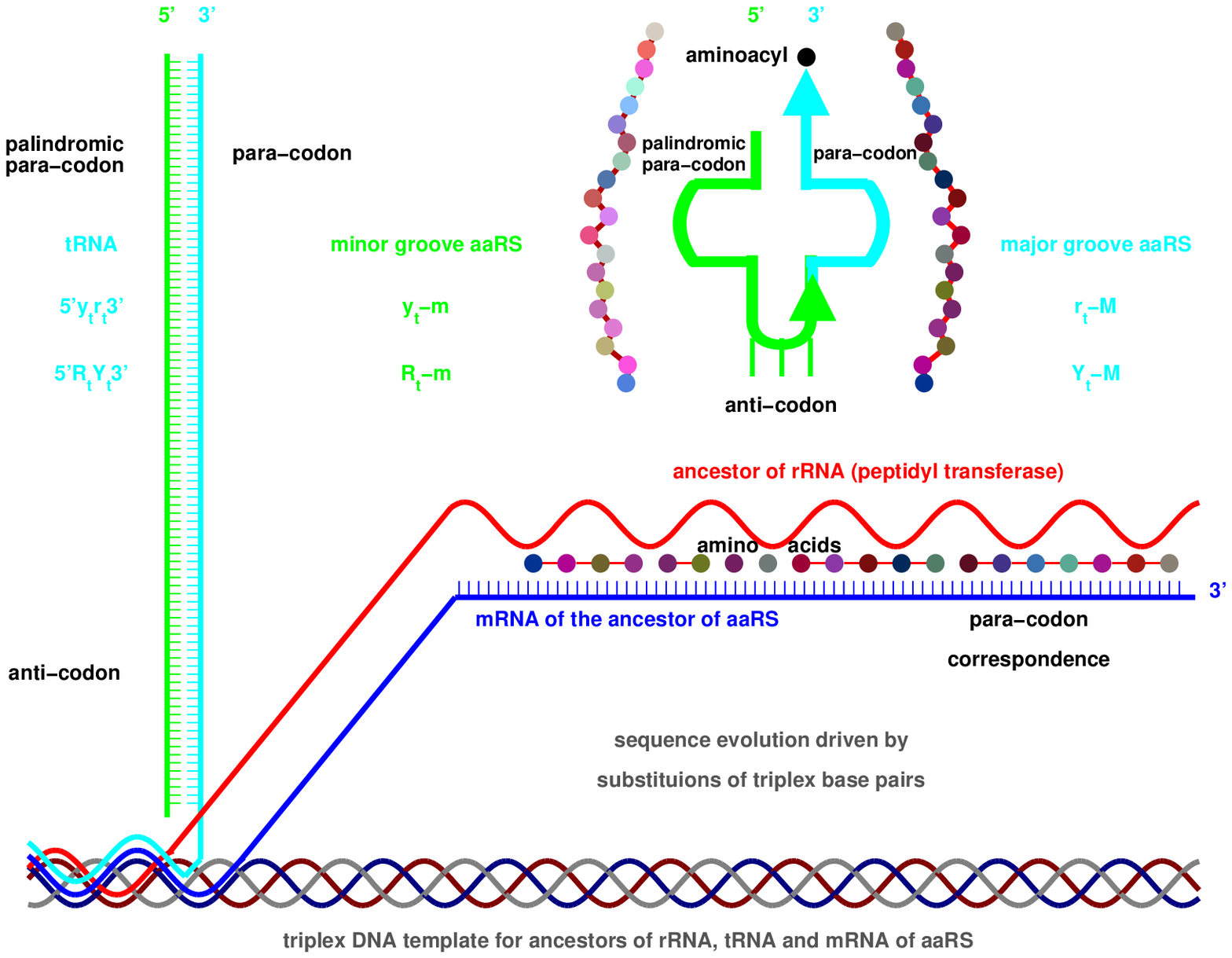}\\
  \vspace{1cm}{\small \bf a}
\end{figure}

\clearpage 
\begin{figure}
 \centering
  \includegraphics[width=18.3cm]{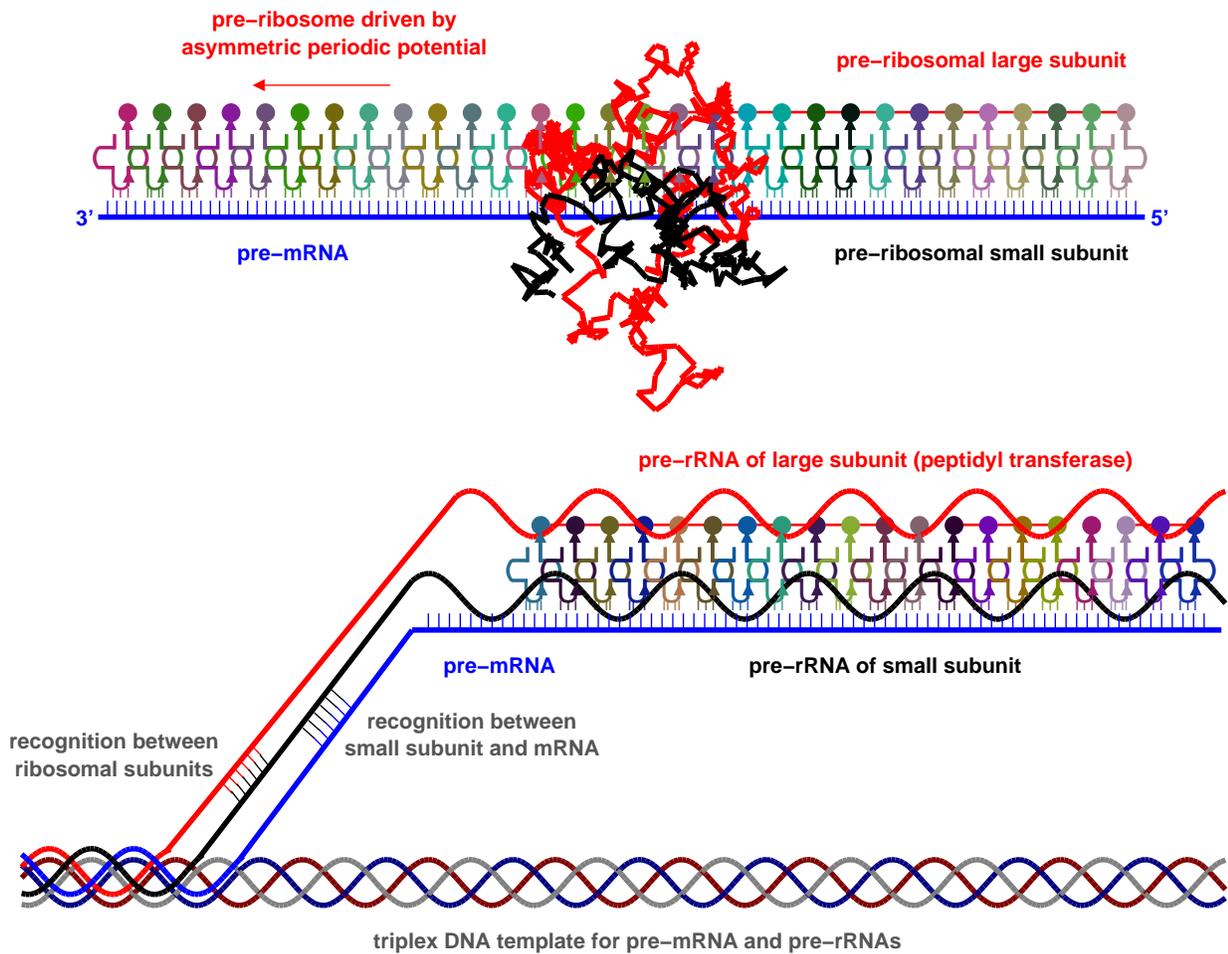}\\
  \vspace{1cm}{\small \bf b}
  \vspace{1cm}\caption{The origin and evolution of the translation mechanism in the triplex2duplex picture. {\bf a} The origin and evolution of four classes of early aaRSs in the junior stage of the primordial translation mechanism in absent of tRNA and ribosome. The first aaRS can be produced through the non-random evolution of the triplex DNA and the corresponding RNAs. At the beginning of the translation mechanism, DNAs are the carrier of information and RNAs develop the functions of life. {\bf b} The origin of ribosome in the senior stage of the primordial translation mechanism. The tRNA, rRNA and mRNA participated in the invention of longer proteins that promoted the evolution of the translation mechanism. The early pre-ribosome can move along the mRNA in absent of elongation factors. The efficiency for producing proteins increased step by step with the evolution of the translation mechanism. The interactions and coevolution among the ribosomal subunits and the tRNAs and mRNAs, as well as the proteins as translation factors, can be explained by the non-random sequence evolution in the triplex2duplex picture.}
\end{figure}

\clearpage
\begin{figure}
 \centering
 \includegraphics[width=12cm]{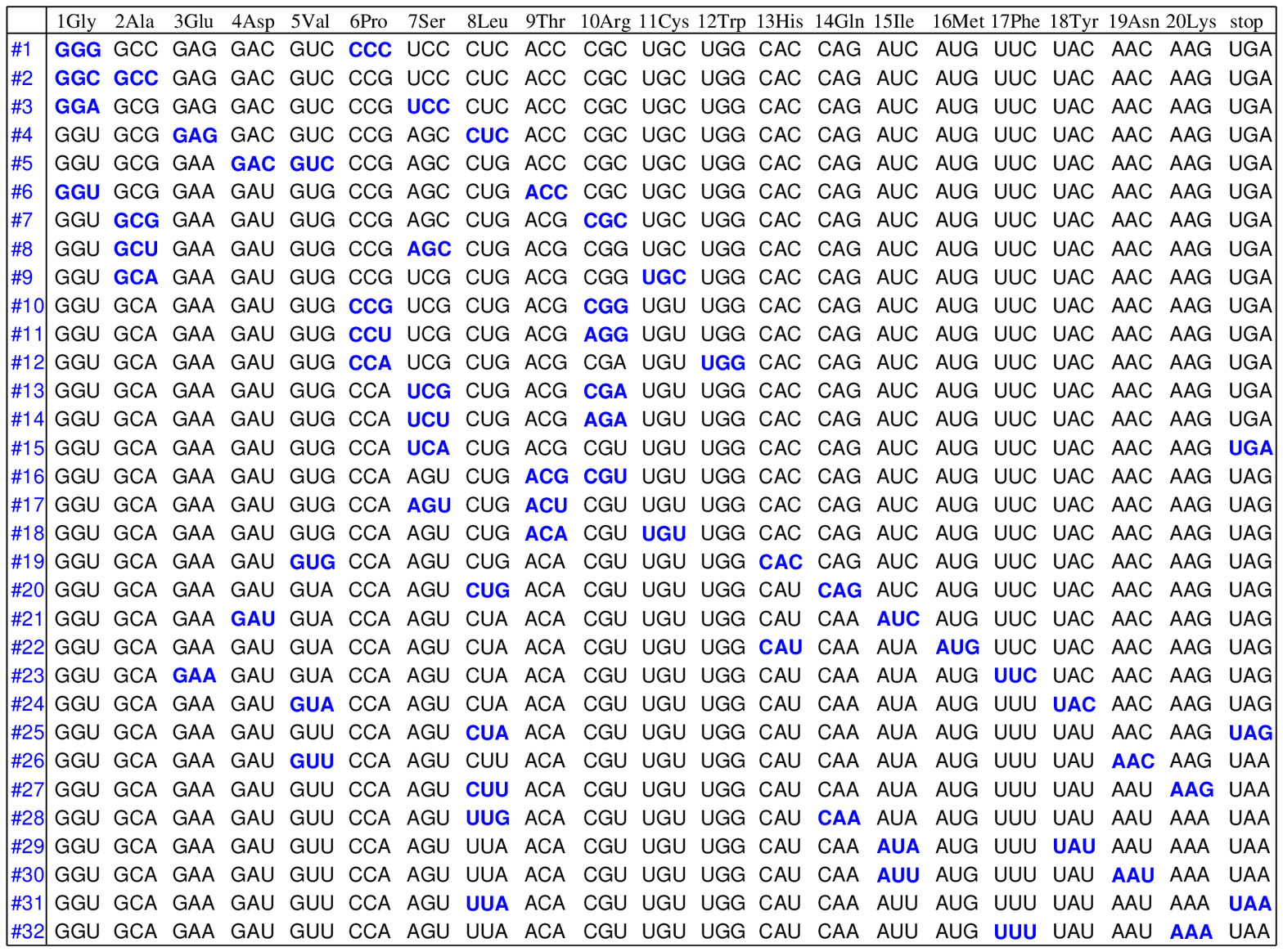}\\
 {\small \bf a}\\
 \includegraphics[width=9cm]{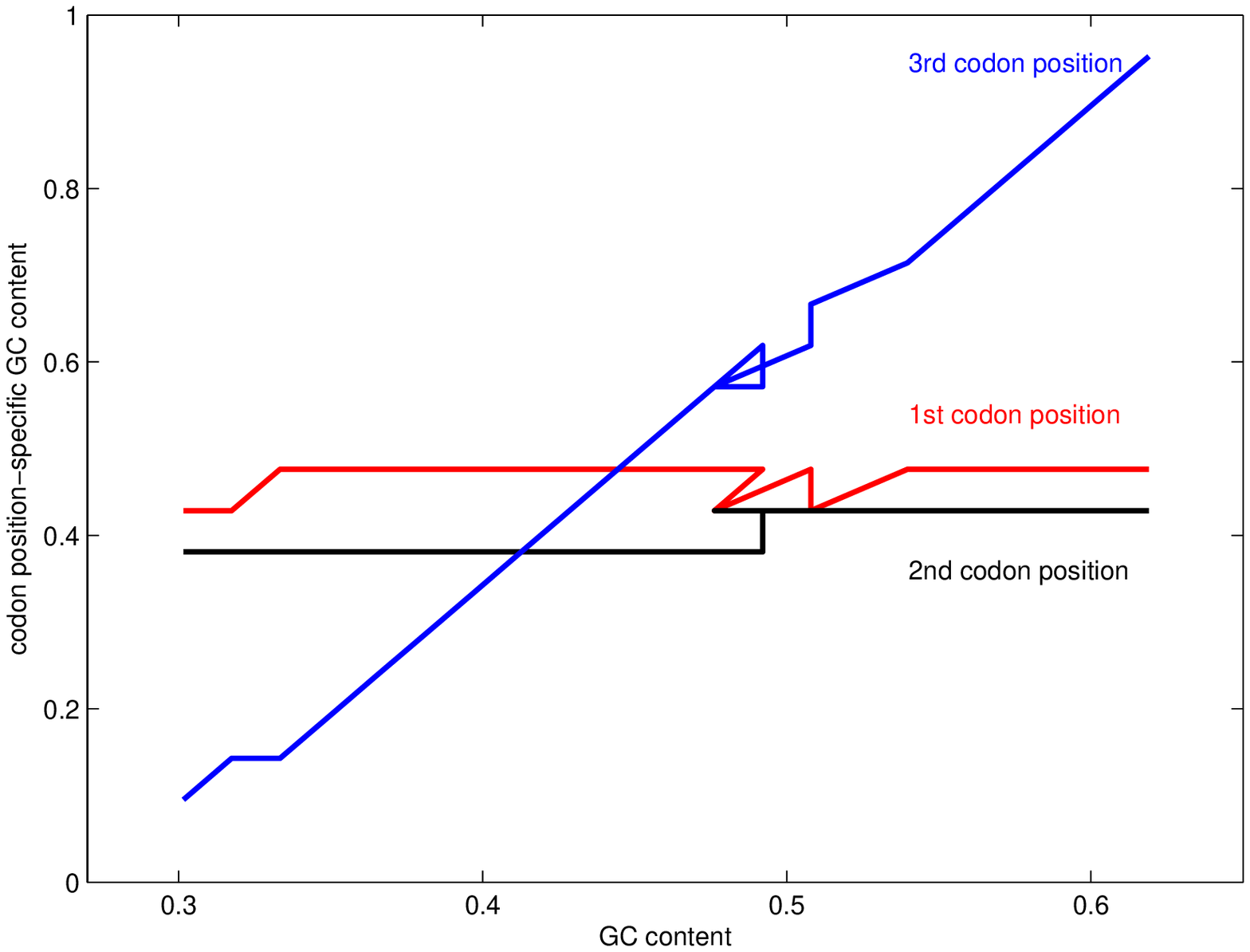}
 \includegraphics[width=9cm]{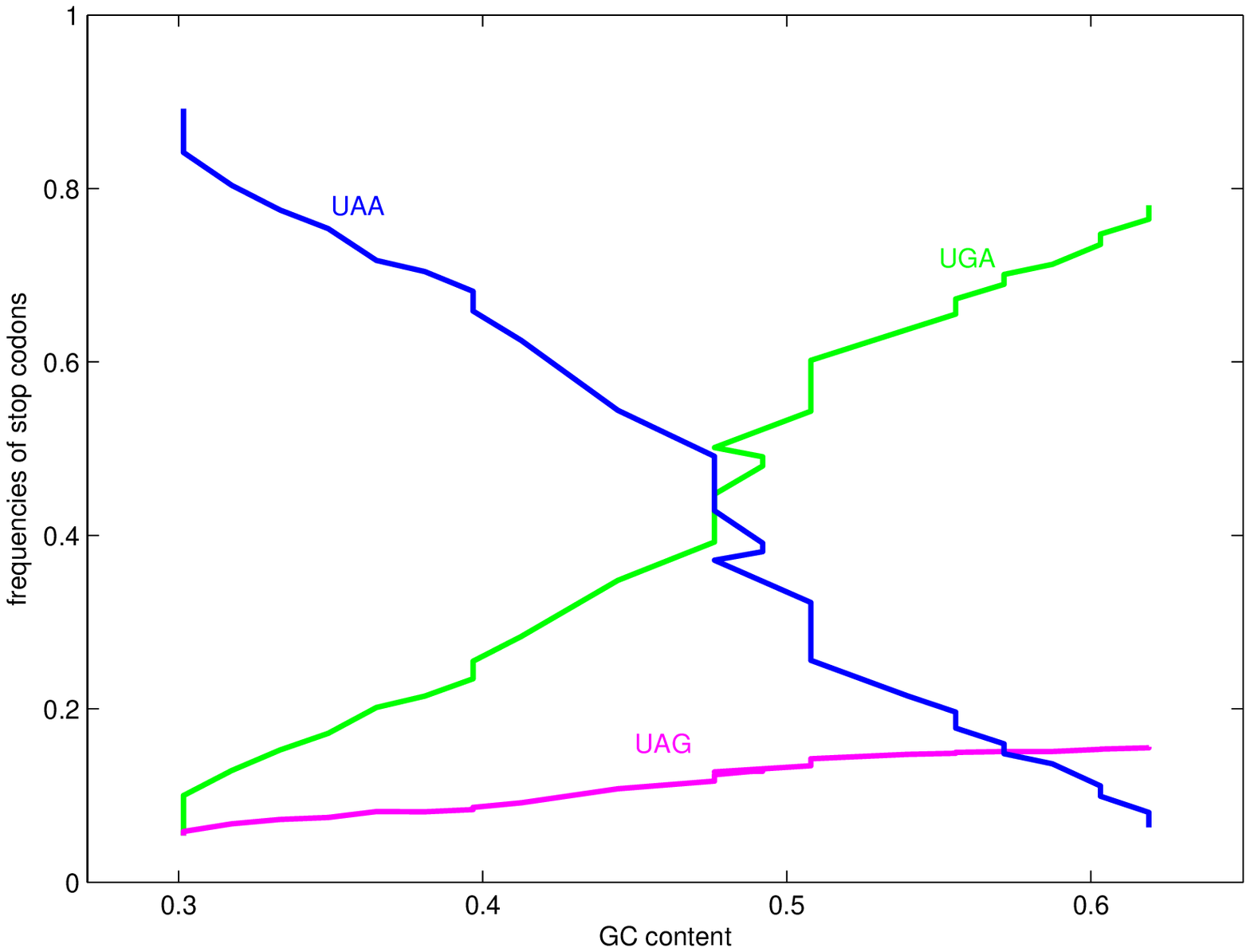}\\
 {\small \bf b}\hspace{9cm} {\small \bf c}\\
 \includegraphics[width=9cm]{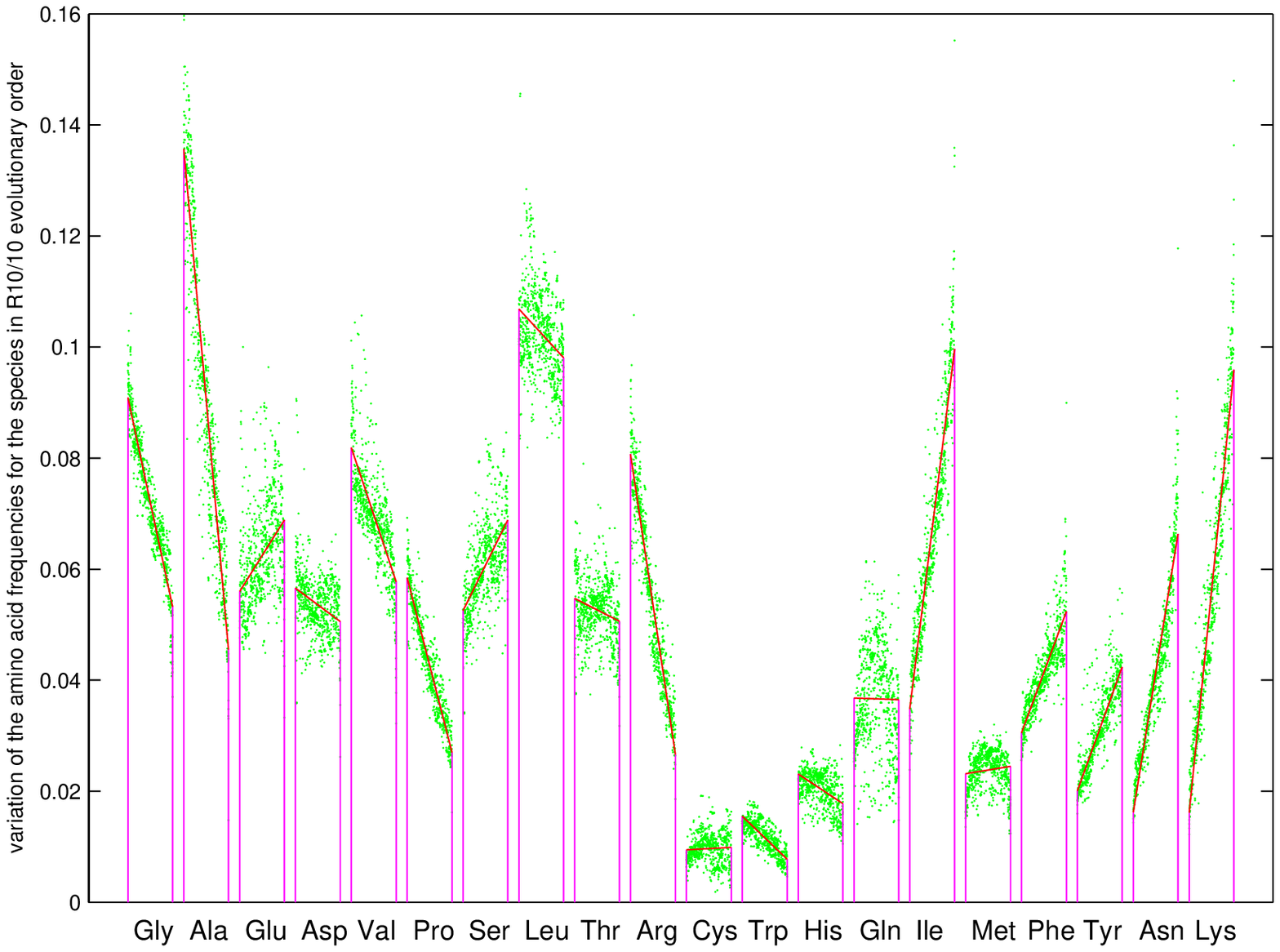}
 \includegraphics[width=9cm]{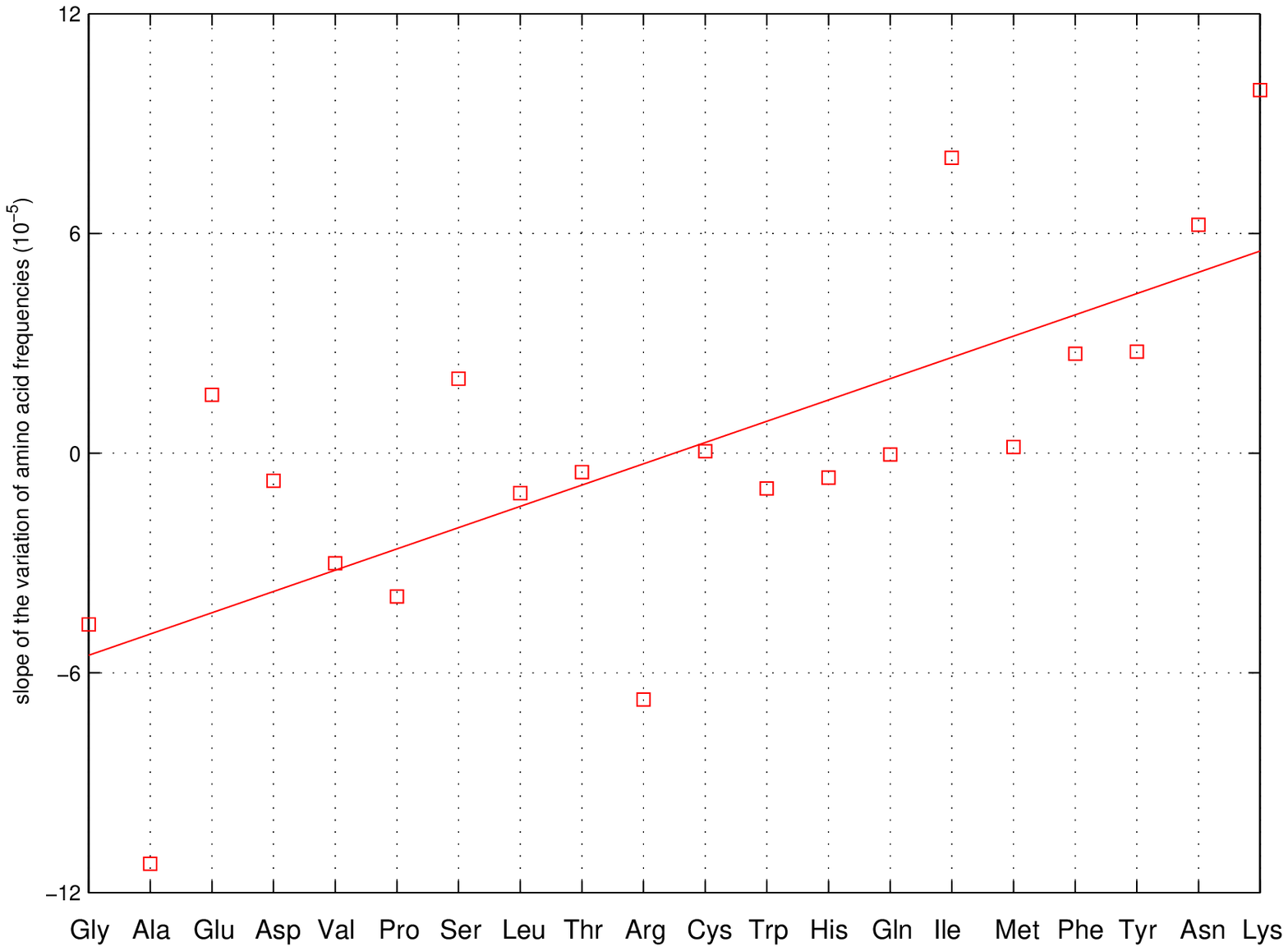}\\
 {\small \bf d}\hspace{9cm} {\small \bf e}\\
\end{figure}

\clearpage 
\begin{figure}
 \centering
 \caption{}
\end{figure}

{The recruitment orders of amino acids and codon pairs on the roadmap are supported by genomic data. The agreement between observations and predictions support  the roadmap theory. {\bf a} The codon pairs and the amino acids are listed in the recruitment orders on the roadmap. The codons that encode the respective amino acids are written in the corresponding stages from $\#1$ to $\#32$ (blue codons), and the remaining positions are filled by the nearby codons respectively. {\bf b} The variation pattern of the codon position GC content with respect to the total GC content can be explained according to the orders of degenerate codons in Fig 9a. Even some detailed features in observation are reproduced in the present figure based on the roadmap Fig 1a, such as the closer distance between 1st and 2nd codon position GC content at low total GC content side. {\bf c} It is observed that the stop codons vary in certain pattern with respect to the total GC content. When the total GC content decreasing, the first stop codon $UGA$ on the roadmap decrease greatly, the second codon $UAG$ on the roadmap decrease slightly, and the third stop codon $UAA$ increase greatly. Such a pattern in observations can be explained by the roadmap. The relationships between the total GC content and the frequencies of the three stop codons are obtained based on the roadmap Fig 1a, respectively, by counting and calculating similarly in Fig 9b, where the variation ranges for the three stop codons have been adjusted according to the observations. A detailed feature of downward and upward leaps of $UGA$ and $UAA$ at the middle total GC content, is reproduced based on the counts in Fig 9a. {\bf d} Explanation of the variation of the amino acid frequencies. The $20$ amino acids are arranged in the recruitment order on the roadmap Fig 1a. The $20$ amino acid frequencies for each of the $803$ species are obtained respectively based on the genomic data in NCBI. And the $803$ amino acid frequencies (green dots) for each of the $20$ amino acids are all arranged properly in the $R_{10/10}$ order, respectively. The variation trend of the amino acid frequencies for each of the $20$ amino acids is obtained by the regression line (denoted in red). Generally speaking, the variation trends for the earlier amino acids tend to decrease, and the variation trends for the latecomers to increase (Fig 9d). {\bf e} Increasing variation rates of the amino acid frequencies. The slope of the regression line of variation of amino acid frequencies for each amino acid is obtained based on the data in Fig 9d. It is observed that the slopes generally vary from negative to positive according to the recruitment order of amino acids on the roadmap. Such a natural result indicates that the recruitment order of the amino acids from $No.1$ to $No.20$ obtained based on the roadmap is considerably reasonable.}

\end{document}